%% file: main.tex
\documentclass[journal,onecolumn]{IEEEtran}
\IEEEoverridecommandlockouts
\usepackage{stmaryrd}
\usepackage[utf8]{inputenc}
\usepackage{amsthm}
\usepackage{amsfonts,amssymb,latexsym,cite}
\usepackage[cmex10]{amsmath}
\usepackage{algorithmic}
\usepackage{array}
\usepackage{bbm}
\usepackage{verbatim}
\usepackage{color,xcolor}
\usepackage{graphicx}
\usepackage{epstopdf}
\usepackage{subcaption}
\usepackage[font={small,it}]{caption}
\usepackage[T1]{fontenc}
\usepackage{url}
\usepackage{enumerate}
\usepackage{enumitem}
\usepackage{hyperref} % ref online resource
\graphicspath{{./figures/}} % location of graphs
\usepackage{caption}
\usepackage[blocks]{authblk}
\usepackage{epsfig}
\usepackage{latexsym}
\usepackage{soul}
\usepackage{thmtools,thm-restate}
\usepackage[margin=1in]{geometry}
\usepackage{etex}
\usepackage{tikz}
\usetikzlibrary{arrows,arrows.meta,automata,shapes,calc,intersections}
\usepackage{pgfplots}
\pgfplotsset{compat=newest}

\newtheorem{definition}{Definition}
\newtheorem{theorem}{Theorem}

\newtheorem{lemma}{Lemma}
\newtheorem{claim}{Claim}

\newtheorem{remark}{Remark}
\allowdisplaybreaks
\newcommand{\upperRomannumeral}[1]{\uppercase\expandafter{\romannumeral#1}}

\newcommand{\C}{\mathcal{C}} 
\newcommand{\T}{T_0} 
\newcommand{\Ts}{T_0^{\ast}(\mathcal{S}_W)}
\newcommand{\Tsb}{T_0^{\ast}(\mathcal{S}_W,\beta)}  
\newcommand{\Tsbt}{\widehat{T}_0^{\ast}(\mathcal{S}_W,\beta, T)}  
\newcommand{\Tst}{\widehat{T}_0^{\ast}(\mathcal{S}_W,T)}  
\newcommand{\B}{\beta}
\newcommand{\n}{n}
\newcommand{\N}{M}
\newcommand{\I}{\mathbb{I}}
\newcommand{\m}{m}
\newcommand{\Pm}{P_{\text{min}}}
\newcommand{\an}{a_n}
\newcommand{\K}{i}

\newcommand{\dtsi}{D_T[\psi_i(t)]}
\newcommand{\s}{S}
\newcommand{\lx}{||\mathbf{x}||_2^2}
\newcommand{\nun}{\nu_n}
\newcommand{\V}{\mathbb{V}}

\newcommand{\x}{\mathbf{x}}
\newcommand{\X}{\mathbf{X}}
\newcommand{\Px}{P_X}
\newcommand{\Dd}{\mathcal{D}_d}
\newcommand{\PP}{\mathbb{P}}
\newcommand{\Pxn}{P_X^{\otimes n}}
\newcommand{\xt}{x(t)}
\newcommand{\Qt}{\widetilde{Q}}
\newcommand{\Qtn}{\widetilde{Q}^{\otimes n}}
\newcommand{\Qa}{Q_a}

\newcommand{\Q}{Q_0}
\newcommand{\Qh}{\widehat{Q}}
\newcommand{\Qn}{Q_0^{\otimes n}}
\newcommand{\fn}{\sqrt{n}}
\newcommand{\h}{\mathbb{H}}
\newcommand{\Hb}{\mathbb{H}_b}
\newcommand{\Pt}{\widetilde{P}}

\newcommand{\Ptn}{\widetilde{P}^{\otimes n}}
\newcommand{\Pa}{P_a}

\newcommand{\Pn}{P_0^{\otimes n}}
\newcommand{\NN}{N}

\newcommand{\wzx}{W_{Z|X}}
\newcommand{\wzxn}{W_{Z|X}^{\otimes n}}
\newcommand{\wyx}{W_{Y|X}}
\newcommand{\wyxn}{W_{Y|X}^{\otimes n}}
\newcommand{\xmt}{x_{ms}(t)}
\newcommand{\Xt}{X(t)}
\newcommand{\hxf}{\widehat{x}(f)}
\newcommand{\jj}{\textup{j}}
\newcommand{\E}{\mathbb{E}}
\newcommand{\D}{\mathbb{D}}
\newcommand{\al}{\alpha}
\newcommand{\U}{U}
\newcommand{\code}{\mathcal{C}}
\newcommand{\phit}{\phi_{\T,\B}(t)}
\newcommand{\phif}{\widehat{\phi}_{\T,\B}(f)}
\newcommand{\phinf}{\widehat{\phi}_{\T,\B}}
\newcommand{\Ltwo}{\mathcal{L}_2}
\newcommand{\bwxt}{B_{\alpha_l W}[x(t)]}
\newcommand{\Y}{\mathbf{Y}}
\newcommand{\y}{\mathbf{y}}
\newcommand{\Z}{\mathbf{Z}}
\newcommand{\z}{\mathbf{z}}
\newcommand{\II}{\mathbb{I}}
\newcommand{\Zc}{\mathbf{Z}^c}
\newcommand{\zc}{\mathbf{z}^c}
\newcommand{\Zl}{\bar{\mathbf{Z}}^d}
\newcommand{\zl}{\bar{\mathbf{z}}^d}
\newcommand{\Aset}{\mathcal{A}_{\gamma}^n}

\newcommand{\ez}[1]{{\color{red}#1}}

\begin{document}
\title{Undetectable Radios: Covert Communication under Spectral Mask Constraints}

\author{Qiaosheng (Eric) Zhang, Matthieu R. Bloch,~\IEEEmembership{Senior~Member,~IEEE,}
        Mayank Bakshi,~\IEEEmembership{Member,~IEEE,}
        Sidharth Jaggi,~\IEEEmembership{Senior~Member,~IEEE}% <-this % stops a space
\thanks{Qiaosheng (Eric) Zhang is with the Department
of Electrical and Computer Engineering, National University of Singapore (e-mail: elezqiao@nus.edu.sg). Matthieu R. Bloch is with the School of Electrical and Computer Engineering, Georgia Institute of Technology (e-mail: matthieu.bloch@ece.gatech.edu). Mayank Bakshi and Sidharth Jaggi are with the Department of Information Engineering, Chinese University of Hong Kong (e-mail: mayank@inc.cuhk.edu.hk, jaggi@ie.cuhk.edu.hk).}
\thanks{This work was supported by NSF under
Grant 1527387. A preliminary version of this work was presented at the 2019 IEEE International Symposium on Information Theory (ISIT), Paris, France.}% <-this % stops a space stops a space
}

\date{}
\maketitle

\begin{abstract}
We consider the problem of covert communication over continuous-time additive white Gaussian noise (AWGN) channels under spectral mask constraints, wherein two legitimate parties attempt to communicate reliably in the presence of an eavesdropper that should be unable to estimate if communication takes place. %In addition to requiring the legitimate receiver to reliably decode, covert communication also requires that the warden is unable to estimate whether or not communication is taking place.
    The spectral mask constraint is imposed to restrict excessive radiation beyond the bandwidth of interest. We develop a communication scheme with theoretical reliability and covertness guarantees based on pulse amplitude modulation (PAM) with Binary Phase Shift Keying (BPSK) and root raised cosine (RRC) carrier pulses. Given a fixed transmission duration $T$ and a spectral mask with bandwidth parameter $W$, we show that one can reliably and covertly transmit $\mathcal{O}(\sqrt{WT})$ bits of information. We characterize the constant behind the $\mathcal{O}$ and show that it is tight under some conditions.% we characterize the constant  and our proposed scheme provides a lower bound on the covert capacity 
\end{abstract}
%\blfootnote{\ez{An up-to-date version can be found at \url{http://personal.ie.cuhk.edu.hk/~zq015/Covert2017_ZBJ.pdf}}}
%Our proposed scheme is in particular of practical interests, and may impact on the design of covert communication systems.


\setcounter{page}{1}

\section{Introduction}
Covert communication considers the scenario in which a transmitter, Alice, wishes to reliably communicate with a legitimate receiver, Bob, while simultaneously hiding the presence of communication from an eavesdropping adversary, referred to as the warden, Willie. %is taking place from a warden, Willie.
Building upon the formalization of the problem in~\cite{bash2013limits}, subsequent studies have progressively refined the information-theoretic analysis of covert communication by considering binary symmetric channels~\cite{CheBJ:13}, discrete memoryless channels (DMCs) and additive white Gaussian noise (AWGN) channels~\cite{7407378,wang2016fundamental,tahmasbi2018first}, multiple-access channels~\cite{arumugam2018covert}, broadcast channels~\cite{arumugam2019embedding,tan2018time,kibloff2019embedding}, channels with state~\cite{lee2018covert, zivari2019keyless}, compound channels~\cite{compound}, and adversarially jammed channels~\cite{zhang2018covert}. The aforementioned works have not only identified a \emph{square-root law} (SRL) for covert communication --- only $\mathcal{O}(\sqrt{n})$ bits of information can be transmitted reliably and covertly over $n$ channel uses --- but also characterized the constant behind the $\mathcal{O}$, which can be interpreted as the \emph{covert capacity}.

While most covert communication studies have focused on discrete-time models, analyzing continuous-time models is legitimate to ensure that all relevant engineering aspects got captured. Perhaps not surprisingly,~\cite{bash2013limits,sobers2017covert} showed that, for strictly band-limited models and with random Gaussian codebooks, the SRL extends to such models, in the sense that $\mathcal{O}(\sqrt{T})$ covert bits can be sent over a transmission duration of $T$. When the bandwidth $W$ is infinite or grows large with time $T$ and with random codebooks,~\cite{wang2018covert, wang2019gaussian, wang2018continuous} showed that the covert capacity is infinite for both AWGN and Poisson channel, i.e., communication is not restricted to the SRL.

%The first extension to continuous-time AWGN channels is provided in~\cite{bash2013limits}, wherein a strictly band-limited system is considered; the follow-up work~\cite{sobers2017covert} further indicates that the SRL extends to such models, in the sense that $\mathcal{O}(\sqrt{T})$ covert bits can be sent in time $T$. Ref.~\cite{wang2018covert} investigates the continuous-time Gaussian channels wherein the bandwidth $W$ is infinite or grows large with time $T$, and shows that the covert capacity of such infinite-bandwidth Gaussian channels is positive (instead of being restricted by the SRL). Ref.~\cite{wang2018continuous} also shows that the covert capacity of infinite-bandwidth Poisson channels is infinite. 

The objective of the present work is to investigate covert communication over continuous-time AWGN channels in a formal and slightly different model, with the following distinctions compared to previous work\cite{bash2013limits, sobers2017covert, wang2018covert}.
\begin{enumerate}
\item We require the input signals to be \emph{strictly time-limited} and \emph{approximately band-limited}, as we aim to clearly characterize the relation between covert throughput and time. The band-limited property is characterized through a \emph{spectral mask} with bandwidth parameter $W$ at the transmitter, which restricts excessive radiation beyond the bandwidth of interest. As precisely defined in Section~\ref{sec:system}, both \emph{peak power} and \emph{integrated power} constraints are imposed on out-of-band emissions.
\item  Although the transmission is time-limited, the warden is allowed to monitor the entire timeline $t \in (-\infty, \infty)$.
\item We adopt Gallager's definition of {\it White Gaussian Noise} (WGN)~\cite[Chapter 7.7]{gallager2008principles} which assumes that the noise is spread over infinite bandwidth. We  elaborate on the rationale behind the model in a supplemental document~\cite{Zhang_supp}. When analyzing the covertness in this model, a sufficient statistic is formally established to preclude any unexpected manipulations of signals by the warden. 
\item Our results hold for fixed codebooks instead of random ones.
\end{enumerate}
Under this model, we then develop a communication scheme with theoretical reliability and covertness guarantees based on pulse amplitude modulation (PAM) with Binary Phase Shift Keying (BPSK) and root raised cosine (RRC) carrier pulses. We show that one can reliably and covertly transmit $\mathcal{O}(\sqrt{WT})$ bits of information and exactly characterize the pre-constant of the scaling. As expected, no shared key between Alice and Bob is needed if Willie's channel is noisier than Bob's channel, while $\mathcal{O}(\sqrt{WT})$ bits of shared key are needed otherwise. We finally highlight some of the key technical contributions of the present work.
\begin{enumerate}
\item The key step to analyze reliability is to convert continuous-time signals to discrete-time signals by applying a {\it matched filter}, after which it suffices to apply relatively standard techniques for discrete-time models. However, it is not \emph{a priori} clear that merely considering the converted discrete-time signals is sufficient from a covertness perspective. In fact, the warden may process continuous-time signals in an arbitrary way (e.g., perform nonuniform sampling, search for discontinuities, etc.) and extract information beyond that contained in the discrete-time signals. Directly analyzing continuous-time signals is, however, rather intricate. We show as part of our analysis (Section~\ref{sec:covert}) that such concerns may be dismissed by proving that the set of converted discrete-time signals forms a sufficient statistic for detection. Hence, in order to analyze covertness, it suffices to study the {variational distance} between the distributions of discrete-time random variables.
\item Our initial analysis shows that, with high probability (w.h.p.), a randomly chosen code $\C$ (a) satisfies the spectral mask constraints, (b) forms a {resolvability code} for Willie's channel (which further implies covertness), and (c) ensures a vanishing {average probability of error} (averaged over both message and shared key). We then develop a key result (Lemma~\ref{lemma:error}) showing that, upon carefully rearranging the codewords in $\C$, the resulting code ensures a vanishing {\it max-average probability of error} (maximized over the shared key and averaged over the message) while preserving the other two properties. %satisfies a stronger achievability criterion --- it
\item Because of technical challenges related to the spectral mask constraints (Section~\ref{sec:converse}), we have not been able to develop a converse for continuous-time AWGN channels; however, if we restrict ourself to the discrete-time model, our converse shows that the BPSK scheme used in this work is optimal.\footnote{The optimality of BPSK for covert communication under KL-divergence metric is attributed to~\cite{wang2019gaussian}.} 
\end{enumerate}

% \begin{figure}
%   \begin{center}
%     \includegraphics[scale=0.35]{RRC.PNG}
%     \caption{RRC pulses with different values of roll-off factor $\B$.}
%     \label{fig:rrc}
%   \end{center}
% \end{figure}

\section{Preliminaries and System Model}
\label{sec:model}

\subsection{Notation}
\label{sec:notation}

Random variables are denoted by upper-case letters (e.g., $X$) while their realizations are denoted by lower-case letters (e.g., $x$). Vectors are implicitly of length $n$ and are denoted by boldface letters (e.g., $\X = [X_1, X_2, \ldots, X_n]$ and $\x = [x_1, x_2, \ldots, x_n]$). Sets are denoted by calligraphic letters (e.g., $\mathcal{X}$).

For any $x \in \mathbb{R}$, $[x]^+$ represents $\max\{x,0\}$, while for any integers $a, b$ such that $a < b$, $\llbracket a, b \rrbracket$ represents the set of integers $\{a,a+1,\ldots, b\}$. All the logarithms $\log$ and exponentials $\exp$ are base $e$. Let $Q(\lambda) \triangleq (1/\sqrt{2\pi}) \int_{\lambda}^{\infty} \exp\left(-u^2/2\right) du$ be the tail distribution function of the standard normal distribution.  For two continuous probability distributions $P$ and $Q$ over the same set $\mathcal{X}$, we respectively define their {\it KL-divergence} and {\it variational distance} as $\mathbb{D}(P \Vert Q) \triangleq \int_{\mathcal{X}}P(x)\log\frac{P(x)}{Q(x)} dx$ and $\mathbb{V}(P,Q) \triangleq \frac{1}{2} \int_{\mathcal{X}} |P(x) - Q(x)| dx$.

\subsection{Root Raised Cosine (RRC) Pulses}
\label{sec:rrc}

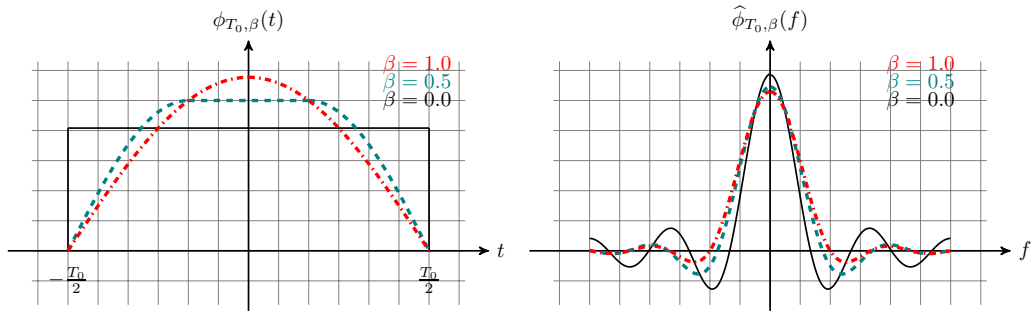
\begin{figure}[h]
  \centering
  \scalebox{0.8}{
    \begin{tikzpicture}[
      thick,
      >=stealth',
      dot/.style = {
        draw,
        fill = white,
        circle,
        inner sep = 0pt,
      minimum size = 4pt,
    }
    ]
    \def\xmin{-4};
    \def\xmax{4};
    \def\ymin{-1};
    \def\ymax{7 / 2};
    \def\yamp{5}
    \def\tzero{6}
    \coordinate (O) at (0,0);
    \draw[step=0.5cm,gray,very thin] (0.9*\xmin,0.9*\ymin) grid (0.9 * \xmax,{0.9*\ymax});
    \draw[->] ({\xmin},0) -- ({\xmax},0) coordinate[label = {right:$t$}] ;
    \draw[->] (0,{\ymin}) -- (0,{\ymax}) coordinate[label = {above: $\phi_{T_0,\beta}(t)$}] ;
    % Plot of RRC for beta=0
    \draw[black, thick, domain=-\tzero/2:\tzero/2] plot (\x, {\yamp*sqrt(1/\tzero)}) {};
    \draw[black, thick] (-\tzero/2,0) -- (-\tzero/2, {\yamp*sqrt(1/\tzero)}) {};
    \draw[black, thick] (\tzero/2,0) -- (\tzero/2, {\yamp*sqrt(1/\tzero)}) {};
    % Plot of RRC for beta=0.5
    \def\bet{0.5}
    \draw[teal, dashed, ultra thick, domain=-(1-\bet)/(1+\bet)*\tzero/2:(1-\bet)/(1+\bet)*\tzero/2] plot (\x, {\yamp*sqrt((1+\bet)/\tzero)}) {}; % node[label={above:$\beta=0.5$}] 
    \draw[teal, dashed, ultra thick, domain=-\tzero/2:-(1-\bet)/(1+\bet)*\tzero/2] plot (\x, {\yamp*sqrt((1+\bet)/(2*\tzero)*(1+cos(deg(3.14*(1+\bet)/(\tzero*\bet)*(-\x-(1-\bet)/(1+\bet)*\tzero/2)))))}) {};
    \draw[teal, dashed, ultra thick, domain=(1-\bet)/(1+\bet)*\tzero/2:\tzero/2] plot (\x, {\yamp*sqrt((1+\bet)/(2*\tzero)*(1+cos(deg(3.14*(1+\bet)/(\tzero*\bet)*(\x-(1-\bet)/(1+\bet)*\tzero/2)))))}) {};
        % Plot of RRC for beta=0.5
    \def\bet{1.0}
    \draw[red, dash dot, ultra thick, domain=-(1-\bet)/(1+\bet)*\tzero/2:(1-\bet)/(1+\bet)*\tzero/2] plot (\x, {\yamp*sqrt((1+\bet)/\tzero)}) ;
    \draw[red, dash dot, ultra thick, domain=-\tzero/2:-(1-\bet)/(1+\bet)*\tzero/2] plot (\x, {\yamp*sqrt((1+\bet)/(2*\tzero)*(1+cos(deg(3.14*(1+\bet)/(\tzero*\bet)*(-\x-(1-\bet)/(1+\bet)*\tzero/2)))))}) {};
    \draw[red, dash dot, ultra thick, domain=(1-\bet)/(1+\bet)*\tzero/2:\tzero/2] plot (\x, {\yamp*sqrt((1+\bet)/(2*\tzero)*(1+cos(deg(3.14*(1+\bet)/(\tzero*\bet)*(\x-(1-\bet)/(1+\bet)*\tzero/2)))))}) {};

    \node (caption1) at (2.8,2.5) {$\beta=0.0$};
    \node (caption1) at (2.8,2.8) {\color{teal} $\beta=0.5$};
    \node (caption1) at (2.8,3.1) {\color{red} $\beta=1.0$};
    \node[label={below:$-\frac{T_0}{2}$}] (caption1) at (-\tzero/2,0) {};
    \node[label={below:$\frac{T_0}{2}$}] (caption1) at (\tzero/2,0) {};
\end{tikzpicture}}
\scalebox{0.8}{
    \begin{tikzpicture}[
    thick,
    >=stealth',
    dot/.style = {
      draw,
      fill = white,
      circle,
      inner sep = 0pt,
      minimum size = 4pt,
    }
    ]
    \def\xmin{-4};
    \def\xmax{4};
    \def\ymin{-1};
    \def\ymax{7 / 2};
    \def\yamp{1.2}
    \def\xamp{0.25}
    \def\tzero{6}
    \coordinate (O) at (0,0);
    \draw[step=0.5cm,gray,very thin] (0.9*\xmin,0.9*\ymin) grid (0.9 * \xmax,{0.9*\ymax});
    \draw[->] ({\xmin},0) -- ({\xmax},0) coordinate[label = {right:$f$}] ;
    \draw[->] (0,{\ymin}) -- (0,{\ymax}) coordinate[label = {above: $\widehat{\phi}_{T_0,\beta}(f)$}] ;
    % Plot of RRC for beta=0
    \def\bet{0.01}
    \draw[black, thick, domain=-3:-0.01,samples=200] plot (\x, {\yamp*4*\bet/3.14*sqrt(\tzero/(1+\bet))*(cos(deg(3.14*\tzero*\xamp*\x))+(1+\bet)*sin(deg((1-\bet)/(1+\bet)*3.14*\tzero*\xamp*\x))/(4*\bet*\tzero*\xamp*\x))/(1-pow(4*\bet*\tzero*\xamp*\x/(1+\bet),2))}) {};
    \draw[black, thick, domain=0.01:3,samples=200] plot (\x, {\yamp*4*\bet/3.14*sqrt(\tzero/(1+\bet))*(cos(deg(3.14*\tzero*\xamp*\x))+(1+\bet)*sin(deg((1-\bet)/(1+\bet)*3.14*\tzero*\xamp*\x))/(4*\bet*\tzero*\xamp*\x))/(1-pow(4*\bet*\tzero*\xamp*\x/(1+\bet),2))}) {};
    % Plot of RRC for beta=0.5
    \def\bet{0.5}
    \draw[teal, dashed, ultra thick, domain=-3:-0.01,samples=100] plot (\x, {\yamp*4*\bet/3.14*sqrt(\tzero/(1+\bet))*(cos(deg(3.14*\tzero*\xamp*\x))+(1+\bet)*sin(deg((1-\bet)/(1+\bet)*3.14*\tzero*\xamp*\x))/(4*\bet*\tzero*\xamp*\x))/(1-pow(4*\bet*\tzero*\xamp*\x/(1+\bet),2))}) {};
    \draw[teal, dashed, ultra thick, domain=0.01:3,samples=100] plot (\x, {\yamp*4*\bet/3.14*sqrt(\tzero/(1+\bet))*(cos(deg(3.14*\tzero*\xamp*\x))+(1+\bet)*sin(deg((1-\bet)/(1+\bet)*3.14*\tzero*\xamp*\x))/(4*\bet*\tzero*\xamp*\x))/(1-pow(4*\bet*\tzero*\xamp*\x/(1+\bet),2))}) {};
    % Plot of RRC for beta=1.0
    \def\bet{1.0}
    \draw[red, dash dot, ultra thick, domain=-3:-0.01,samples=100] plot (\x, {\yamp*4*\bet/3.14*sqrt(\tzero/(1+\bet))*(cos(deg(3.14*\tzero*\xamp*\x))+(1+\bet)*sin(deg((1-\bet)/(1+\bet)*3.14*\tzero*\xamp*\x))/(4*\bet*\tzero*\xamp*\x))/(1-pow(4*\bet*\tzero*\xamp*\x/(1+\bet),2))}) {};
    \draw[red, dash dot, ultra thick, domain=0.01:3,samples=100] plot (\x, {\yamp*4*\bet/3.14*sqrt(\tzero/(1+\bet))*(cos(deg(3.14*\tzero*\xamp*\x))+(1+\bet)*sin(deg((1-\bet)/(1+\bet)*3.14*\tzero*\xamp*\x))/(4*\bet*\tzero*\xamp*\x))/(1-pow(4*\bet*\tzero*\xamp*\x/(1+\bet),2))}) {};

    \node (caption1) at (2.5,2.5) {$\beta=0.0$};
    \node (caption1) at (2.5,2.8) {\color{teal} $\beta=0.5$};
    \node (caption1) at (2.5,3.1) {\color{red} $\beta=1.0$};
    \end{tikzpicture}}
    \caption{RRC pulses with different values of roll-off factor $\B$.}
    \label{fig:rrc}
\end{figure}

In the time domain, an RRC pulse is defined by 
\begin{align*}
&\phit \triangleq \begin{cases} \sqrt{\frac{1+\B}{\T}}, & 0 \le |t| \le \frac{1-\B}{1+\B}\frac{\T}{2}; \\
\sqrt{\frac{1+\cos\left[\frac{\pi(1+\B)}{\T\B}\left(|t|-\frac{1-\B}{1+\B}\frac{\T}{2}\right)\right]}{2\T/(1+\B)}}, & \frac{1-\B}{1+\B}\frac{\T}{2} < |t| \le \frac{\T}{2}.
\end{cases}
\end{align*}
Note that $\phit$ is uniquely determined by the {\it pulse duration} $\T$ and {\it roll-off factor} $\beta \in [0,1]$. The pulse duration $\T$ controls the length of the support of $\phit$, i.e., it is non-zero only when $t\in [-\T/2, \T/2]$. Also, note that $\phit$ is flat when $0 \le |t| \le \frac{1-\B}{1+\B}\frac{\T}{2}$, and the length and height of this flat interval depend on the roll-off factor $\B$ (a smaller $\B$ implies a longer and lower flat interval). Correspondingly, the spectrum~\cite[Chapter 11.3]{lapidoth2009foundation} of an RRC is given by 
\begin{align*}
&\phif \triangleq \begin{cases}
\frac{\sqrt{\T}(4\B + (1-\B)\pi)}{2\pi \sqrt{1+\B}}, &f = 0; \\ \frac{4\B\sqrt{\T}}{\pi\sqrt{1+\B}}\cdot \frac{\cos(\pi\T f) + \frac{(1+\B)\sin\left(\frac{1-\B}{1+\B}\pi\T f\right)}{4\B\T f}}{1 - \left(\frac{4\B\T f}{1+\B}\right)^2}, & f \ne 0.
\end{cases}
\end{align*}
Note that we swap the roles of time and frequency domain compared to standard use of RRC pulses. The advantages are two-fold: (a) pulses are strictly time-limited; and (b) the energy/power decays fast enough in the frequency domain. One may design alternative carrier pulses that outperform RRC pulses but this optimization is outside the scope of the present paper. RRC pulses with different values of $\B$ are illustrated in Fig.~\ref{fig:rrc}.

\subsection{White Gaussian Noise (WGN)} \label{sec:wgn} 
Given an $\Ltwo$ {\it function} $g(t)$ (i.e., $\int_{-\infty}^{\infty}|g(t)|^2 dt < \infty$), and a stochastic process $\{W(t),t \in \mathbb{R}\}$ which sample functions are real-valued $\Ltwo$ functions, the random variable $V \triangleq \int_{-\infty}^{\infty}W(t)g(t) dt$ is called a {\it linear functional} of $\{W(t),t \in \mathbb{R}\}$. 

\begin{definition}[WGN~\cite{gallager2008principles}] \label{def:wgn}
	The WGN $\{W(t),t \in \mathbb{R}\}$ with intensity $N_0/2$ is a generalized zero-mean Gaussian process such that (a) for any $t_1,t_2 \in \mathbb{R}$, the autocovariance $\E(W(t_1)W(t_2))$ is $(N_0/2)\delta(t_2-t_1)$, where $\delta(\cdot)$ is the impulse function, and (b) for any set of $\Ltwo$ functions $\{g_i(t)\}$, the linear functionals $\{V_i = \int_{-\infty}^{\infty}W(t)g_i(t) dt\}$ are jointly Gaussian, and satisfy $\E(V_iV_j) = \frac{N_0}{2} \int_{-\infty}^{\infty}g_i(t)g_j(t)dt, \ \forall i,j$.
\end{definition}

	\begin{remark}
			The WGN at any time $t$ is a Gaussian variable with infinite variance; that is, the WGN is not a well-defined random process. However, as discussed in~\cite[Chapter 7.7]{gallager2008principles}, ``WGN is not viewed in terms of random variables at each epoch of time. Rather, it is viewed as a generalized zero-mean random process (in the same sense as $\delta(t)$ is viewed as a generalized function) for which the properties (a) and (b) in Definition~\ref{def:wgn} are satisfied.''  
	\end{remark}

If $\{g_i(t) \}$ is a set of orthonormal functions, we have $\E(V_i V_j) = \frac{N_0}{2} \mathbbm{1}\{i = j\}$ and each $V_i$ is a Gaussian random variable with zero mean and variance $N_0/2$. Hence, the linear functionals $\{V_i\}$ are independent and identically distributed (i.i.d.). This property is critical in our achievability scheme --- we construct the transmitted signals in terms of an orthonormal basis (e.g., time-shifted RRC pulses), thus we can represent the WGN in terms of the same orthonormal basis, and the resulting ``noise variables'' are i.i.d. Gaussian variables.

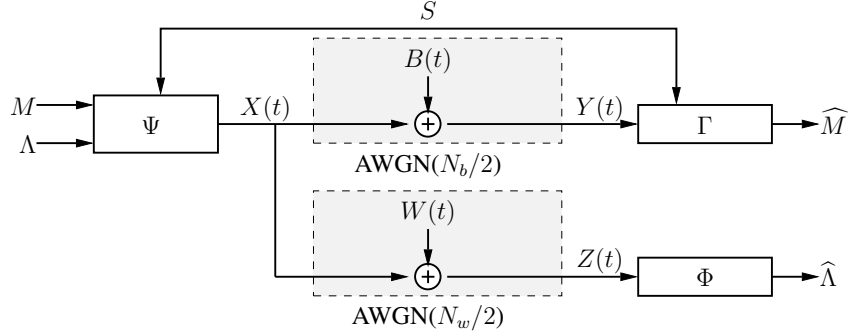
\begin{figure}
  \begin{center}
    \scalebox{0.8}{\input{system-model.tex}}
    \caption{System diagram.}
    \label{fig:system}
  \end{center}
\end{figure}

\subsection{System Model} \label{sec:system}

\noindent{1) \underline{Shared key}}:
Prior to communication, Alice and Bob share a {\it secret key} $\s$, which is uniformly distributed in $\llbracket 1,K\rrbracket$ and unknown to Willie. 

\noindent{2) \underline{Encoder}}:
Given a fixed and publicly known $T > 0$, Alice uses her encoder $\Psi(\cdot)$ to encode the {\it transmission status} $\Lambda \in \{0,1\}$, the {\it message}\footnote{With a small abuse of notation, we denote the size of the message by $M$ as well.} $M \in \{0\} \cup \llbracket 1,\N\rrbracket$, and the shared key $\s \in \llbracket 1,K\rrbracket$ into a continuous-time signal $X(t), t \in [0,T]$. When Alice is silent ($\Lambda = 0$), her message is required to be $0$ and her transmission must satisfy $X(t) = 0, t \in [0,T]$. When Alice is active ($\Lambda = 1$), her message is uniformly distributed in $\llbracket 1,\N\rrbracket$. For each message $m$ and key $s$, she encodes the message-key pair $(m,s)$ into a codeword $X_{ms}(t)$. The {\it sub-codebook index by $s$} is the collection of codewords $\C_s \triangleq \{X_{ms}(t)\}_{m=1}^{\N}$. The {\it codebook} $\C$ is the union of all the sub-codebooks, i.e., $\C = \cup_{s \in \llbracket1,K\rrbracket} \C_s$.

\noindent{3) \underline{Noise}}:
The channel noise from Alice to Bob is modeled as an {\it additive white Gaussian noise} (AWGN) $B(t)$ with intensity $N_b/2$ while the channel noise from Alice to Willie is modeled as another independent AWGN $W(t)$ with intensity $N_w/2$.

\noindent{4) \underline{Willie's estimator}}:
Willie observes $Z(t) = X(t)+ W(t)$ and uses an estimator $\Phi(\cdot)$ to perform a hypothesis test on $Z(t)$. The {\it null hypothesis} $H_0$ corresponds to $\Lambda=0$ while the {\it alternative hypothesis} $H_1$ corresponds to $\Lambda=1$. Willie's estimation is denoted $\widehat{\Lambda} = \Phi(Z(t))$. Let $P_{\text{FA}}(\Phi) \triangleq \PP(\widehat{\Lambda}=1|\Lambda = 0)$ and $P_{\text{MD}}(\Phi) \triangleq \PP(\widehat{\Lambda}=0|\Lambda = 1)$ denote the probability of {\it false alarm} and the probability of {\it missed detection} of the estimator $\Phi$, respectively. We use a hypothesis testing metric to measure covertness.
\begin{definition}[Covertness] \label{def:covert}
	The communication is guaranteed to be $(1-\delta)$-covert if 
	\begin{align}
	\min_{\Phi} \left\{P_{\emph{FA}}(\Phi) + P_{\emph{MD}}(\Phi) \right\} \ge 1 - \delta,
	\end{align}
	where $\Phi$ is minimized over all possible estimators.
\end{definition}

\noindent{5) \underline{Decoder}}:
Bob observes $Y(t) = X(t)+ B(t)$. His decoder $\Gamma(\cdot)$ takes $Y(t)$ and the shared key $\s$ as input and produces a message reconstruction $\widehat{M}$. The {\it max-average probability of error} (maximized over the shared key and averaged over the message) is defined as 
\begin{align}
P_{\text{err}} \triangleq \max_{s \in \llbracket1,K\rrbracket}\PP(M \ne \widehat{M}|\s=s,\Lambda = 1)  + \PP(\widehat{M} \ne 0 | \Lambda = 0). \label{eq:error}
\end{align}

\noindent{6) \underline{Spectral mask}}: The spectral mask at the transmitter is provided {\it a priori}, and the code $\C$ must fit into the spectral mask. Let $E_{x_{ms}(t)}(f) = |\widehat{x}_{ms}(f)|^2$ be the {\it Energy Spectral Density} (ESD) of the codeword $x_{ms}(t)$, and $\widehat{E}(f)$ be the ESD of the code $\C$, where 
\begin{align}
\widehat{E}(f) \triangleq \frac{1}{\N K} \sum_{m=1}^\N \sum_{s=1}^K E_{x_{ms}(t)}(f). \label{eq:esd}
\end{align}
\begin{definition}[Spectral mask] \label{def:mask} Let $l \in \mathbb{N}^{\ast}$ be the number of constraints, $W > 0$ be the bandwidth of interest, $\{\U_\K\}_{\K=1}^l$, $\{\al_\K\}_{\K=1}^l$, and $\{\eta_\K \}_{\K=1}^l$ be non-decreasing real-valued sequences, and $V_\K = 10^{-\frac{U_\K}{10}}$ for each $\K$. A code $\C$ with ESD $\widehat{E}(f)$ is said to fit into the spectral mask $\mathcal{S}(W, \{\U_\K\}_{\K=1}^l, \{\al_\K\}_{\K=1}^l, \{\eta_\K\}_{\K=1}^l)$ if for every $\K \in \llbracket 1,l \rrbracket$:
	\begin{enumerate}
		\item its $U_\K$-\emph{dB} bandwidth is at most $\al_\K W$, i.e., $\forall f \ge \al_\K W, \ \ \widehat{E}(f) < V_{\K} \big[\widehat{E}(f)\big]_{\max};$
		\item the energy allocated in $f \in [-\al_\K W, \al_\K W]$ satisfies $\int_{-\al_\K W}^{\al_\K W} \widehat{E}(f) df \ge \eta_\K \int_{-\infty}^{+\infty} \widehat{E}(f) df$.
	\end{enumerate}
\end{definition}
For example, if $(U_1,\al_1,\eta_1) = (3,1,0.8)$, the $3$-dB bandwidth should be at most $W$ and $\int_{-W}^{W} \widehat{E}(f) df$ should be at least $0.8 \big( \int_{-\infty}^{+\infty} \widehat{E}(f) df\big)$. For notational convenience, we abbreviate the spectral mask $\mathcal{S}(W, \{\U_\K\}_{\K=1}^l, \{\al_\K\}_{\K=1}^l, \{\eta_\K\}_{\K=1}^l)$ as $\mathcal{S}_W$, when the arguments are clear from the context. 

\noindent{7) \underline{Throughput pair and covert capacity}}: For any spectral mask $\mathcal{S}_W$, noise parameters $N_w, N_b > 0$, and covertness parameter $\delta$, a {\it throughput pair} $(r,r_K)$ is said to be achievable if there exists a sequence of code with increasing support $T$ such that
\begin{align*}
&\liminf_{T \to \infty} \frac{\log \N}{\sqrt{T}} \ge r, \quad \limsup_{T \to \infty} \frac{\log K}{\sqrt{T}} \le r_K, \\
&\lim_{T \to \infty} P_{\text{err}} = 0, \qquad \lim_{T \to \infty} \min_{\Phi} \left\{P_{\text{FA}}(\Phi) + P_{\text{MD}}(\Phi) \right\} \ge 1 - \delta,
\end{align*}  
and the ESD $\widehat{E}(f)$ of the code fits into the spectral mask $\mathcal{S}_W$. The {\it covert capacity} is the supremum of $r$ over all achievable throughput pairs.

\section{Achievability Schemes and Main results} \label{sec:achievability}
Our communication scheme relies on a PAM scheme, wherein the information bits are carried over a train of RRC pulses. Recall that the duration of the RRC pulse $\phit$ equals $\T$, thus the effective {\it blocklength} of the transmission is $n \triangleq \lfloor T/\T \rfloor$. Without loss of generality, we drop the floor operator for simplicity, and write $n = T/\T$ by assuming $\T$ divides $T$. We specify the exact choice of $n$ in Lemma~\ref{lemma:blocklength}. The sequence of $n$ RRC carrier pulses used for PAM is given by $\{g_i(t)\}_{i=1}^n$, where 
\begin{align}
g_i(t) \triangleq \phi_{\T,\B}\left(t-\left(i-0.5\right)\T\right)
\end{align}
is the time-shifted version of $\phit$, and is non-zero only when $t \in [(i-1)\T,i\T]$. For a length-$n$ {\it discrete-time codeword} (DT-codeword) $\x = [x_1,x_2, \ldots,x_n]$, the corresponding {\it continuous-time codeword} (CT-codeword) $\xt$ takes the form 
\begin{align}
\xt = \sum_{i=1}^{\n}x_i g_i(t). \label{eq:codeword}
\end{align}

We use a {\it random coding argument} with BPSK to show the existence of good codes. Each $x_i$ is chosen from $\{-\an,\an\}$, where $\an$ scales as $\mathcal{O}(n^{-1/4})$ for covert communication and the exact value will be specified in Section~\ref{sec:proof}. For $n \in \mathbb{N}^{\ast}$, let $\Px$ be the distribution on $\mathcal{X} = \{-\an,\an \}$ such that 
\begin{align}
\Px(\an) = \Px(-\an) = 1/2, \label{eq:px}
\end{align}   
and $\Pxn = \prod_{i=1}^n \Px$ be the $n$-letter product distribution of $\Px$. For each message-key pair $(m,s)$, its DT-codeword $\x_{\m s} = [x_{\m s,1}, x_{\m s,2}, \ldots, x_{\m s,n}]$ is generated independently according to $\Pxn$, and its CT-codeword $x_{ms}(t)$ equals $\sum_{i=1}^n x_{ms,i}g_i(t)$.

\subsection{Optimal Effective Blocklength}
One important step in the analysis is to convert continuous-time AWGN channels to discrete-time AWGN channels. We wish to maximize the number of discrete-time channel uses to achieve the highest possible throughput; however, the spectral mask $\mathcal{S}_W$ imposes constraints on the ESD $\widehat{E}(f)$ of the code $\C$. These constraints imply that the pulse duration $\T$ of $\{g_i(t)\}$ cannot be arbitrarily small, hence the blocklength $n$ cannot be arbitrarily large. 

Though the ESD $\widehat{E}(f)$ has intricate dependencies with respect to the codewords in the code (as defined in~\eqref{eq:esd}), the {\it ensemble-averaged ESD} $\widetilde{E}(f)$, which is averaged over the codeword generation process $\Pxn$, has a relatively simple form, i.e.,
\begin{align}
\widetilde{E}(f) \triangleq \E_{\Pxn}(E_{X(t)}(f)) = \an^2 n \cdot |\phif|^2. \label{eq:average}
\end{align}
In fact, Lemma~\ref{lemma:esd} below shows that, with high probability over the code design, the ESD $\widehat{E}(f)$ of a randomly chosen code $\C$ is tightly concentrated around $\widetilde{E}(f)$. The proofs of~\eqref{eq:average} and Lemma~\ref{lemma:esd} are provided in Appendix~\ref{appendix:esd}. 
\begin{lemma} \label{lemma:esd}
	With probability at least $1 - 2n^2e^{-\sqrt{MK}/2}$ over the code design, a randomly chosen code $\C$ satisfies
	\begin{align*}
	\forall f \in \mathbb{R}, \ \left|\widetilde{E}(f) - \widehat{E}(f)\right| \le \frac{\widetilde{E}(f) \cdot n}{(\N K)^{1/4}}.
	\end{align*}
\end{lemma}

Suppose the code $\C$, which is required to fit into the spectral mask $\mathcal{S}_W$, is such that its ESD $\widehat{E}(f)$ exactly equals the ensemble-averaged ESD $\widetilde{E}(f)$. Then, the minimum value of $\T$ is given by the optimal value $\Ts$ of the optimization problem ({P1}) defined as follows:  
\begin{equation*}
\begin{aligned}
(\text{P1}) \quad & \underset{\T,\B}{\text{min}}
& & \T
& &\\
& \text{s.t.}
& & \forall f \ge \al_\K W, \ |\phinf(f)|^2 < V_{\K} |\phinf(0)|^2, 
& & \K \in \llbracket 1, l \rrbracket;\\
&
& & \int_{-\al_\K W}^{\al_\K W} |\phif|^2 df \ge \eta_\K,  
& & \K \in \llbracket 1, l \rrbracket;\\
&
& & \T > 0, \ 0 \le \B \le 1.
\end{aligned} 
\end{equation*} 
The optimization in ({P1}) is over both the pulse duration $\T$ and roll-off factor $\B$, while the constraints follow from the definition of the spectral mask $\mathcal{S}_W$, the expression of $\widetilde{E}(f)$ in~\eqref{eq:average}, and the fact that 
\begin{align}
\int_{-\infty}^{\infty} |\phif|^2 df = \int_{-\infty}^{\infty} |\phit|^2 dt = 1.
\end{align}

In general, the ESD $\widehat{E}(f)$ of a specific code may not be equal to $\widetilde{E}(f)$ but Lemma~\ref{lemma:esd} ensures that, with high probability, it does not deviate from $\widetilde{E}(f)$ by a factor of $(1 \pm n/(MK)^{1/4})$. Note that $MK$ is a function of $n$, and $n$ depends on the time $T$ and pulse duration $\T$. We denote $MK = f(n)$, and let 
\begin{align}
u(T,\T) \triangleq \frac{T/\T}{[f(T/\T)]^{1/4}} = \frac{n}{(MK)^{1/4}}
\end{align}
be the slackness parameter. The optimization problem ({P2}), which is obtained by adding slackness $u(T,\T)$ to ({P1}), takes the form
\begin{equation*}
\begin{aligned}
(\text{P2}) \quad & \underset{\T,\B}{\text{min}}
& & \T
& &\\
& \text{s.t.}
& & \forall f \ge \al_\K W, \frac{|\phinf(f)|^2 (1-u(T,\T))}{|\phinf(0)|^2 (1+u(T,\T))} < V_{\K} , 
& & \K \in \llbracket 1, l \rrbracket;\\
&
& & \int_{-\al_\K W}^{\al_\K W} |\phif|^2 df \ge \frac{\eta_\K}{1-u(T,\T)},  
& & \K \in \llbracket 1, l \rrbracket;\\
&
& & \T > 0, \ 0 \le \B \le 1.
\end{aligned} 
\end{equation*}
The optimal value of ({P2}) is denoted by $\Tst$ and explicitly depends on $T$. By choosing $\T = \Tst$, Lemma~\ref{lemma:esd} implies that w.h.p., a randomly chosen code $\C$ fits into the spectral mask. Since ({P1}) and ({P2}) only differ in the slackness $u(T,\T)$ (which goes to zero as $T$ goes to infinity), we have the following lemma. 
\begin{lemma} \label{lemma:final}
	For sufficiently large $T$, we have
\begin{align}
\lim_{T \to \infty} \Tst = \Ts. \label{eq:kun1}
\end{align}
\end{lemma}
The proof of Lemma~\ref{lemma:final} is provided in Appendix~\ref{appendix:limit}.
We now take a detour to establish connections between $\Ts$ and the bandwidth parameter $W$. Given any $\B \in [0,1]$,  the optimization problem (P$1_\B$), which differs from ({P1}) only in the third constraint, i.e., only $\T > 0$ (without $0 \le \B \le 1$) is active, is defined as follows:
\begin{equation*}
\begin{aligned}
(\text{P}1_\B) \quad & \underset{\T}{\text{min}}
& & \T
& &\\
& \text{s.t.}
& & \forall f \ge \al_\K W, \ |\phinf(f)|^2 < V_{\K} |\phinf(0)|^2, 
& & \K \in \llbracket 1, l \rrbracket;\\
&
& & \int_{-\al_\K W}^{\al_\K W} |\phif|^2 df \ge \eta_\K,  
& & \K \in \llbracket 1, l \rrbracket;\\
&
& & \T > 0.
\end{aligned} 
\end{equation*} 
Let $\Tsb$ be the optimal value of (P$1_\B$), and it
operationally represents the minimum value of $\T$ if the RRC carrier pulses are restricted to a fixed roll-off factor $\B$. If we merely increase the bandwidth $W$ (with the other arguments $\{\U_\K\}_{\K=1}^l, \{\al_\K\}_{\K=1}^l, \{\eta_\K\}_{\K=1}^l$ invariant), the value of $\Tsb$ decreases since the spectral mask with a broader range of frequency typically allows a faster change of signals in the time domain. Lemma~\ref{lemma:obs}, which is formally proved in Appendix~\ref{appendix:obs}, formalizes this intuition by proving that the product of $W$ and $\Tsb$ is a constant.
\begin{lemma} \label{lemma:obs}
	For any roll-off factor $\B \in [0,1]$ and any $W_1,W_2 > 0$, we have
	\begin{align}
	W_1\cdot T_0^{\ast}(\mathcal{S}_{W_1},\B) = W_2\cdot T_0^{\ast}(\mathcal{S}_{W_2},\B). 
	\end{align}
\end{lemma}      
Thus, we define a constant $c(\B)$ as
\begin{align}
c(\B) \triangleq W\cdot \Tsb,  \label{eq:cb}
\end{align} 
where $W$ can be chosen as an arbitrary positive value. Therefore, we have
\begin{align}
\Ts = \min_{\B \in [0,1]}\Tsb = \frac{1}{W}\min_{\B \in [0,1]} c(\B). \label{eq:kun2}
\end{align}    
By combining~\eqref{eq:kun1},~\eqref{eq:kun2}, and the fact that $n = T/\Tst$, we have the following Lemma, which establishes a bridge connecting the continuous-time channel (characterized by the time $T$ and the spectral mask $\mathcal{S}_W$) and the discrete-time channel (characterized by the blocklength $n$).  
\begin{lemma} \label{lemma:blocklength}
	For every $\xi > 0$, there exists $T(\xi)$ such that for all $T > T(\xi)$, by setting the blocklength $n = (1-\xi)\cdot \frac{ WT}{\min_{\B \in [0,1]} c(\B)}$, the probability of a randomly chosen code fitting into the spectral mask $\mathcal{S}_W$ is at least $1 - \exp\big(-\exp(\mathcal{O}(\sqrt{WT}))\big)$. Further, as $T$ goes to infinity,
	\begin{align}
	\lim_{T \to \infty} n = \lim_{T \to \infty} \frac{(1-\xi)\cdot WT}{\min_{\B \in [0,1]} c(\B)} = \frac{WT}{\min_{\B \in [0,1]} c(\B)}.
	\end{align} 
\end{lemma}
\noindent{\it Proof:} Equations~\eqref{eq:kun1} and~\eqref{eq:kun2} imply that for every $\xi' > 0$, there exists a $T'(\xi')$ such that for all $T > T'(\xi')$,  
\begin{align*}
\frac{1}{W}\min_{\B \in [0,1]} c(\B) \le \Tst \le \frac{1}{W}\min_{\B \in [0,1]} c(\B) + \xi',
\end{align*}
hence we can set 
\begin{align}
n = \frac{T}{\frac{1}{W}\min_{\B \in [0,1]} c(\B) + \xi'} = \frac{WT}{\min_{\B \in [0,1]} c(\B)} \left(1 - \frac{\xi' W}{\min_{\B \in [0,1]} c(\B)+ \xi' W}\right). \label{eq:xi}
\end{align}
We simplify~\eqref{eq:xi} by choosing $\xi = \frac{\xi' W}{\min_{\B \in [0,1]} c(\B)+ \xi' W}$. For every $\xi > 0$, there exists a $T(\xi)$, where $T(\xi) = T'\left(\frac{\xi' W}{\min_{\B \in [0,1]} c(\B)+ \xi' W}\right)$, such that for all $T > T(\xi)$, if we set   
$n = \frac{(1-\xi)\cdot WT}{\min_{\B \in [0,1]} c(\B)}$,  the probability of a randomly chosen code fitting into the spectral mask is at least 
\begin{align*}
1 - 2n^2 \exp(-\sqrt{MK}/2) = 1 - \exp\left(-\exp(\mathcal{O}(\sqrt{n}))\right) = 1 - \exp\left(-\exp(\mathcal{O}(\sqrt{WT}))\right).
\end{align*}
This completes the proof of Lemma~\ref{lemma:blocklength}. \qed

\begin{remark}
	One may define the {\it power spectral density} (PSD) averaged over the distribution $\Pxn$ as 
	\begin{align}
	S_{X}(f) \triangleq \lim_{T \to \infty} \E_{\Pxn}\left(\frac{|\widehat{X}(f)|^2}{T}\right) &= \lim_{T \to \infty} \frac{\widetilde{E}(f)}{T} \stackrel{T = n\T}{=}  \lim_{n \to \infty} \frac{\an^2 \cdot |\phif|^2}{\T}.
	\end{align}
	However, $S_X(f) = \lim_{n \to \infty} \frac{\an^2 \cdot |\phif|^2}{\T} = 0$ for every $f \in \mathbb{R}$ since the amplitude $\an$ is a decaying function of $n$ in the covert setting. This is not surprising since the time-averaged power $P$ of any CT-codeword $x(t)$ goes to zero as $T$ goes to infinity, i.e., $P \triangleq \lim_{T \to \infty} \frac{1}{T} \int_{0}^T |x(t)|^2 dt = \lim_{T \to \infty} \frac{\an^2 n}{T} = \lim_{n \to \infty}\frac{\an^2}{\T} = 0.$
	Since the objective of analyzing the frequency domain is to concentrate the energy of transmissions in the spectral mask, it suffices to study the distribution of energy through the ESD. 
\end{remark}

\subsection{Main Results}
Based on the coding and modulation scheme above, we present a lower bound on the covert capacity by proving an achievable throughput pair.
\begin{theorem}\label{thm:result}
	For any spectral mask $\mathcal{S}_W$, noise parameters $N_w, N_b > 0$, and covertness parameter $\delta \in (0,1)$, the throughput pair $(r,r_K)$ with   
	\begin{align}
	& r = \frac{N_w}{N_b} \sqrt{\frac{2W}{\min_{\B \in [0,1]}c(\B)}}\cdot Q^{-1}\left(\frac{1-\delta}{2}\right), \ \ r_K = \left(1-\frac{N_w}{N_b}\right)^+ \sqrt{\frac{2W}{\min_{\B \in [0,1]}c(\B)}}\cdot Q^{-1}\left(\frac{1-\delta}{2}\right), \label{eq:thm2}
	\end{align} 
	is achievable.
\end{theorem}

\section{Proof of Theorem~\ref{thm:result}} \label{sec:proof}
We set the blocklength to be 
\begin{align}
n = (1-\xi)\cdot \frac{WT}{\min_{\B \in [0,1]} c(\B)}, \label{eq:n}
\end{align}
for an arbitrarily small $\xi >0$. Furthermore, we set the amplitude $\an$ to be 
\begin{align} \Big(\frac{2}{n}\Big)^{1/4}\sqrt{Q^{-1}\Big(\frac{1-\delta}{2}\Big)N_w}\cdot \Big(1 - n^{-1/8}\Big), \label{eq:an}
\end{align}
and the message size and key size to satisfy
\begin{align}
&\log \N = \Big(1 - \frac{1}{\log n}\Big) \frac{\an^2 n}{N_b}, \ \ \log K = \Big[\Big(1+ \frac{1}{\log n}\Big) - \Big(1- \frac{1}{\log n}\Big)\frac{N_w}{N_b} \Big]^+ \frac{\an^2 n}{N_w}.
\end{align}
The above choices of $n, \an, \log \N$, and $\log K$ lead to the achievable throughput pair in Theorem~\ref{thm:result}. Recall that with the value of $n$ chosen as per~\eqref{eq:n}, Lemma~\ref{lemma:blocklength} ensures that, for sufficiently large $T$, a random code $\C$ fits into the spectral mask with high probability. In Sub-sections~\ref{sec:covert} and~\ref{sec:reliability}, we prove that, with high probability, a randomly chosen code $\C$ guarantees covertness and a small probability of error $P_{\text{err}}$, respectively.

\subsection{Analysis of Covertness} \label{sec:covert}
\noindent{\bf \underline{1) Hypothesis testing and sufficient statistics:}}
Recall that if we map the noise $W(t)$ to any set of orthonormal basis, the resulting coefficients are i.i.d. Gaussian variables. A specific orthonormal basis can be chosen in such a way that the first $n$ functions are time-shifted RRC pulses $\{g_i(t) \}_{i=1}^n$, and the remaining ones form a set of orthonormal functions $\{g_i(t)\}_{i=n+1}^{L}$ that satisfy\footnote{Note that the total number of functions $L$ in this orthonormal basis can be either finite or infinite.}
\begin{itemize}
	\item Each of the function in $\{g_i(t)\}_{i=n+1}^{L}$ is orthogonal to the space spanned by $\{g_i(t) \}_{i=1}^n$. 
	\item $\{g_i(t)\}_{i=1}^{n}$ together with $\{g_i(t)\}_{i=n+1}^{L}$ span the $\Ltwo$ space.
\end{itemize}
Therefore, we have 
\begin{align}
W(t) = \sum_{i = 1}^L W_i g_i(t),
\end{align}
where $\{W_i\}_{i=1}^{L}$ are i.i.d. random variables distributed according to $ \mathcal{N}(0,N_w/2)$. The channel input $X(t) = \sum_{i=1}^n X_i g_i(t)$, where $X_i = 0$ under the null hypothesis $H_0$, and $X_i \in \{-\an,\an\}$ under the alternative hypothesis $H_1$. By mapping Willie's received signal $Z(t) = X(t) + W(t)$ to the orthonormal basis $\{g_i(t)\}_{i=1}^L$, we have $Z(t) = \sum_{i=1}^L Z_i g_i(t)$, where 
\begin{align}
Z_i \triangleq \int_{-\infty}^{\infty} Z(t)g_i(t) dt = \begin{cases}
X_i + W_i, \ i \in \llbracket 1, n \rrbracket, \\
W_i, \ i \in \llbracket n+1, L \rrbracket.
\end{cases} \label{eq:10}
\end{align}  

Let $\Z$ be the length-$n$ random vector consisting of $(Z_1, \ldots, Z_n)$, and $\Zc$ be the random vector consisting of $(Z_{n+1}, \ldots, Z_L)$. Intuitively, $\Zc$ does not play any role in the hypothesis test, since the probability densities of $\Zc$ only depend on the i.i.d. Gaussian noise $(W_{n+1},\ldots,W_L)$. Lemma~\ref{lemma:sufficient} below formalizes the above intuition by showing that the random vector $\Z$ forms a sufficient statistic for Willie to determine $H_0$ or $H_1$. The definition of sufficient statistics when observing a stochastic process, which is adapted from~\cite[Chapter 26.3]{lapidoth2009foundation}, is provided here for reference.
\begin{definition}[Sufficient Statistics] \label{def:sufficient}
	The random vector $\Z$ is a sufficient statistic for estimating the transmission status $\Lambda \in \{0,1\}$ based on the stochastic process $Z(t)$ if:
	\begin{enumerate}
		\item $\Z$ is computable from the stochastic process $Z(t)$;
		\item For any finite number of samples $d \in \mathbb{N}$ of the observations $\Zl = (Z(t_1), \ldots, Z(t_d))$, $\Lambda-\Z-\Zl$ forms a Markov chain for any prior on $\Lambda$. 
	\end{enumerate} 
\end{definition}

\begin{lemma} \label{lemma:sufficient}
	The random vector $\Z = (Z_1, \ldots, Z_n)$ forms a sufficient statistic for determining $H_0$ and $H_1$.
\end{lemma}
\noindent{\it Proof:} The first condition in Definition~\ref{def:sufficient} is satisfied since each element of $\Z$ can be computed via 
\begin{align}
Z_i = \int_{-\infty}^{\infty} Z(t)g_i(t) dt, \ \forall i \in \llbracket 1, n\rrbracket.
\end{align} 
For any finite number of samples $d \in \mathbb{N}$ of the observations $\Zl = \zl$, any realization $\z$ of the random vector $\Z$, and any transmission status $\Lambda \in \{0,1\}$, we have 
\begin{align}
f_{\Zl|\Z,\Lambda}\left(\zl|\z,\Lambda \right)
&= \int f_{\Zc|\Z,\Lambda}\left(\zc|\z,\Lambda \right) f_{\Zl|\Z,\Zc,\Lambda}\left(\zl|\z,\zc,\Lambda \right) d\zc  \\
&= \int f_{\Zc|\Z}\left(\zc|\z \right) f_{\Zl|\Z,\Zc,\Lambda}\left(\zl|\z,\zc,\Lambda \right) d\zc \label{eq:visa1} \\
&= \int f_{\Zc|\Z}\left(\zc|\z \right) f_{\Zl|\Z,\Zc}\left(\zl|\z,\zc \right) d\zc  \label{eq:visa2}\\
&= f_{\Zl|\Z}\left(\zl|\z \right),
\end{align}
which implies $\Lambda-\Z-\Zl$ forms a Markov chain.
Equation~\eqref{eq:visa1} holds since $\Zc = (Z_{n+1},\ldots, Z_{L})$ is independent of the channel inputs $\{X_i\}_{i=1}^n$, hence, independent of the transmission status $\Lambda$. Equation~\eqref{eq:visa2} holds since $Z(t)$ is completely determined by $\z$ and $\zc$, and $\zl = (z(t_1), \ldots, z(t_d))$ only depends on $Z(t)$.  \qed

\vspace{3pt}
\noindent{\bf \underline{2) Distributions of Interest:}}
The fact that $\Z$ is a sufficient statistic for determining $H_0$ or $H_1$ implies that there exists a hypothesis test based on $\Z$ that is as good as\footnote{Also, no hypothesis test based on $\Z$ outperforms the optimal hypothesis test based on $Z(t)$~\cite[Theorem 20.11.5]{lapidoth2009foundation}.} the optimal hypothesis test based on $Z(t)$~\cite[Proposition 20.12.3]{lapidoth2009foundation}. We denote the optimal test based on $\Z$ by $\Phi^{\ast}(\Z)$. A classical result in hypothesis testing~\cite[Theorem 13.1.1]{lehmann2006testing} shows that the optimal test $\Phi^{\ast}(\Z)$ satisfies 
\begin{align}
P_{\text{FA}}(\Phi^{\ast}) + P_{\text{MD}}(\Phi^{\ast}) = 1 - \V\left(Q_{\Z|H_1}, Q_{\Z|H_0}\right),
\end{align}
where $Q_{\Z|H_1}$ and $Q_{\Z|H_0}$ are distributions on $\Z$ conditioned on $H_1$ and $H_0$ respectively. Hence, by recalling the definition of covertness (Definition~\ref{def:covert}), the communication is said to be $(1-\delta)$-covert if 
\begin{align}
\V(Q_{\Z|H_1},Q_{\Z|H_0}) \le \delta. \label{eq:covert}
\end{align} 

From now on, it suffices to study the converted discrete-time Gaussian channels instead of the continuous-time AWGN channels. The discrete-time Gaussian channels is described by $(\mathcal{X},\wzx,\mathcal{Z})$, where $\mathcal{X}$ is chosen to be $\{-\an,0,\an\}$, $\mathcal{Z} = \mathbb{R}$, and $\wzx$ is the channel transition probability given by
\begin{align}
\wzx(z|x) = \frac{1}{\sqrt{\pi N_w}}\exp\left(-\frac{(z-x)^2}{N_w}\right), \label{eq:wzx}
\end{align}
since $Z_i = X_i + W_i$ for every $i \in \llbracket1,n\rrbracket$, and $W_i \sim \mathcal{N}(0,N_w/2)$. We respectively define the output distributions of $Z$ when $X$ is equal to $0,\an,-\an$, as
\begin{align}
&\Q(z) \triangleq \wzx(z|0), \ \Qa(z) \triangleq \wzx(z|\an), \ Q_{-a}(z) \triangleq \wzx(z|-\an), \ \forall z \in \mathbb{R}.
\end{align}   
Further, we define the output distribution of $Z$ induced by $P_X$ (defined in~\eqref{eq:px}) and $W_{Z|X}$ as 
\begin{align}
\Qt(z) \triangleq \sum_{x \in \{-\an,\an\}}\Px(x) \wzx(z|x) =  \frac{1}{2\sqrt{\pi N_w}}\exp\left(-\frac{(z-\an)^2}{N_w}\right) + \frac{1}{2\sqrt{\pi N_w}}\exp\left(-\frac{(z+\an)^2}{N_w}\right), \ \forall z \in \mathbb{R}. \notag
\end{align} 
The channel transition probability corresponding to $n$ channel uses is denoted by $\wzxn \triangleq \prod_{i=1}^n \wzx$ while the $n$-letter product distributions corresponding to $\Q$ and $\Qt$ are respectively denoted by $\Qn \triangleq \prod_{i=1}^n \Q$ and $\Qtn \triangleq \prod_{i=1}^n \Qt$. 

Note that $Q_{\Z|H_0}$ --- the distribution of $\Z$ under $H_0$ --- equals $\Qn$, since the channel inputs $X_i = 0$ for every channel use when Alice is silent. On the other hand, the distribution of $\Z$ under $H_1$ depends on the specific code $\C$ used by Alice and Bob, and is given by
\begin{align}
Q_{\Z|H_1}(\z) \triangleq \frac{1}{\N K}\sum_{m=1}^{\N}\sum_{s=1}^{K}\wzxn(\z|\x_{ms}), \ \forall \z \in \mathbb{R}^n.  
\end{align}   
For notational convenience we use $\Qh(\z)$ to represent the distribution of $\Z$ under $H_1$, i.e.,
\begin{align}
\Qh(\z) \triangleq Q_{\Z|H_1}(\z), \ \forall \z \in \mathbb{R}^n.
\end{align} 

\vspace{3pt}
\noindent{\bf \underline{3) Bounding $\V(\Qh,\Qn)$:}} 
In the following, we show that with high probability over the code design, $\V(\Qh,\Qn) \le \delta$, hence the communication is $(1-\delta)$-covert. By the {\it triangle inequality}, we have 
\begin{align}
\V(\Qh,\Qn) \le \V(\Qtn,\Qn) + \V(\Qh, \Qtn).
\end{align}
Lemmas~\ref{lemma:distance1} and~\ref{lemma:distance2} below bound the two terms $\V(\Qtn,\Qn)$ and $\V(\Qh, \Qtn)$ respectively. The proof of Lemma~\ref{lemma:distance1} relies on the {\it Berry-Esseen Theorem}, and we recall the theorem below for completeness. 

\vspace{3pt}
\begin{theorem}[Berry-Esseen Theorem]
	Suppose $X_1, \ldots, X_n$ are $n$ independent random variables, then we have 
	\begin{align}
	\left|\PP \left(\sum_{i=1}^n (X_i - \E(X_i)) \ge \lambda \sigma\right) - Q(\lambda)\right| \le \frac{6S}{\sigma^3},
	\end{align}
	where $\lambda > 0$, $\sigma^2 = \sum_{i=1}^n \emph{Var}(X_i)$, and $S = \sum_{i=1}^n \E(|X_i - \E(X_i)|^3)$.
\end{theorem}

\begin{lemma} \label{lemma:distance1}
By setting $\an = \left(\frac{2}{n}\right)^{1/4}\sqrt{Q^{-1}\left(\frac{1-\delta}{2}\right)N_w}\cdot \left(1 - n^{-1/8}\right)$, for large enough $n$, we have 
\begin{align}
\V(\Qtn,\Qn) =  \delta - \mathcal{O}(n^{-1/8}).
\end{align}
\end{lemma}
\noindent{\it Proof:} The proof techniques are adapted from~\cite[Lemma 8]{tahmasbi2018first} for DMCs but are specialized to Gaussian channels. Note that 
\begin{align}
\V\left(\Qtn,\Qn\right) &= \PP_{\Qtn}\left(\Qtn(\Z) \ge \Qn(\Z)\right) - \PP_{\Qn}\left(\Qtn(\Z) \ge \Qn(\Z)\right) \\
&= \PP_{\Qtn}\left(\sum_{i=1}^n \log \frac{\Qt(Z_i)}{\Q(Z_i)} \ge 0  \right) - \PP_{\Qn}\left(\sum_{i=1}^n \log \frac{\Qt(Z_i)}{\Q(Z_i)} \ge 0  \right).
\end{align}
As proved in Appendix~\ref{appendix:berry}, for every $i \in \llbracket1,n\rrbracket$,
\begin{align}
&\mu_0 \triangleq \E_{\Q}\left(\log \frac{\Qt(Z_i)}{\Q(Z_i)} \right) = -\frac{\an^4}{N_w^2} + \mathcal{O}(\an^6), \label{eq:jia1} \\
&\sigma_0^2 \triangleq \text{Var}_{\Q}\left(\log \frac{\Qt(Z_i)}{\Q(Z_i)} \right) = \frac{2\an^4}{N_w^2} + \mathcal{O}(\an^6), \label{eq:jia2}\\
&s_0 \triangleq \E_{\Q}\left(\left|\log \frac{\Qt(Z)}{\Q(Z)} - \mu_0\right|^3 \right) = \mathcal{O}(\an^6), \label{eq:jia3}
\end{align} 
hence,
\begin{align}
\sigma_{\Q}^2 \triangleq \sum_{i=1}^n \sigma^2_0 = \frac{2\an^4n}{N_w^2} + \mathcal{O}\left(n^{-1/2}\right), \ \ S_{\Q} \triangleq \sum_{i=1}^n s_0 = \mathcal{O}\left(n^{-1/2}\right).
\end{align}
By the Berry-Esseen Theorem, we have 
\begin{align}
\PP_{\Qn}\left(\sum_{i=1}^n \log \frac{\Qt(Z_i)}{\Q(Z_i)} \ge 0  \right) &= \PP_{\Qn}\left(\sum_{i=1}^n \log \frac{\Qt(Z_i)}{\Q(Z_i)} \ge n\mu_0 + \sigma_{\Q}\left(-\frac{n\mu_0}{\sigma_{\Q}}\right) \right) \\
&\le Q\left(-\frac{n\mu_0}{\sigma_{\Q}}\right) + \frac{6S_{\Q}}{\sigma_{\Q}^3} = Q\left(\frac{\an^2}{N_w}\sqrt{\frac{n}{2}}\right) + \mathcal{O}\left(n^{-1/4}\right).
\end{align} 
Similarly, we also show in Appendix~\ref{appendix:berry} that for every $i \in \llbracket1,n\rrbracket$,
\begin{align}
&\widetilde{\mu} \triangleq \E_{\Qt}\left(\log \frac{\Qt(Z_i)}{\Q(Z_i)} \right) = \frac{\an^4}{N_w^2} + \mathcal{O}(\an^6),  \label{eq:jia4}\\
&\widetilde{\sigma}^2 \triangleq \text{Var}_{\Qt}\left(\log \frac{\Qt(Z_i)}{\Q(Z_i)} \right) = \frac{2\an^4}{N_w^2} + \mathcal{O}(\an^6),\label{eq:jia5}\\
&\widetilde{s} \triangleq \E_{\Qt}\left(\left|\log \frac{\Qt(Z)}{\Q(Z)} - \widetilde{\mu}\right|^3 \right) = \mathcal{O}(\an^6). \label{eq:jia6}
\end{align} 
By defining 
\begin{align}
\sigma_{\Qt}^2 \triangleq \sum_{i=1}^n \widetilde{\sigma}^2 = \frac{2\an^4n}{N_w^2} + \mathcal{O}\left(n^{-1/2}\right), \ \ S_{\Qt} \triangleq \sum_{i=1}^n \widetilde{s} = \mathcal{O}\left(n^{-1/2}\right),
\end{align}
the Berry-Esseen Theorem yields 
\begin{align}
\PP_{\Qtn}\left(\sum_{i=1}^n \log \frac{\Qt(Z_i)}{\Q(Z_i)} \ge 0  \right) &= \PP_{\Qtn}\left(\sum_{i=1}^n \log \frac{\Qt(Z_i)}{\Q(Z_i)} \ge n\widetilde{\mu} + \sigma_{\Qt}\left(-\frac{n\widetilde{\mu}}{\sigma_{\Qt}}\right) \right) \\
&\le Q\left(-\frac{n\widetilde{\mu}}{\sigma_{\Qt}}\right) + \frac{6S_{\Qt}}{\sigma_{\Qt}^3} = 1 - Q\left(\frac{\an^2}{N_w}\sqrt{\frac{n}{2}}\right) + \mathcal{O}\left(n^{-1/4}\right).
\end{align}
Finally, by setting $\an = \left(\frac{2}{n}\right)^{1/4}\sqrt{Q^{-1}\left(\frac{1-\delta}{2}\right)N_w}\cdot \left(1 - n^{-1/8}\right)$, for large enough $n$, we have
\begin{align}
\V\left(\Qtn,\Qn\right) &= 1 - 2Q\left(\frac{\an^2}{N_w}\sqrt{\frac{n}{2}}\right) + \mathcal{O}\left(n^{-1/4}\right) \\
&= 1 - 2Q\left(Q^{-1}\left(\frac{1-\delta}{2}\right)\left(1 - \frac{1}{\log n}\right)^2\right) + \mathcal{O}\left(n^{-1/4}\right) \\
& = 1 - 2Q\left(Q^{-1}\left(\frac{1-\delta}{2}\right)\right) - \sqrt{\frac{2}{\pi}}\int_{Q^{-1}\left(\frac{1-\delta}{2}\right)\left(1-n^{-1/8}\right)^2}^{Q^{-1}\left(\frac{1-\delta}{2}\right)} \exp\left(-\frac{u^2}{2}\right) du + \mathcal{O}\left(n^{-1/4}\right) \label{eq:q}\\
& = \delta - \mathcal{O}(n^{-1/8}),
\end{align}
where~\eqref{eq:q} follows from the definition of the $Q$-function.
\qed

\begin{lemma} \label{lemma:distance2}
	With probability at least $1 - \exp\left(-\mathcal{O}\left(n^{-1/4}\right)\right) $ over the code design, a randomly chosen code $\C$ satisfies $\V(\Qh,\Qtn) \le n^{-1/4}$.
\end{lemma}
\noindent{\it Proof:} Our proof directly follows from~\cite[Lemma 5]{7407378} and~\cite[Theorem VII.1]{cuff2013distributed}, which states that for any channel $(\mathcal{X},\wzx,\mathcal{Z})$ and any $\tau > 0$,
\begin{align}
\E\left(\V\left(\Qh,\Qtn\right)\right) \le \PP_{\wzxn \Pxn}\left(\log \frac{\wzxn(\Z|\X)}{\Qn(\Z)} > \tau\right) + \frac{1}{2}\sqrt{\frac{e^{\tau}}{\N K}}. \label{eq:lemma}
\end{align}  
When $X = \an$, the random variable $\log \frac{\wzx(Z|\an)}{\Q(Z)}$ with $Z \sim \mathcal{N}(\an,N_w/2)$ is distributed according to  $\mathcal{N}\left(\frac{\an^2}{N_w},\frac{2\an^2}{N_w}\right)$, since 
\begin{align}
\log \frac{\wzx(Z|\an)}{\Q(Z)} = \frac{2\an}{N_w}Z - \frac{\an^2}{N_w} \label{eq:new1}.
\end{align}
Similarly, when $X = -\an$, the random variable $\log \frac{\wzx(Z|-\an)}{\Q(Z)}$ with $Z \sim \mathcal{N}(-\an,N_w/2)$ is also distributed according to  $\mathcal{N}\left(\frac{\an^2}{N_w},\frac{2\an^2}{N_w}\right)$, since 
\begin{align}
\log \frac{\wzx(Z|-\an)}{\Q(Z)} = -\frac{2\an}{N_w}Z - \frac{\an^2}{N_w}. \label{eq:new2}
\end{align}
Note that both random variables are sub-Gaussian with parameter $2\an^2/N_w$.
We now consider the first term of~\eqref{eq:lemma} as follows.
\begin{align}
\PP_{\wzxn \Pxn}\left(\log \frac{\wzxn(\Z|\X)}{\Qn(\Z)} > \tau\right)
&=\sum_{\x \in \{-\an,\an\}^n} \Pxn(\x) \cdot \PP_{\wzxn}\left(\sum_{i=1}^n\log \frac{\wzx(Z_i|x_i)}{\Q(Z_i)} > \tau\right) \label{eq:wei1}
\end{align}
For any $\x \in \{-\an,\an\}^n$, because of~\eqref{eq:new1} and~\eqref{eq:new2}, we have 
\begin{align}
\E_{\wzxn}\left(\sum_{i=1}^n\log \frac{\wzx(Z_i|x_i)}{\Q(Z_i)}\right) = \sum_{i=1}^n \E_{W_{Z|X=x_i}}\left(\log \frac{\wzx(Z_i|x_i)}{\Q(Z_i)}\right) = \frac{\an^2}{N_w},
\end{align}
and by setting $\tau = (1+n^{-1/8})\an^2n/N_w$, the Hoeffding's inequality guarantees
\begin{align}
\PP\left(\sum_{i=1}^n \log\frac{\wzx(Z_i|x_i)}{\Q(Z_i)} > (1+n^{-1/8})\frac{\an^2n}{N_w} \right) \le \exp\left(-\frac{\an^2 n^{3/4}}{4 N_w}\right). \label{eq:wei2}
\end{align}
By plugging~\eqref{eq:wei2} into~\eqref{eq:wei1}, the first term of~\eqref{eq:lemma} can be bounded from above as  
\begin{align}
\PP_{\wzxn \Pxn}\left(\log \frac{\wzxn(\Z|\X)}{\Qn(\Z)} >  (1+n^{-1/8})\frac{\an^2 n}{N_w} \right) \le \exp\left(-\frac{ \an^2 n^{3/4}}{4 N_w}\right) = \exp\left(-\mathcal{O}\left(n^{1/4}\right)\right). \label{eq:term1}
\end{align} 
Since $\log \N + \log K \ge \left(1+ \frac{1}{\log n}\right)\frac{\an^2n}{N_w}$ (regardless of the ratio $N_w/N_b$), the second term of~\eqref{eq:lemma} satisfies
\begin{align}
\frac{1}{2}\sqrt{\frac{e^{\tau}}{\N K}} \le \frac{1}{2}\sqrt{\frac{\exp\left((1+n^{-1/8})\frac{\an^2n}{N_w}\right)}{\exp\left(\left(1+ \frac{1}{\log n}\right)\frac{\an^2n}{N_w} \right)}} = \exp\left(-\mathcal{O}\left(\frac{\sqrt{n}}{\log n}\right)\right). \label{eq:term2}
\end{align}
Combining~\eqref{eq:term1} and~\eqref{eq:term2}, we have
\begin{align}
\E\left(\V\left(\Qh,\Qtn\right)\right) \le \PP_{\wzxn \Pxn}\left(\log \frac{\wzxn(\Z|\X)}{\Qn(\Z)} > \tau\right) + \frac{1}{2}\sqrt{\frac{e^{\tau}}{\N K}}  \le  \exp\left(-\mathcal{O}\left(n^{1/4}\right)\right).
\end{align}
Finally, by the Markov's inequality, with probability at least $1 - \exp\left(-\mathcal{O}(n^{1/4})\right) $, a randomly chosen code $\C$ satisfies $\V(\Qh,\Qtn) \le n^{-1/4}$.
\qed 

Combining Lemmas~\ref{lemma:distance1} and~\ref{lemma:distance2}, we prove that for sufficiently large $n$, 
\begin{align}
\V(\Qh,\Qn) \le \delta - \mathcal{O}(n^{-1/8}) + n^{-1/4} \le \delta.
\end{align}
By converting $n$ to $T$ via Lemma~\ref{lemma:blocklength}, we conclude that with probability at least $1 - \exp\left(-\mathcal{O}\left((WT)^{1/4}\right)\right)$, a randomly chosen code $\C$ satisfies $\V(\Qh,\Qn) \le \delta$.

\subsection{Analysis of Reliability}\label{sec:reliability}

The legitimate receiver Bob receives $Y(t) = X(t) + B(t)$. Recall that $X(t) = \sum_{i=1}^n X_i g_i(t)$ and $B(t) = \sum_{i=1}^n B_i g_i(t) + \sum_{i=n+1}^L B_i g_i(t)$, where $\{B_i\}_{i=1}^L$ are i.i.d. random variables distributed according to $ \mathcal{N}(0,N_b/2)$. By mapping $Y(t)$ to $\{g_i(t)\}_{i=1}^n$, Bob obtains
\begin{align}
Y_i \triangleq \int_{-\infty}^{\infty}Y(t) g_i(t) dt = \int_{-\infty}^{\infty}X(t) g_i(t) dt + \int_{-\infty}^{\infty}B(t) g_i(t) dt = X_i + B_i, \ \forall i \in \llbracket1,n\rrbracket,
\end{align}     
and thus, he converts the continuous-time AWGN channel to a discrete-time Gaussian channel with $n$ channel uses. We denote the corresponding discrete-time Gaussian channel by $(\mathcal{X},\wyx,\mathcal{Y})$, where $\mathcal{X}$ is chosen to be $\{-\an,0,\an\}$, $\mathcal{Y} = \mathbb{R}$, and 
\begin{align}
\wyx(y|x) = \frac{1}{\sqrt{\pi N_b}} \exp\left(-\frac{(y-x)^2}{N_b}\right).
\end{align}
When $X$ is equal to $0, \an, -\an$, we respectively define the output distributions of $Y$ as 
\begin{align}
P_0(y) \triangleq \wyx(y|0), \ \Pa(y) \triangleq \wyx(y|\an), \ P_{-a}(y) \triangleq \wyx(y|-\an), \ \forall y \in \mathbb{R}.
\end{align}
Also, we define the distribution of $Y$ induced by $\Px$ and $\wyx$ as 
\begin{align}
\Pt(y) \triangleq \sum_{x \in \{-\an,\an\}}\Px(x) \wyx(y|x) =  \frac{1}{2\sqrt{\pi N_b}}\exp\left(-\frac{(y-\an)^2}{N_b}\right) + \frac{1}{2\sqrt{\pi N_b}}\exp\left(-\frac{(y+\an)^2}{N_b}\right), \ \forall y \in \mathbb{R}. \notag
\end{align}
Similar to Section~\ref{sec:covert}, we use $\wyxn$, $\Pn$, and $\Ptn$ to denote the $n$-letter product distributions $\prod_{i=1}^n \wyx$, $\prod_{i=1}^n P_0$, and $\prod_{i=1}^n \Pt$, respectively.

We now turn to analyze the reliability of our achievability scheme. Let 
\begin{align}
\Aset \triangleq \left\{(\x,\y): \log\frac{\wyxn(\y|\x)}{\Pn(\y)} > \gamma \right\},
\end{align}
where $\gamma > 0$ will be specified later. 
Based on the length-$n$ vector $\y$ and the shared key $s$, Bob's decoder operates as follows:
\begin{itemize}
	 \item if there exists a unique codeword $\x_{ms}$ such that $(\x_{ms},\y) \in \Aset$, output $\widehat{M}=m$ and $\widehat{\Lambda}=1$,
	 \item if no codeword in $\{\x_{ms}\}_{m=1}^M$ satisfies $(\x_{ms},\y) \in \Aset$, output $\widehat{M} = 0$ and $\widehat{\Lambda} = 0$,
	 \item otherwise, declare an error.
\end{itemize}	 
Instead of studying the desired max-average probability of error $P_{\text{err}}$ (defined as per~\eqref{eq:error}) directly, we first show that with high probability, a randomly chosen code ensures a decaying {\it average probability of error}  
\begin{align*}
P_{\text{err}}^{\text{(avg)}} \triangleq \E_{\s}\big\{\PP(M \ne \widehat{M}|\s,\Lambda = 1) + \PP(\widehat{M} \ne 0 | \Lambda = 0)\big\},
\end{align*}
which is averaged over both the message and the shared key. 

\begin{lemma} \label{lemma:reliability}
	For any $\gamma > 0$, there exists a constant $C_1 > 0$ such that
	\begin{align}
	\E\left(P_{\emph{err}}^{\emph{(avg)}}\right) \le \PP_{\wyxn \Pxn}\left(\log \frac{\wyxn(\Y|\X)}{\Pn(\Y)} \le \gamma \right) + C_1 \N e^{-\gamma} \label{eq:lemma1}
	\end{align} 
\end{lemma}
Lemma~\ref{lemma:reliability} is adapted from~\cite[Lemma 3]{7407378} for DMCs, and the proof can be found in Appendix~\ref{appendix:reliability}. Similar to~\eqref{eq:new1} and~\eqref{eq:new2} in Sub-section~\ref{sec:covert}, when $X = \an$ and $X = -\an$, the corresponding random variables $\log \frac{\wyx(Y|\an)}{P_0(Y)}$ with $Y \sim \mathcal{N}(\an,N_b)$ and $\log \frac{\wyx(Y|-\an)}{P_0(Y)}$ with $Y \sim \mathcal{N}(-\an,N_b)$ are both distributed according to $\mathcal{N}\left(\frac{\an^2}{N_b},\frac{2\an^2}{N_b}\right)$. Hence, both random variables are sub-Gaussian with parameter $2\an^2/N_b$. Note that
\begin{align}
\PP_{\wyxn \Pxn}\left(\log \frac{\wyxn(\Y|\X)}{\Pn(\Y)} \le \gamma \right)
&=\sum_{\x \in \{-\an,\an\}^n} \Pxn(\x) \cdot \PP_{\wyxn}\left(\sum_{i=1}^n\log \frac{\wyx(Y_i|x_i)}{P_0(Y_i)} \le \gamma \right) \label{eq:new5}
\end{align}
For any $\x \in \{-\an,\an\}^n$, we have  
\begin{align}
\E_{\wyxn}\left(\sum_{i=1}^n\log \frac{\wyx(Y_i|x_i)}{P_0(Y_i)}\right) = \sum_{i=1}^n \E_{W_{Y|X=x_i}}\left(\log \frac{\wyx(Y_i|x_i)}{P_0(Y_i)}\right) = \frac{\an^2 n}{N_b},
\end{align}
and by setting $\gamma = (1-n^{-1/8})\frac{\an^2 n}{N_b}$, the Hoeffding's inequality guarantees 
\begin{align}
\PP_{\wyxn}\left(\sum_{i=1}^n \log\frac{\wyx(Y_i|x_i)}{P_0(Y_i)} \le (1-n^{-1/8})\frac{\an^2n}{N_b} \right) \le \exp\left(-\frac{ \an^2 n^{3/4}}{4 N_b}\right). \label{eq:rep1}
\end{align}
By plugging~\eqref{eq:rep1} into~\eqref{eq:new5}, we have  
\begin{align}
\PP_{\wyxn \Pxn}\left(\log \frac{\wyxn(\Y|\X)}{\Pn(\Y)} \le  (1-n^{-1/8})\frac{\an^2n}{N_b} \right) \le \exp\left(-\frac{ \an^2 n^{3/4}}{4N_b}\right) = \exp\left(-\mathcal{O}(n^{1/4})\right).
\end{align} 
Since the value of $\log \N$ is set to be $\left(1 -\frac{1}{\log n}\right)\frac{\an^2 n}{N_b}$, the second term of~\eqref{eq:lemma1} is given by 
\begin{align}
C_1 \N e^{-\gamma} = C_1\exp\left(\left(1 -\frac{1}{\log n}\right)\frac{\an^2 n}{N_b} - \left(1-n^{-1/8}\right)\frac{\an^2 n}{N_b} \right) = \exp\left(-\mathcal{O}\left(\frac{\sqrt{n}}{\log n}\right)\right).
\end{align}
Therefore, we have
\begin{align}
\E\left(P_{\text{err}}^{\text{(avg)}}\right) \le \PP_{\wyxn \Pxn}\left(\log \frac{\wyxn(\Y|\X)}{\Pn(\Y)} \le \gamma \right) + \N e^{-\gamma}(1+2n) \le \exp\left(-\mathcal{O}(n^{1/4})\right).
\end{align}
The Markov's inequality ensures that with probability at least $1 - \exp\left(-\mathcal{O}(n^{1/4})\right)$, a randomly chosen code $\C$ guarantees 
\begin{align}
P_{\text{err}}^{\text{(avg)}} \le \exp\left(-\mathcal{O}(n^{1/4})\right). \label{eq:ave}
\end{align}

Though~\eqref{eq:ave} is weak in the sense that it only guarantees a small $P_{\text{err}}^{\text{(avg)}}$, we also develop a novel result showing that given a code $\C$ with small $P_{\text{err}}^{\text{(avg)}}$ and covertness guarantee, we can construct another code $\C'$ with small $P_{\text{err}}$ and simultaneously preserve the covertness property by merely rearranging the codewords in $\C$ (without expurgating existing codewords or adding new codewords).
\begin{lemma} \label{lemma:error}
	Suppose a code $\C$ contains $K$ sub-codes of size $M$, and guarantees $\V(\Qh,\Qn) \le \delta$, and $P_{\emph{err}}^{\emph{(avg)}}(\C) \le \epsilon_n$.
	Then, there exists a code $\C'$ containing $K'$ sub-codes of size $M'$ such that $\V(\Qh,\Qn) \le \delta$, and $P_{\emph{err}}(\C') \le \epsilon'_n$. In particular, $\lim_{n \to \infty} \epsilon_n = \lim_{n \to \infty} \epsilon'_n = 0$, $\lim_{n \to \infty} \frac{M'}{M} = 1$, and $\lim_{n \to \infty} \frac{K'}{K} = 1$.
\end{lemma} 
\noindent{\it Proof:} We partition the $K$ sub-codes of the code $\C$ into two groups --- {\it good sub-codes} and {\it bad sub-codes}. A sub-code $\C_s$ is good if its probability of error satisfies
\begin{align}
P_{\text{err}}(\C_s) \triangleq  \PP(M \ne \widehat{M}|\s=s,\Lambda = 1) + \PP(\widehat{M} \ne 0 | \Lambda = 0) \le \sqrt{\epsilon_n}, 
\end{align} 
and is bad otherwise. Note that $P_{\text{err}}^{\text{(avg)}}(\C) = \E_{\s}\left(P_{\text{err}}(\C_\s)\right)$. Since $\E_{\s}\left(P_{\text{err}}(\C_\s)\right) \le \epsilon_n$, the Markov's inequality guarantees that the number of bad sub-codes is at most $\sqrt{\epsilon_n} K$. Let 
\begin{align}
K' \triangleq (1-\sqrt{\epsilon_n})K. 
\end{align}
Without loss of generality, we assume the first $K'$ sub-codes $\C_1, \C_2, \ldots, \C_{K'}$ are good, and the last $(K-K')$ sub-codes $\C_{K'+1}, \ldots, \C_K$ are bad. Let $\C_{\mathcal{B}} \triangleq \cup_{s \in \llbracket K'+1,K\rrbracket} \C_s$ be the union of the codewords in the bad sub-codes. The size of $\C_{\mathcal{B}}$ is $\sqrt{\epsilon_n} MK$. 

We now construct a new code $\C'$ that contains $K'$ sub-codes $\C'_1, \C'_2, \ldots, \C'_{K'}$.   
\begin{enumerate}
	\item Let $\widehat{\epsilon}_n = \frac{\sqrt{\epsilon_n}}{1-\sqrt{\epsilon_n}}$. We arbitrarily partition $\C_{\mathcal{B}}$ into $K'$ disjoint subsets $\C_{\mathcal{B}}^{(1)}, \ldots, \C_{\mathcal{B}}^{(K')}$ of equal size. The size of each $\C_{\mathcal{B}}^{(s)}$ ($s \in \llbracket 1,K' \rrbracket$) is 
	\begin{align}
	\frac{|\C_{\mathcal{B}}|}{K'} = \frac{\sqrt{\epsilon_n} MK}{(1-\sqrt{\epsilon_n})K} \triangleq \widehat{\epsilon}_n M.
	\end{align}
	\item Let $\C'_s = \C_s \cup \C_{\mathcal{B}}^{(s)}$ for every $s \in \llbracket1,K'\rrbracket$; hence, the size of each sub-code $\C'_s$ is 
	\begin{align}
	M' \triangleq (1+\widehat{\epsilon}_n)M.
	\end{align} 
	\item For every $s \in \llbracket 1,K' \rrbracket$, the decoding region for $\C'_s$ is exactly the same as that for $\C_s$. We do not assign any decoding region for codewords from $\C_{\mathcal{B}}^{(s)}$, thus the probability of decoding error when $\x \in \C_{\mathcal{B}}^{(s)}$ is transmitted equals one. However, since the probability of transmitting a codeword from $\C_{\mathcal{B}}^{(s)}$ is negligible, the average probability of error of $\C'_s$ is bounded from above as 
	\begin{align}
	P_{\text{err}}(\C'_s) \le \frac{|\C_s|}{|\C'_s|}\cdot P_{\text{err}}(\C_s) + \frac{|\C_{\mathcal{B}}^{(s)}|}{|\C'_s|} \cdot 1 &\le P_{\text{err}}(\C_s) + \frac{\widehat{\epsilon}_n M}{(1+\widehat{\epsilon}_n)M} \\
	&\le \sqrt{\epsilon_n} + \widehat{\epsilon}_n, \label{eq:ge}
	\end{align}
	where~\eqref{eq:ge} holds since $\C_s$ (for $s \in \llbracket 1,K' \rrbracket$) is a good sub-code with probability of error at most $\sqrt{\epsilon_n}$.
	By setting $\epsilon'_n \triangleq \sqrt{\epsilon_n} + \widehat{\epsilon}_n$, we have 
	\begin{align}
	P_{\text{err}}(\C') = \max_{s \in \llbracket1,K'\rrbracket}P_{\text{err}}(\C'_s) \le \sqrt{\epsilon_n} + \widehat{\epsilon}_n = \epsilon'_n.
	\end{align}  
\end{enumerate}
We construct $\C'$ by merely rearranging the codewords in $\C$, without expurgating existing codewords or adding new codewords. Hence, if the code $\C$ satisfies $\V(\Qh,\Qn) \le \delta$, the new code $\C'$ also ensures $\V(\Qh,\Qn) \le \delta$.
\qed

\section{Converse}
\label{sec:converse}
In order to establish the necessity of a square-root law, one needs to show how the covertness constraint implies that the energies (spread over all the frequencies) of a substantial fraction of codewords are at most $\mathcal{O}(\sqrt{WT})$. However, it is difficult for Willie to directly measure the energy of the transmitted codeword since it is perturbed by additive noise with infinite energy. Instead, a plausible approach for Willie is to focus on a specific bandwidth over which the energy of noise is finite (e.g., $f \in [-\al_i W, \al_i W]$), and show that 
\begin{enumerate}[label=(\alph*)]
    \item there exists a substantial fraction of codewords satisfying the property that the energy allocated in the bandwidth of interest is at most $\mathcal{O}(\sqrt{WT})$ (due to the covertness constraint);
    \item for codewords satisfying the property in (a), their energies (spread over all the frequencies) are also upper bounded by $\mathcal{O}(\sqrt{WT})$ since the spectral mask requires the energy of the codebook to be concentrated. 
\end{enumerate}
While one can prove statement (a), statement (b) may not hold since (i) the spectral mask defined in Definition~\ref{def:mask} only imposes an energy concentration constraint on the codebook instead of every single codeword; and (ii) the energy of each codeword may significantly differ.\footnote{Indeed, suppose one codeword in the codebook has infinite energy and its energy is completely allocated in the bandwidth of interest, and all the remaining codewords satisfy the property in (a). In this case, statement (b) may not be true since, regardless of the energy allocations of other codewords,  the single ``very heavy'' codeword ensures the codebook satisfies the energy concentration constraint imposed by the spectral mask.}

Consequently, we have not been able to provide a full converse for the continuous-time channel. Nevertheless, upon introducing the additional constraint that the ratio between the maximum energy and minimum energy of codewords is  bounded in the definition of spectral mask, we are able to verify that our achievability\footnote{Note that our achievability is still valid since all the codewords have the same energy.} is order-optimal by proving a converse for this continuous-time channel. The result is provided in Appendix~\ref{appendix:converse} and its proof relies on the use of \emph{prolate spheroidal wave functions}~\cite{slepian1961prolate}. Alternatively, we restrict ourselves here to the converted discrete-time AWGN channels characterized by $\mathcal{N}(0,N_w/2)$ and $\mathcal{N}(0,N_b/2)$, and show that the BPSK scheme used in Section~\ref{sec:achievability} is optimal.

\begin{theorem} \label{thm:converse}
Let $\delta \in (0,1)$. For any sequence of codes with $\lim\limits_{n \to \infty} P_{\emph{err}} = 0$ and $\lim\limits_{n \to \infty} \V(\Qh, \Qn) \le \delta$, we have
\begin{align}
\lim_{n \to \infty} \frac{\log M}{\sqrt{n}} \le \frac{\sqrt{2} N_w}{N_b}Q^{-1}\Big(\frac{1-\delta}{2}\Big). \label{eq:converse1}
\end{align}
\end{theorem}

Replacing $\frac{WT}{\min_{\B \in [0,1]} c(\B)}$ by $n$ in Theorems~\ref{thm:result}, one can verify that equation~\eqref{eq:converse1} implies that BPSK is optimal under the variational distance metric. The proof techniques are leveraged from~\cite{tahmasbi2018first, zhang2018covert} --- in particular, Lemmas~\ref{lemma:converse1} and~\ref{lemma:converse2} below are analogous to Lemmas 11 and 12 in~\cite{tahmasbi2018first} (which are derived for DMCs). We provide the detailed proofs in the following.

Consider any code $\C$ consisting of $|\C|$ length-$n$ codewords $\x$. For any codeword $\x \in \C$, let $\lx \triangleq \sum_{i=1}^n x_i^2$ be the power of $\x$, and $\Pm \triangleq \min_{\x \in \C} \lx$ be the minimum power among the code $\C$.
\begin{lemma} \label{lemma:converse1}
	For any code $\C$ with minimum power $P_{\emph{min}}$, the variational distance between $\Qh$ and $\Qn$ is bounded from below as 
	\begin{align}
	\V\left(\Qh,\Qn\right)  \ge 1 - 2Q\left(\frac{P_{\emph{min}}}{\sqrt{2n}N_w}\right) - \frac{P_{\emph{min}}^2}{\sqrt{\pi}N_w^2 n^{3/2}} - \frac{\nu_1+\nu_2}{\sqrt{n}},
	\end{align}
	where $\nu_1$ and $\nu_2$ are constants specified later.
\end{lemma}
\noindent{\it Proof:}
Willie uses a {\it power detector} $\Phi_\tau$ with threshold $\tau$ to produce an estimation $\widehat{\Lambda}$ of Alice's transmission status. The decision rule is given by 
\begin{align}
\widehat{\Lambda} = \Phi_\tau(Z(t)) = \begin{cases}
1, \ \text{ if } \sum_{i=1}^n Z_i^2 \ge \tau, \\
0, \ \text{ if } \sum_{i=1}^n Z_i^2 < \tau,
\end{cases}
\end{align} 
where $Z_i = \int_{-\infty}^{\infty}Z(t)g_i(t) dt$, $\forall i \in \llbracket1,n\rrbracket$. 
Under the null hypothesis $H_0$, for every $i \in \llbracket1,n\rrbracket$, $Z_i = W_i$ is a Gaussian random variable with zero mean and variance $N_w/2$, hence
\begin{align}
&\E_{H_0}(Z_i^2) = \E_{H_0}(W_i^2) = \frac{N_w}{2},\\
&\text{Var}_{H_0}(Z_i^2) = \E_{H_0}(W_i^4) - \left(\E_{H_0}(W_i^2)\right)^2 = \frac{3}{4}N_w^2 - \frac{1}{4}N_w^2 = \frac{N_w^2}{2}, \\
&S_{H_0} = \sum_{i=1}^n \E_{H_0}\left(\left|Z_i^2 - \E_{H_0}(Z_i^2) \right|^3 \right) = \mathcal{O}(n).
\end{align}
By using the Berry-Esseen Theorem, the probability of false alarm of $\Phi_\tau$ is bounded from above as 
\begin{align}
P_{\text{FA}}(\Phi_\tau) = \PP_{H_0}\left(\sum_{i=1}^n Z_i^2 \ge \tau \right) &\le Q\left(\frac{\tau - n\E_{H_0}(Z_i^2)}{\sqrt{n\text{Var}_{H_0}(Z_i^2)}}\right) + \frac{6S_{H_0}}{\left(n\text{Var}_{H_0}(Z_i^2)\right)^{3/2}} \\
&\le Q\left(\frac{\tau - nN_w/2}{N_w \sqrt{n/2}}\right) + \frac{\nu_1}{\sqrt{n}},
\end{align}
for some properly chosen constant $\nu_1$.
Under the alternative hypothesis $H_1$, we have $Z_i = X_i + W_i$. For any given $X_i = x$ ($x\in \mathbb{R}$), the expectation and variance of $Z_i$ are respectively given by  
\begin{align}
&\E_{x}(Z_i^2) = \E(Z_i^2|X_i=x) = x^2 + \frac{N_w}{2},\\
&\text{Var}_{x}(Z_i^2) = \text{Var}(Z_i^2|X_i=x) = \E_{x}(Z_i^4) - \left(\E_{x}(Z_i^2)\right)^2 = 2x^2N_w + \frac{N_w^2}{2}, \\
&S_{x} = \sum_{i=1}^n \E_{x}\left(\left|Z_i^2 - \E_{x}(Z_i^2) \right|^3 \right) = \mathcal{O}(n).
\end{align}
By using the Berry-Esseen Theorem, we can also bound the probability of missed detection from above as 
\begin{align}
P_{\text{MD}}(\Phi_\tau) = \PP_{H_1}\left(\sum_{i=1}^n Z_i^2 < \tau \right) &= \frac{1}{|\C|}\sum_{\x \in \C} \PP\left(\sum_{i=1}^n Z_i^2 < \tau | \X = \x\right)  \\
&\le \frac{1}{|\C|}\sum_{\x \in \C} Q\left(\frac{-\tau + \sum_{i=1}^n \E_{x_{i}}\left(Z_i^2\right)}{\sqrt{\sum_{i=1}^n \text{Var}_{x_{i}}\left(Z_i^2\right)}}\right) + \frac{6S_x}{\left(\sum_{i=1}^n \text{Var}_{x_{i}}\left(Z_i^2\right)\right)^{3/2}}\\
&\le \frac{1}{|\C|}\sum_{\x \in \C} Q\left(\frac{-\tau+\frac{nN_w}{2}+\lx}{\sqrt{\frac{nN_w^2}{2}+2\lx N_w}}\right) + \frac{\nu_2}{\sqrt{n}} \\
&\le Q\left(\frac{-\tau+\frac{nN_w}{2}+\Pm}{\sqrt{\frac{nN_w^2}{2}+2\Pm N_w}}\right) + \frac{\nu_2}{\sqrt{n}},
\end{align}
for some properly chosen constant $\nu_2$.
By choosing the threshold $\tau$ to be $\frac{nN_w}{2}+\frac{\Pm}{2}$, we have 
\begin{align}
&P_{\text{FA}}(\Phi_\tau) \le Q\left(\frac{\Pm}{\sqrt{2n}N_w}\right)  + \frac{\nu_1}{\sqrt{n}} \ \text{ and } \ P_{\text{MD}}(\Phi_\tau) \le  Q\left(\frac{\Pm}{\sqrt{2n}N_w}\right) + \frac{\Pm^2}{\sqrt{\pi}N_w^2n^{3/2}} + \frac{\nu_2}{\sqrt{n}}.
\end{align}
Therefore, the variational distance between $\Qh$ and $\Qn$ is bounded from below as 
\begin{align}
\V\left(\Qh,\Qn\right) \ge 1 - P_{\text{FA}}(\Phi_\tau) -P_{\text{MD}}(\Phi_\tau) \ge 1 - 2Q\left(\frac{\Pm}{\sqrt{2n}N_w}\right) - \frac{\Pm^2}{\sqrt{\pi}N_w^2 n^{3/2}} - \frac{\nu_1+\nu_2}{\sqrt{n}}.
\end{align}
\qed

\begin{lemma} \label{lemma:converse2}
	Let $A = \sqrt{2}N_w Q^{-1}\left(\frac{1-\delta}{2}-\frac{2\nu^2}{\sqrt{\pi n}N_w^2} - \gamma\right)$, where $\gamma \in [0,1]$ and $\nu > 0$ is a constant specified later. For any code $\C$ (of size $|\C|$) satisfying $\V(\Qh,\Qn) \le \delta$, there exists a ``low-power'' sub-code $\C^{(l)} \subseteq \C$ satisfying (i) $|\C^{(l)}| \ge \gamma |\C|,$ and (ii) $\forall \x \in \C^{(l)}, \ \lx \le A\sqrt{n}.$

\end{lemma}
\noindent{\it Proof:}
We first partition the code $\C$ into ``low-power'' sub-code $\C^{(l)}$ and ``high-power'' sub-code $\C^{(h)}$, where
\begin{align}
\C^{(l)} \triangleq \{\x \in \C: \sum_{i=1}^n x_i^2 \le A\sqrt{n} \},\ \ \C^{(h)} \triangleq \{\x \in \C: \sum_{i=1}^n x_i^2 > A\sqrt{n} \} =  \C \setminus \C^{(l)}.
\end{align}
The output distributions induced by $\C^{(l)}$ and $\C^{(h)}$ are respectively denoted by 
\begin{align}
\Qh^{(l)}(\z) = \frac{1}{\left|\C^{(l)}\right|}\sum_{\x \in \C^{(l)}} \wzxn(\z|\x), \ \ \Qh^{(h)}(\z) = \frac{1}{\left|\C^{(h)}\right|}\sum_{\x \in \C^{(h)}} \wzxn(\z|\x).
\end{align}  
By the triangle inequality, we have
\begin{align}
\V\left(\Qh,\Qn\right) \ge \frac{\left|\C^{(h)}\right|}{\left|\C\right|}\V\left(\Qh^{(h)},\Qn\right) - \frac{\left|\C^{(l)}\right|}{\left|\C\right|}\V\left(\Qh^{(l)},\Qn\right).
\end{align} 
Since the code $\C$ guarantees $\V(\Qh,\Qn) \le \delta$, one can show that 
\begin{align}
\delta \ge \V\left(\Qh,\Qn\right) &\ge \frac{\left|\C^{(h)}\right|}{\left|\C\right|}\V\left(\Qh^{(h)},\Qn\right) - \frac{\left|\C^{(l)}\right|}{\left|\C\right|}\V\left(\Qh^{(l)},\Qn\right) \\
&= \V\left(\Qh^{(h)},\Qn\right) - \frac{\left|\C^{(l)}\right|}{\left|\C\right|}\left(\V\left(\Qh^{(h)},\Qn\right) + \V\left(\Qh^{(l)},\Qn\right)\right) \\
&\ge \V\left(\Qh^{(h)},\Qn\right) - \frac{2\left|\C^{(l)}\right|}{\left|\C\right|} \label{eq:ham0} \\
&\ge \left(1 - 2Q\left(\frac{A\sqrt{n}}{\sqrt{2n}N_w}\right) - \frac{A^2n}{\sqrt{\pi}N_w^2n^{3/2}} - \frac{\nu_1+\nu_2}{\sqrt{n}} \right) - \frac{2\left|\C^{(l)}\right|}{\left|\C\right|} \label{eq:ham}\\
&\ge \delta + 2\gamma + \frac{4\nu^2}{\sqrt{\pi n}N_w^2} - \frac{A^2}{\sqrt{\pi n}N_w^2}-\frac{\nu_1+\nu_2}{\sqrt{n}} - \frac{2\left|\C^{(l)}\right|}{\left|\C\right|}\label{eq:ham2} \\
&\ge \delta + 2\gamma - \frac{2\left|\C^{(l)}\right|}{\left|\C\right|}, \label{eq:ham3}
\end{align}
where~\eqref{eq:ham0} holds since the variational distance between any two distributions is upper bounded by one, and~\eqref{eq:ham} follows from Lemma~\ref{lemma:converse1}. Inequality~\eqref{eq:ham3} is obtained by choosing $\nu$ to satisfy $\frac{4\nu^2}{\sqrt{\pi n}N_w^2} - \frac{A^2}{\sqrt{\pi n}N_w^2} - \frac{\nu_1+\nu_2}{\sqrt{n}} \ge 0.$
Hence, the size of $\C^{(l)}$ is bounded from below as $|\C^{(l)}| \ge \gamma |\C|.$
\qed

We now consider any code $\C$ consisting of $K$ sub-codes $\C_s$ ($s\in \llbracket1,K\rrbracket$, each of size $M$) that ensures
\begin{itemize} 
\item (covertness) $\V(\Qh,\Qn) \le \delta,$
\item (reliability) $P_{\text{err}}(\C) = \max\limits_{s \in \llbracket1,K\rrbracket}\left\{\PP(M \ne \widehat{M}|\s=s,\Lambda = 1) + \PP(\widehat{M} \ne 0 | \Lambda = 0)\right\} \le \epsilon_n$, where  $\lim\limits_{n \to \infty} \epsilon_n = 0$.
\end{itemize}
From Lemma~\ref{lemma:converse2}, there exists a ``low-power'' sub-code $\C^{(l)} \subseteq \C$ satisfying (i) $|\C^{(l)}| \ge \gamma MK$, and (ii) every codeword $\x$ in $\C^{(l)}$ satisfies $\lx \le \sqrt{2n}N_w Q^{-1}\left(\frac{1-\delta}{2}-\frac{2\nu^2}{\sqrt{\pi n}N_w^2} - \gamma\right)$.
For every $s \in \llbracket 1,K \rrbracket$, we define the intersection between $\C_s$ and $\C^{(l)}$ as 
\begin{align}
\C_s^{(l)} \triangleq \C_s \cap \C^{(l)}.
\end{align}  
Note that there must exist a sub-code $\C_s$ ($s \in \llbracket 1,K \rrbracket$) satisfying $\C_s^{(l)} \ge \gamma M$. Since the average probability of error for $\C_s$ is at most $\epsilon_n$, the Markov's inequality yields that for $\C_s^{(l)}$, the average probability of error satisfies $P_{\text{err}}(\C_s^{(l)})\le \epsilon_n/\gamma$. We choose $\gamma = \max\{\sqrt{\epsilon_n},\exp(-n^{\frac{1}{2}-\epsilon})\}$, for any sufficiently small $\epsilon >0$, to guarantee that $P_{\text{err}}(\C_s^{(l)})$ goes to zero asymptotically as $n$ grows without bound. Let $\widetilde{M}$ be the (uniformly distributed) random variable that corresponds to the message in $\C_s^{(l)}$. By standard information inequalities, we have 
\begin{align}
\log(|\C_s^{(l)}|) = \h(\widetilde{M})    &= \I(\widetilde{M};\Y S) + \h(\widetilde{M}|\Y S) \label{eq:con1}\\
&= \I(\widetilde{M};\Y|S) + \h(\widetilde{M}|\Y S) \\
&\le \I(\widetilde{M}S;\Y) + \h(\widetilde{M}|\Y S) \\
&\le \I(\X;\Y) + (\epsilon_n/\gamma) \log(|\C_s^{(l)}|) + \Hb(\epsilon_n/\gamma) \\
&\le \sum_{i=1}^n \I(X_i;Y_i) + (\epsilon_n/\gamma)\sqrt{\epsilon_n}\log(|\C_s^{(l)}|) + 1 \\
&\le n\I(\widetilde{X}; \widetilde{Y}) + (\epsilon_n/\gamma)\log(|\C_s^{(l)}|) + 1, \label{eq:con2}
\end{align}
where $P_{\widetilde{X}}(a) = \frac{1}{n}\sum\limits_{i=1}^n P_{X_i}(a) = \frac{1}{n}\sum\limits_{i=1}^n \frac{1}{|\C_s^{(l)}|}\sum\limits_{\x \in \C_s^{(l)}}\mathbbm{1}\left\{x_i = a \right\}$ and $P_{\widetilde{X}\widetilde{Y}}(a,b) = P_{\widetilde{X}}(a) \wzx(b|a)$. Note that 
\begin{align}
\I(\widetilde{X}; \widetilde{Y}) = h(\widetilde{Y}) - h(\widetilde{Y}|\widetilde{X}) = h(\widetilde{Y}) - h(\widetilde{B}),
\end{align}
where $\widetilde{B}$ is a Gaussian random variable with zero mean and variance $N_b/2$. Since 
\begin{align}
\E(\widetilde{Y}^2) = \E(\widetilde{X}^2) + \E(\widetilde{B}^2) + 2\E(\widetilde{X})\E(\widetilde{B}) = \E(\widetilde{X}^2) + \frac{N_b}{2} \le \frac{A}{\sqrt{n}} + \frac{N_b}{2},
\end{align}
we have 
\begin{align}
\I(\widetilde{X}; \widetilde{Y}) = h(\widetilde{Y}) - h(\widetilde{B}) &\le \frac{1}{2}\log \left(2\pi e \left(\frac{A}{\sqrt{n}}+\frac{N_b}{2}\right)\right) - \frac{1}{2}\log \left(2\pi e \frac{N_b}{2} \right) \\
&= \sqrt{\frac{2}{n}}\frac{N_w}{N_b}Q^{-1}\left(\frac{1-\delta}{2}\right) + o(n^{-1/2}).
\end{align}
Therefore, 
\begin{align}
\log(|\C_s^{(l)}|) \le \frac{\sqrt{2n}\frac{N_w}{N_b}Q^{-1}\left(\frac{1-\delta}{2}\right) + o(\sqrt{n}) + 1}{1- (\epsilon_n/\gamma)}.
\end{align}
Note that $\log(1/\gamma) = o(\sqrt{n})$ since we set $\gamma = \max\{\sqrt{\epsilon_n},\exp(-n^{\frac{1}{2}-\epsilon})\}$. Since $\log M + \log \gamma = \log(\gamma M) \le \log(|\C_s^{(l)}|)$, we obtain
\begin{align}
\lim_{n \to \infty} \frac{\log M}{\sqrt{n}} \le  \lim_{n \to \infty} \frac{\log(|\C_s^{(l)}|)-\log \gamma}{\sqrt{n}} \le \frac{\sqrt{2} N_w}{N_b}Q^{-1}\left(\frac{1-\delta}{2}\right),  
\end{align}
which matches the achievability result.

\newcommand{\Yb}{\bar{Y}}
\newcommand{\Xb}{\bar{X}}
\newcommand{\Qb}{\bar{Q}}

\section{Extension to KL-divergence Metric}
In addition to variational distance, another widely used  covertness metric in the literature is the KL-divergence. For completeness, we also consider covert communication under KL-divergence metric, in which the code $\C$ is required to satisfy 
\begin{align}
\D(\Qh || \Qn) \le \delta.
\end{align}
For any spectral mask $\mathcal{S}_W$, noise parameters $N_w, N_b > 0$, and covertness parameter $\delta$, a {\it throughput pair} $(r,r_K)$ is said to be achievable under KL-divergence metric if there exists a sequence of code with increasing support $T$ such that
\begin{align*}
&\liminf_{T \to \infty} \frac{\log M}{\sqrt{T}} \ge r, \quad \limsup_{T \to \infty} \frac{\log K}{\sqrt{T}} \le r_K, \\
&\lim_{T \to \infty} P_{\text{err}} = 0, \quad \lim_{T \to \infty} \D(\Qh || \Qn) \le \delta,
\end{align*}  
and the ESD $\widehat{E}(f)$ fits into the spectral mask. Theorem~\ref{thm:result2} below presents a lower bound on the covert capacity under KL-divergence metric, based on a PAM scheme with BPSK and RRC carrier pulses.   

\begin{theorem}\label{thm:result2}
	For any spectral mask $\mathcal{S}_W$, noise parameters $N_w, N_b > 0$, and covertness parameter $\delta \in (0,1)$, the throughput pair $(r,r_K)$ with   
	\begin{align}
	& r = \frac{N_w}{N_b} \sqrt{\frac{\delta W}{\min_{\B \in [0,1]}c(\B)}}, \ \ r_K = \left(1-\frac{N_w}{N_b}\right)^+ \sqrt{\frac{\delta W}{\min_{\B \in [0,1]}c(\B)}},
	\end{align} 
	is achievable under KL-divergence metric.
\end{theorem}

Again, if we consider the converted discrete-time AWGN channels characterized by $\mathcal{N}(0,N_w/2)$ and $\mathcal{N}(0,N_b/2)$, we have the following converse result. Substituting $n$ by $\frac{WT}{\min_{\B \in [0,1]} c(\B)}$, one can verify the optimality of BPSK under KL-divergence metric.   

\begin{theorem} \label{thm:kl_converse}
	Let $\delta \in (0,1)$. For any sequence of codes with $\lim\limits_{n \to \infty} P_{\emph{err}} = 0$ and $\lim\limits_{n \to \infty} \D(\Qh || \Qn) \le \delta$, we have
	\begin{align}
	\lim_{n \to \infty} \frac{\log M}{\sqrt{n}} \le \frac{\sqrt{\delta} N_w}{N_b}.  
	\end{align}
\end{theorem}
We provide detailed proofs of Theorems~\ref{thm:result2} and~\ref{thm:kl_converse} in Subsections~\ref{sec:nus1} and~\ref{sec:nus2}, respectively. 

%Since covert communication under KL-divergence metric has been well studied (see, for example,~\cite{7407378, wang2016fundamental}), we skip the detailed proofs here, and refer the interested readers to the full version~\cite{Zhang_2019_full}. It is worth highlighting that standard proof techniques do not directly apply to the discrete-time AWGN channels with infinite alphabet size, and several new techniques are developed in~\cite{Zhang_2019_full} to resolve the technical difficulties.

\subsection{Proof of Theorem~\ref{thm:result2}}  \label{sec:nus1}
In the following, the blocklength $n$ is kept the same as in Section~\ref{sec:proof}, hence Lemma~\ref{lemma:blocklength} guarantees that a randomly chosen code $\C$ fits in the spectral mask with high probability. The value of $\an$ is chosen differently from~\eqref{eq:an} in Section~\ref{sec:proof}, and is given by 
\begin{align}
\an = \left(\frac{\delta N_w^2}{n}\right)^{1/4}\left(1 - n^{-1/9} \right).
\end{align}
The sizes of message and shared key are given by 
\begin{align}
&\log M = \left(1 - \frac{1}{\log n}\right) \frac{\an^2 n}{N_b}, \\
&\log K = \left(\left(1+ \frac{1}{\log n}\right) - \left(1- \frac{1}{\log n}\right)\frac{N_w}{N_b} \right)^+ \frac{\an^2 n}{N_w}.
\end{align}
The analysis of reliability is independent of the covertness metric, hence the reliability part can be proved by applying Lemmas~\ref{lemma:reliability} and~\ref{lemma:error} in Section~\ref{sec:reliability} directly, and we omit the details here. We now turn to analyze covertness. Note that unlike variational distance, KL-divergence does not satisfy triangle inequality. Instead, we have 
\begin{align}
\D(\Qh || \Qn) = \D(\Qtn || \Qn) + \D(\Qh || \Qtn) +  \int_{\z}\left(\Qh(\z)-\Qtn(\z)\right)\log \frac{\Qtn(\z)}{\Qn(\z)} d\z. \label{eq:kl}
\end{align}
With the help of Lemmas~\ref{lemma:kl1}-\ref{lemma:kl3} below, one can show that there exists a code $\C$ satisfying $\D(\Qh || \Qn) \le \delta$. We prove Lemmas~\ref{lemma:kl1} and~\ref{lemma:kl2} in the following, and defer the proof of Lemma~\ref{lemma:kl3} to Appendix~\ref{appendix:kl3}. 

\begin{lemma} \label{lemma:kl1}
	By setting $\an = \left(\frac{\delta N_w^2}{n}\right)^{1/4}\left(1 - n^{-1/9} \right)$, for large enough $n$, we have 
	\begin{align}
	\D(\Qtn ||\Qn) \le \delta - \mathcal{O}\left(n^{-1/9}\right).
	\end{align}
\end{lemma}

\begin{lemma} \label{lemma:kl2}
	For sufficiently large $n$, there exists a code $\C$ such that 
	\begin{align}
	\D(\Qh ||\Qtn) \le \exp\left(-\mathcal{O}\left(\frac{n^{1/8}}{\log n}\right)\right).
	\end{align}
\end{lemma}

\begin{lemma} \label{lemma:kl3}
	With high probability over code design, a randomly chosen code $\C$ satisfies 
	\begin{align}
	\int_{\z}\left(\Qh(\z)-\Qtn(\z)\right)\log \frac{\Qtn(\z)}{\Qn(\z)} d\z \le \exp\left(-\mathcal{O}(\sqrt{n})\right).
	\end{align}
\end{lemma}

\subsubsection{Proof of Lemma~\ref{lemma:kl1}} 
The proof of Lemma~\ref{lemma:kl1} is initially developed by Wang~\cite{wang2019gaussian}, and we repeat the details here for completeness. 
By the chain rule of KL-divergence and the fact that both $\Qtn$ and $\Qn$ are product distributions, we have
\begin{align}
\D(\Qtn || \Qn) = n \D(\Qt || Q_0). 
\end{align}
We now calculate the KL-divergence between two single-letter distributions as follows:
\begin{align}
\D(\Qt || Q_0) &= \int_{-\infty}^{\infty}\Qt(z) \log \frac{\Qt(z)}{Q_0(z)} dz \\
&= \int_{-\infty}^{\infty}\Qt(z) \log\left(\frac{1}{2}\exp\left(-\frac{\an^2 - 2\an z}{N_w}\right) + \frac{1}{2}\exp\left(-\frac{\an^2 + 2\an z}{N_w}\right) \right) dz \\
& = -\frac{\an^2}{N_w} + \int_{-\infty}^{\infty}\Qt(z) \log\left(\frac{1}{2}\exp\left(\frac{2\an z}{N_w}\right) + \frac{1}{2}\exp\left(\frac{- 2\an z}{N_w}\right) \right) dz \\
& \le -\frac{\an^2}{N_w} + \int_{-\infty}^{\infty}\Qt(z)\left[\frac{1}{2}\left(\frac{2\an z}{N_w}\right)^2 - \frac{1}{12}\left(\frac{2\an z}{N_w}\right)^4 + \frac{1}{45}\left(\frac{2\an z}{N_w}\right)^6 \right] dz, \label{eq:taylor}
\end{align} 
where the last step follows from Taylor series expansion. With some calculations, we obtain
\begin{align}
&\frac{1}{2} \int_{-\infty}^{\infty} \Qt(z) \left(\frac{2\an z}{N_w}\right)^2 dz = \frac{\an^2}{N_w} + \frac{2 \an^4}{N_w^2},\\
&-\frac{1}{12} \int_{-\infty}^{\infty} \Qt(z) \left(\frac{2\an z}{N_w}\right)^4 dz = -\frac{\an^4}{N_w^2} - \frac{4 \an^6}{N_w^3} - \frac{4\an^8}{3N_w^4}, \\
& \frac{1}{45} \int_{-\infty}^{\infty} \Qt(z) \left(\frac{2\an z}{N_w}\right)^6 dz = \frac{8\an^6}{3N_w^3} + \frac{16\an^8}{N_w^4} + \frac{32\an^{10}}{3N_w^5} + \frac{64\an^{12}}{45N_w^6}.
\end{align} 
Since $\an$ is set to $\left(\frac{\delta N_w^2}{n}\right)^{1/4}\left(1 - n^{-1/9} \right)$, for sufficiently large $n$, we have 
\begin{align}
\D(\Qtn || \Qn) = n \D(\Qt || Q_0) \le \frac{n\an^4}{N_w^2} \le \delta - \mathcal{O}\left(n^{-1/9}\right). 
\end{align}

\subsubsection{Proof of Lemma~\ref{lemma:kl2}} 
The technique of bounding $\D(\Qh ||\Qtn)$ used in~\cite{7407378} does not work for Gaussian channels, since it requires the output alphabet to be finite. To circumvent this challenge, we turn to study the largest exponent of $\D(\Qh ||\Qtn)$ which is induced by a code generated according to $\Pxn$. We define the largest exponent as 
\begin{align}
E_{\mathrm{KL}}(\Px,\wzx, R) \triangleq \lim_{n \to \infty} \max_{\C_n: \log M_n = nR} -\frac{1}{n} \log \D(\Qh ||\Qtn),
\end{align}
where $\Px$ and $\wzx$ are defined in~\eqref{eq:px} and~\eqref{eq:wzx} respectively. Hayashi~\cite{hayashi2006general} provides a lower bound on $E_{\mathrm{KL}}(\Px,\wzx, R)$, based on {\it Renyi-divergence} of order $\rho$ and an optimization over all $\rho \in [0,1]$, which is repeated as follows. 
\begin{claim} 
	Let 
	\begin{align}
	f(\rho) \triangleq \rho R - \log \left(\sum_{x \in \{-\an,\an\}}\Px(x) \int_{-\infty}^{\infty} \wzx(z|x)^{1+\rho} \Qt(z)^{-\rho} dz \right).
	\end{align}
	For any $R > 0$, 
	\begin{align}
	E_{\mathrm{KL}}(\Px,\wzx, R) \ge \max_{\rho \in [0,1]} f(\rho). 
	\end{align}  
\end{claim} 

By the {\it Taylor Theorem}, we have
\begin{align}
f(\rho) = f(0) + f'(0)\cdot \rho + \frac{f''(0)}{2}\rho^2 + \frac{f'''(b)}{6}\rho^3 \text{ for some } 0 \le b \le \rho.
\end{align}
Note that
\begin{align}
&f'(\rho) = R - \frac{\sum_{x}\Px(x)\int_{-\infty}^{\infty} \wzx(z|x)^{1+\rho}\Qt(z)^{-\rho} \log\left(\frac{\wzx(z|x)}{\Qt(z)}\right)dz}{\sum_{x}\Px(x)\int_{-\infty}^{\infty} \wzx(z|x)^{1+\rho}\Qt(z)^{-\rho} dz}, \\
& f''(\rho) = \left(f'(\rho)\right)^2 - \frac{\sum_{x}\Px(x)\int_{-\infty}^{\infty} \wzx(z|x)^{1+\rho}\Qt(z)^{-\rho} \log^2\left(\frac{\wzx(z|x)}{\Qt(z)}\right)dz}{\sum_{x}\Px(x)\int_{-\infty}^{\infty} \wzx(z|x)^{1+\rho}\Qt(z)^{-\rho} dz},
\end{align}   
and one can then show that 
\begin{align}
&f(0) = 0, \\ 
&f'(0) = R - \II(X;Z), \\
&f''(0) = \II(X;Z)^2 - \E_{\Px\wzx}\left(\log^2\left(\frac{\wzx(Z|X)}{\Qt(Z)}\right)\right),
\end{align} 
where random variables $(X,Z) \sim \Px \wzx$ in $\II(X;Z)$. By the symmetry of $\Qa(Z)$ and $Q_{-a}(Z)$, we have 
\begin{align}
\II(X;Z) = \D(\Qa || \Qt) \le \frac{\an^2}{N_w} = \mathcal{O}(\an^2). \label{eq:milk1}
\end{align}
Moreover, we have 
\begin{align}
\E_{\Px\wzx}\left(\log^2\left(\frac{\wzx(Z|X)}{\Qt(Z)}\right)\right) = \mathcal{O}(\an^2). \label{eq:milk2}
\end{align}
The proof of~\eqref{eq:milk1} and~\eqref{eq:milk2} can be found in Appendix~\ref{sec:calculation}. One can also check that $f'''(\rho)$ is a continuous function from $[0,1] \mapsto \mathbb{R}$, hence there exists a constant $B$ such that $|f'''(b)| < B$ for all $b \in [0,1]$. We fix $\rho = n^{-3/8}$ in the following analysis.

\begin{itemize}
	
	\item Case 1: When $N_w \ge N_b$, we set $R = (1 + \frac{1}{\log n}) \frac{\an^2}{N_w}$. Hence, we have 
	\begin{align}
	E_{\mathrm{KL}}(\Px,\wzx, R)  &\ge f(\rho = n^{-3/8}) \label{eq:case11} \\
	&= [R - \II(X;Z)]\rho + \frac{\rho^2}{2}\left[\II(X;Z)^2 - \E_{\Px\wzx}\left(\log^2\left(\frac{\wzx(Z|X)}{\Qt(Z)}\right)\right)\right] + \frac{f'''(b)}{6}\rho^3  \\
	&= \mathcal{O}\left(\frac{1}{n^{7/8}(\log n)}\right), \label{eq:case12}
	\end{align} 
	which implies for sufficiently large $n$, there exists a code $\C$ of size $\log \N + \log K = (1 + \frac{1}{\log n}) \frac{n\an^2}{N_w}$ such that 
	\begin{align}
	\D(\Qh || \Qtn) \le \exp\left(-nE_{\mathrm{KL}}(\Px,\wzx, R)\right) = \exp\left(-\mathcal{O}\left(\frac{n^{1/8}}{\log n}\right)\right).
	\end{align} 
	
	\item Case 2: When $N_w < N_b$, we set $R = (1 - \frac{1}{\log n}) \frac{\an^2}{N_b}$. Similar to equations~\eqref{eq:case11}-\eqref{eq:case12}, we have
	\begin{align}
	E_{\mathrm{KL}}(\Px,\wzx, R) \ge \mathcal{O}\left(n^{-7/8}\right).
	\end{align}
	Hence, for sufficiently large $n$, there exists a code of size $\log \N = (1 - \frac{1}{\log n}) \frac{\an^2}{N_b}$ such that
	\begin{align}
	\D(\Qh || \Qtn) \le \exp\left(-nE_{\mathrm{KL}}(\Px,\wzx, R)\right) = \exp\left(-\mathcal{O}(n^{1/8})\right).
	\end{align}
\end{itemize}
Combining case 1 and case 2, we complete the proof of Lemma~\ref{lemma:kl2}.

\subsection{Proof of Theorem~\ref{thm:kl_converse}} \label{sec:nus2}
As shown in~\cite{wang2016fundamental, hou_thesis, 7407378}, the KL-divergence between $\Qh$ and $\Qn$ is bounded from below as    
\begin{align}
\D(\Qh || \Qn) \ge n \D(\Qb ||Q_0), \ \text{ where } \Qb(z) = \frac{1}{n\N K} \sum_{m=1}^{\N} \sum_{s=1}^K \sum_{i=1}^n Q_{x_{\m s,i}}(z), \forall z \in \mathbb{R}.
\end{align}
Hence, for any code $\C$ satisfying $\D(\Qh || \Qn) \le \delta$, we have
\begin{align}
\frac{\delta}{n} \ge \frac{\D(\Qh || \Qn)}{n} \ge \D(\Qb || Q_0) &= \int_{-\infty}^{\infty}\Qb(z) \log \frac{\Qb(z)}{Q_0(z)} dz \\
&= - h(\Qb) + \E_{\Qb}\left(\log \frac{1}{Q_0(Z)}\right) \\
&= - h(\Qb) + \frac{1}{2}\log(\pi N_w) + \E_{\Qb}\left(\frac{Z^2}{N_w}\right)  \\
&= - h(\Qb) + \frac{1}{2}\log(\pi N_w) + \frac{\E(\Xb^2)}{N_w} + \frac{1}{2}, \label{eq:d1}
\end{align}
where the last step follows since $\E_{\Qb}(Z^2) = \E(\Xb^2) + \E(W^2) + 2\E(\Xb)\E(W) = \E(\Xb^2) + N_w/2$, where $\Xb$ is distributed according to 
\begin{align}
P_{\Xb}(x) = \frac{1}{n\N K} \sum_{m=1}^{\N} \sum_{s=1}^K \sum_{i=1}^n \mathbbm{1}\left\{x_{ms,i}=x \right\}, \ \forall x \in \mathbb{R}.
\end{align}
Given $\E(Z^2)$, the differential entropy $h(\Qb)$ is bounded from above as 
\begin{align}
h(\Qb) \le \frac{1}{2}\log \left[2\pi e \left(\E(\Xb^2) + \frac{N_w}{2}\right)\right]. \label{eq:d2}
\end{align} 
By combining~\eqref{eq:d1} and~\eqref{eq:d2}, we obtain
\begin{align}
\frac{\delta}{n} &\ge -\frac{1}{2}\log\left[\frac{2\pi e \E(\Xb^2) + \pi e N_w}{\pi N_w}\right] + \frac{\E(\Xb^2)}{N_w} + \frac{1}{2} \\
&= - \frac{1}{2}\log\left(\frac{2\E(\Xb^2)+N_w}{N_w}\right) - \frac{1}{2} + \frac{\E(\Xb^2)}{N_w} + \frac{1}{2}\\
&= \frac{\E(\Xb^2)}{N_w} - \frac{1}{2}\log\left(1 + \frac{2\E(\Xb^2)}{N_w}\right) \\
&\ge \frac{\E(\Xb^2)}{N_w} - \frac{1}{2}\left(\frac{2\E(\Xb^2)}{N_w} - \frac{4(\E(\Xb^2))^2}{2N_w^2} + \frac{8(\E(\Xb^2))^3}{3N_w^3} \right) \\
& = \frac{(\E(\Xb^2))^2}{N_w^2} - \frac{4(\E(\Xb^2))^3}{3N_w^3},
\end{align} 
which implies
\begin{align}
\E(\Xb^2) \le N_w \sqrt{\frac{\delta}{n}} + o(n^{-1/2}). 
\end{align}
We now turn to analyze the channel $\wyx$ (between Alice and Bob) with Gaussian noise $B_i \sim \mathcal{N}(0,N_b/2)$. Let $\Yb$ be the random variable distributed according to 
\begin{align}
P_{\Yb}(y) = \sum_{x} P_{\Xb}(x)\wyx(y|x), \ \forall y \in \mathbb{R}.
\end{align}
Note that 
\begin{align}
&\E(\Yb^2) = \E(\Xb^2) + \E(B^2) + 2\E(\Xb)\E(B) \le N_w \sqrt{\frac{\delta}{n}} + \frac{N_b}{2} + o(n^{-1/2}),
\end{align}
hence the mutual information $\I(X;Y)$ is bounded from above as
\begin{align}
\I(\Xb;\Yb) = h(\Yb) - h(B) &\le \frac{1}{2}\log\left(2\pi e \left(N_w\sqrt{\frac{\delta}{n}} + \frac{N_b}{2} +o(n^{-1/2})\right)\right) - \frac{1}{2}\log\left(2\pi e \frac{N_b}{2}\right) \\
& =\frac{1}{2} \log\left(1+ \frac{2N_w}{N_b}\sqrt{\frac{\delta}{n}} + o(n^{-1/2})\right) \\
&= \frac{N_w}{N_b}\sqrt{\frac{\delta}{n}} + o\left(n^{-1/2}\right).
\end{align} 
For a sequence of codes with $P_{\text{err}} = \epsilon_n$ and $\D(\Qh ||\Qn) \le \delta$, we use standard information inequalities (similar to equations~\eqref{eq:con1} to~\eqref{eq:con2}) to bound $\log \N$ from above as
\begin{align}
\log \N &\le \I(\X;\Y) + \Hb(\epsilon_n) + \epsilon_n \log \N \\
& \le n \I(\Xb;\Yb) + \Hb(\epsilon_n) + \epsilon_n \log \N,
\end{align}
which further implies that as $n$ goes to infinity,
\begin{align}
\lim_{n \to \infty} \frac{\log \N}{\sqrt{n}} \le \sqrt{n}\cdot \I(\Xb;\Yb) \le \frac{\sqrt{\delta}N_w}{N_b}. 
\end{align}

\section{Conclusion}

This work studies covert communication over continuous-time AWGN channels under spectral mask constraints. We develop a PAM communication scheme with BPSK and RRC carrier pulses, which is proved to be capable to transmit $\mathcal{O}(\sqrt{WT})$ bits of information (with pre-constant exactly characterized) reliably and covertly, given a fixed time $T$ and a spectral mask with bandwidth parameter $W$. The critical step of analyzing covertness and reliability is to convert the continuous-time AWGN channels to discrete-time AWGN channels via matched filters. 

In addition, we have also provided tight converse results under discrete-time models, while the converse under continuous-time AWGN channels with spectral mask constraints is still missing. One fertile avenue for future research is to show that the throughput of our scheme is order-optimal (i.e., one cannot transmit $\omega(\sqrt{WT})$ bits reliably and covertly), by providing an upper bound on covert capacity under the continuous-time models.

\section*{Acknowledgement}
Qiaosheng (Eric) Zhang would like to thank Mehrdad Tahmasbi, Vincent Y. F. Tan, and Lei Yu for their valuable suggestions.

\appendices

\section{} \label{appendix:esd}
Let $\jj$ be the imaginary unit such that $\jj^2 = -1$. For any $\xt = \sum_{i=1}^{\n}x_i g_i(t)$, the Fourier Transform and ESD of $\xt$ are respectively given by 
\begin{align*}
&\hxf \triangleq \int_{0}^{T} x(t) e^{-\jj 2\pi f t} dt = \phif \cdot \sum_{i=1}^{\n}x_i e^{-\jj 2\pi(i-\frac{1}{2})\T f}, \\
&E_{\xt}(f) \triangleq |\hxf|^2 = |\phif|^2 \cdot \left|\sum_{i=1}^{\n}x_i e^{-\jj 2\pi(i-\frac{1}{2})\T f}\right|^2 = |\phif|^2 \cdot \left(\an^2n + \an^2 \sum_{i_1 \ne i_2} \cos(2\pi(i_1-i_2)\T f) \frac{x_{i_1}}{x_{i_2}} \right).
\end{align*}
Hence, the ensemble-averaged ESD $\widetilde{E}(f)$ is given by
\begin{align*}
\widetilde{E}(f) = \E_{\Pxn}\left(E_{\Xt}(f)\right) &= \sum_{\x \in \{-\an,\an\}^n} \Pxn(\x)\cdot |\phif|^2 \left(\an^2n + \an^2 \sum_{i_1 \ne i_2} \cos(2\pi(i_1-i_2)\T f) \frac{x_{i_1}}{x_{i_2}} \right) \\
&= |\phif|^2 \left(\an^2n + \an^2 \sum_{i_1 \ne i_2} \cos(2\pi(i_1-i_2)\T f) \cdot \frac{1}{2^n} \left(\sum_{\x:x_{i_1}=x_{i_2}} \frac{x_{i_1}}{x_{i_2}}+\sum_{\x:x_{i_1}\ne x_{i_2}} \frac{x_{i_1}}{x_{i_2}}\right) \right) \\
&= \an^2 n \cdot |\phif|^2.
\end{align*}

For a specific code $\C$, the ESD $\widehat{E}(f)$, first defined in~\eqref{eq:esd}, is
\begin{align*}
\widehat{E}(f) &= \frac{1}{\N K}\sum_{\m=1}^{\N} \sum_{s=1}^K E_{\xmt}(f)\\
& = \an^2 n |\phif|^2 + \frac{\an^2}{\N K} |\phif|^2 \sum_{i_1 \ne i_2} \cos(2\pi(i_1-i_2)\T f) \sum_{\m = 1}^{\N} \sum_{s=1}^K \frac{x_{ms,i_1}}{x_{ms,i_2}}.
\end{align*}

\noindent{\it Proof of Lemma~\ref{lemma:esd}:}
Since $\left|\frac{X_{ms,i_1}}{X_{ms,i_2}}\right| \le 1$ by our code construction, and 
\begin{align}
\E\left(\sum_{\m = 1}^{\N} \sum_{s=1}^K \frac{X_{ms,i_1}}{X_{ms,i_2}} \right) = \sum_{\m = 1}^{\N} \sum_{s=1}^K \E_{\Pxn}\left(\frac{X_{ms,i_1}}{X_{ms,i_2}}\right) = 0,
\end{align}
the Hoeffding's inequality yields that for any $t \ge 0$, 
\begin{align}
\PP\left(\left|\frac{1}{MK}\sum_{\m = 1}^{\N} \sum_{s=1}^K \frac{X_{ms,i_1}}{X_{ms,i_2}}\right| \ge t \right) \le 2 \exp\left(-\frac{\N K t^2}{2}\right).
\end{align}
By setting $t = (\N K)^{-1/4}$ and taking a union bound over all $(i_1, i_2)$-pairs, where $i_1,i_2 \in \llbracket 1,n \rrbracket$ and $i_1 \ne i_2$, one can show that with probability at least $1 - 2n^2\exp(-\sqrt{\N K}/2)$, a randomly chosen code $\C$ satisfies
\begin{align}
\left|\sum_{\m=1}^{\N} \sum_{s=1}^K \frac{x_{ms,i_1}}{x_{ms,i_2}}\right| \le (\N K)^{3/4}, \ \ \forall i_1,i_2 \in \llbracket 1,n \rrbracket,  i_1 \ne i_2. \label{eq:uptown}
\end{align}
Therefore, the ESD $\widehat{E}(f)$ of the code $\C$ satisfying~\eqref{eq:uptown} is tightly concentrated around $\widetilde{E}(f)$, i.e.,
\begin{align}
\widehat{E}(f) &\le \an^2 n |\phif|^2 + \frac{\an^2}{\N K} |\phif|^2 \sum_{i_1 \ne i_2} \cos(2\pi(i_1-i_2)\T f) \cdot (\N K)^{3/4} \\
&\le \an^2 n |\phif|^2 + \frac{\an^2}{\N K} |\phif|^2 n^2 \cdot (\N K)^{3/4} \\
&= \widetilde{E}(f) \left(1+ \frac{n}{(\N K)^{1/4}}\right), \ \forall f \in \mathbb{R}; \label{eq:hate1} \\
\widehat{E}(f) &\ge \an^2 n |\phif|^2 \left(1- \frac{n}{(\N K)^{1/4}}\right) = \widetilde{E}(f) \left(1- \frac{n}{(\N K)^{1/4}}\right), \ \forall f \in \mathbb{R}. \label{eq:hate2}
\end{align}  
\qed

\section{Proof of Lemma~\ref{lemma:final}} \label{appendix:limit}
Similar to the relationship between the optimization problems (P1) and (P$1_\B$), we also define an optimization problem (P$2_\B$) for each $\B \in [0,1]$, which differs from (P$2$) only in the third constraint, as 
\begin{equation*}
\begin{aligned}
(\text{P}2_\B) \quad & \underset{\T}{\text{min}}
& & \T
& &\\
& \text{s.t.}
& & \forall f \ge \al_\K W, \frac{|\phinf(f)|^2 (1-u(T,\T))}{|\phinf(0)|^2 (1+u(T,\T))} < V_{\K} , 
& & \K \in \llbracket 1, l \rrbracket;\\
&
& & \int_{-\al_\K W}^{\al_\K W} |\phif|^2 df \ge \frac{\eta_\K}{1-u(T,\T)},  
& & \K \in \llbracket 1, l \rrbracket;\\
&
& & \T > 0.
\end{aligned} 
\end{equation*}
The optimal value of (P$2_\B$) is denoted by $\Tsbt$, and note that $\Tst = \min_{\B \in [0,1]} \Tsbt$. 

\begin{claim}\label{claim:limit}
	Let $\B \in [0,1]$. For every $\varepsilon > 0$, there exists a $T_{\varepsilon,\B}' > 0$ such that for all $T > T'_{\varepsilon,\B}$, 
	\begin{align}
	\left| \Tsbt - \Tsb \right| < \varepsilon. \label{eq:eason}
	\end{align}
\end{claim}
\newcommand{\Tabb}{T_0^{\ast}}
\newcommand{\phivar}{\phi_{\Tabb+\varepsilon,\B}}
\newcommand{\phiwb}{\phi_{\Tabb,\B}}
\newcommand{\phiratio}{\phi_{\frac{\Tabb + \varepsilon}{\Tabb}\Tabb,\B}}
\newcommand{\phihvar}{\widehat{\phi}_{\Tabb+\varepsilon,\B}}
\newcommand{\phihwb}{\widehat{\phi}_{\Tabb,\B}}
\newcommand{\phihratio}{\widehat{\phi}_{\left(\frac{\Tsb + \varepsilon}{\Tsb}\right)\Tsb,\B}}
\noindent{\it Proof:} We first note that $\Tsbt \ge \Tsb$, since $u(T,\T) > 0$ and $\Tsbt$ is always a feasible solution in (P$1_\B$). As a consequence, the statement in~\eqref{eq:eason} is equivalent to the following statements:
\begin{align}
\left| \Tsbt - \Tsb \right| < \varepsilon &\iff \Tsbt < \Tsb + \varepsilon \notag  \\
& \iff \Tsb + \varepsilon \text{ is a feasible solution of (P}2_\B\text{)}. \label{eq:equal}
\end{align}
For notational convenience we abbreviate $\Tsb$ as $\Tabb$. One can verify that 
\begin{align*}
\phivar(t) = \phiratio(t) = \sqrt{\frac{\Tabb}{\Tabb+\varepsilon}}\cdot \phiwb\left(\frac{\Tabb}{\Tabb+\varepsilon} \cdot t\right),
\end{align*}
and in the frequency domain, we have 
\begin{align*}
\phihvar(f) = \sqrt{\frac{\Tabb+\varepsilon}{\Tabb}}\cdot \phihwb\left(\frac{\Tabb+\varepsilon}{\Tabb} \cdot f\right).
\end{align*}

We now show that for every $\varepsilon > 0$, there exists a $T_{\varepsilon,\B}' > 0$ such that for all $T > T_{\varepsilon,\B}'$, $\Tabb + \varepsilon$ is a feasible solution of (P$2_\B$). Note that the optimization (P$2_\B$) explicitly depends on $T$.

\begin{enumerate}
	\item \underline{\bf (The first constraint)} Since $\Tabb$ is a feasible solution of (P$1_\beta$), it satisfies that for every $\K \in \llbracket 1, l \rrbracket$,
	\begin{align}
	\forall f \ge \al_\K W, \ |\phihwb(f)|^2 < V_\K \cdot |\phihwb(0)|^2,
	\end{align}
	hence there exists a $\gamma_{\K} > 0$ such that 
	\begin{align}
	\forall f \ge \alpha_\K W, \quad |\phihwb(f)|^2 \le V_\K \cdot |\phihwb(0)|^2 - \gamma_{\K}.
	\end{align}
	It is also worth noting that $\forall f \ge \al_\K W$,
	\begin{align}
	\left|\phihvar(f)\right|^2 &=\frac{\Tabb+\varepsilon}{\Tabb}\left|\phihwb\left(f\cdot \frac{\Tabb+\varepsilon}{\Tabb}\right)\right|^2 \\
	&\le \frac{\Tabb+\varepsilon}{\Tabb}\left(V_\K \left|\phihwb\left(0\right)\right|^2 - \gamma_\K\right) = V_\K \left|\phihvar(0)\right|^2 \left(1- \frac{(\Tabb + \varepsilon)\gamma_\K}{\Tabb  V_\K \left|\phihvar(0)\right|^2}\right). \label{eq:xingfu}
	\end{align}
	Since $\lim_{T \to \infty}u(T, \Tabb+\varepsilon) = 0$ and $u(T, \Tabb+\varepsilon) > 0$, we have that for every $\tau_{1\K} > 0$, there exists a $T'_{\tau_{1\K},\varepsilon,\B} > 0$ such that for all $T > T'_{\tau_{1\K},\varepsilon,\B}$, $0 < u(T, \Tabb+\varepsilon) < \tau_{1\K}$, which further implies
	\begin{align*}
	\frac{1 - u(T, \Tabb+\varepsilon)}{1 + u(T, \Tabb+\varepsilon)} > \frac{1 - \tau_{1\K}}{1 + \tau_{1\K}}.
	\end{align*} 
	By choosing $\tau_{1\K}$ to satisfy
	\begin{align}
	\frac{1- \tau_{1\K}}{1+ \tau_{1\K}} = 1- \frac{(\Tabb + \varepsilon)\gamma_\K}{\Tabb \left(10^{-\frac{U_\K}{10}}\right) \left|\phihvar(0)\right|^2}, \label{eq:xingfu2}
	\end{align}
	one can show that for all $T > T'_{\tau_{1\K},\varepsilon,\B}$, the term in~\eqref{eq:xingfu} satisfies
	\begin{align}
	1- \frac{(\Tabb + \varepsilon)\gamma_\K}{\Tabb \cdot V_\K \left|\phihvar(0)\right|^2} = \frac{1- \tau_{1\K}}{1+ \tau_{1\K}} < \frac{1 - u(T, \Tabb+\varepsilon)}{1 + u(T, \Tabb+\varepsilon)},
	\end{align} 
	and thus,
	\begin{align}
	\forall f \ge \al_\K W, \ \left|\phihvar(f)\right|^2 < V_\K \left|\phihvar(0)\right|^2 \cdot \frac{1 - u(T, \Tabb+\varepsilon)}{1 + u(T, \Tabb+\varepsilon)}.
	\end{align}
	This implies that $\Tabb +\varepsilon$ satisfies the first constraint in (P$2_{\beta}$).
	\item {\bf (The second constraint)} We first note that for every $\K \in \llbracket 1, l \rrbracket$, $\Tabb$ satisfies $\int_{-\al_\K W}^{\al_\K W}|\phihwb(f)|^2 df \ge \eta_\K.$
	Let $\sigma \triangleq \frac{\Tabb+\varepsilon}{\Tabb}$ and $\theta_\K \triangleq (\int_{-\al_\K W \sigma}^{\al_\K W\sigma}|\phihwb(f)|^2 df)/(\int_{-\al_\K W }^{\al_\K W}|\phihwb(f)|^2 df)$. One may check that $\theta_\K > 1$. Note that for every $\K \in \llbracket 1, l \rrbracket$,
	\begin{align}
	\int_{-\al_\K W}^{\al_\K W}\left|\phihvar(f) \right|^2 df &= \frac{\Tabb+\varepsilon}{\Tabb} \int_{-\al_\K W}^{\al_\K W}\left|\phihwb\left(\frac{\Tabb+\varepsilon}{\Tabb}f\right) \right|^2 df \\
	& = \int_{-\al_\K W \frac{\Tabb+\varepsilon}{\Tabb}}^{\al_\K W\frac{\Tabb+\varepsilon}{\Tabb}}\left|\phihwb(f') \right|^2 df'  = \theta_\K \cdot \int_{-\al_\K W }^{\al_\K W}\left|\phihwb(f') \right|^2 df' \ge \theta_\K \eta_\K.
	\end{align}
	Since $\lim_{T \to \infty}u(T, \Tabb+\varepsilon) = 0$ and $u(T, \Tabb+\varepsilon) > 0$, for every $\tau_{2\K} > 0$, there exists a $T'_{\tau_{2\K},\varepsilon,\B} > 0$ such that for all $T > T'_{\tau_{2\K},\varepsilon,\B}$, $0 < u(T, \Tabb+\varepsilon) < \tau_{2\K}$, which further implies 
	\begin{align*}
	1 - u(T, \Tabb+\varepsilon) > 1 - \tau_{2\K}.
	\end{align*} 
	By choosing $\tau_{2\K}$ to satisfy $1 - \tau_{2\K} = 1/\theta_\K$, we have that for all $T > T'_{\tau_{2\K},\varepsilon,\B}$, 
	\begin{align}
	\left(\int_{-\al_\K W}^{\al_\K W}\left|\phihvar(f) \right|^2 df\right) \cdot  (1 - u(T,\Tabb+\varepsilon)) &\ge \theta_\K \eta_\K (1 - u(T,\Tabb+\varepsilon))  > \theta_\K \eta_\K (1 - \tau_{2\K}) = \eta_\K.
	\end{align} 
\end{enumerate}
Finally, for every $\varepsilon > 0$, we define $T'_{\varepsilon,\B} \triangleq \max_{j \in \{1,2\}, \K \in \llbracket 1, l \rrbracket} T'_{\tau_{j\K},\varepsilon,\B}$, thus for all $T > T'_{\varepsilon,\B}$, $\Tabb + \varepsilon$ is a feasible solution of the optimization problem (P$2_\B$), which, by~\eqref{eq:equal}, is equivalent to saying that $| \Tsbt - \Tabb | < \varepsilon$. This completes the proof of Claim~\ref{claim:limit}. \qed

By defining $T'_{\varepsilon} \triangleq \max_{\B \in [0,1]} T'_{\varepsilon,\B}$ for every $\varepsilon > 0$, one can prove that for every $\varepsilon > 0$, for all $T > T'_{\varepsilon}$, 
\begin{align}
\left|\min_{\B \in [0,1]} \Tsbt - \min_{\B \in [0,1]} \Tabb \right| < \varepsilon,
\end{align}
which eventually leads to
\begin{align}
\lim_{T \to \infty} \min_{\B \in [0,1]} \Tsbt &= \min_{\B \in [0,1]} \Tsb, \\
\lim_{T \to \infty} \Tst &= \Ts.
\end{align}

\section{Proof of Lemma~\ref{lemma:obs}} \label{appendix:obs}
Consider a fixed roll-off factor $\B\in [0,1]$. To prove Lemma~\ref{lemma:obs}, it suffices to show that for any $c > 0$, 
\begin{align}
\T^{\ast}(\mathcal{S}_{c W}, \B) = \frac{1}{c} \T^{\ast}(\mathcal{S}_W, \B), \label{eq:kappa}
\end{align}  
since~\eqref{eq:kappa} implies that for any $W, \widetilde{W} > 0$, where $\widetilde{W} = c W$ for some $c$, 
\begin{align}
\T^{\ast}(\mathcal{S}_{\widetilde{W}}, \B) \cdot \widetilde{W} = \T^{\ast}(\mathcal{S}_{c W}, \B) \cdot c W = \frac{1}{c} \T^{\ast}(\mathcal{S}_W, \B) \cdot c W = \T^{\ast}(\mathcal{S}_W, \B) \cdot W.
\end{align} 
Note that $\T^{\ast}(\mathcal{S}_{c W}, \B)$ is the optimal value of the following optimization problem:
\begin{equation}
\begin{aligned}
(\text{P}1'_\B) \quad  \ \ & \underset{\T}{\text{min}}
& & \T
& &\\
& \text{s.t.}
& & \forall f \ge \al_\K c W, \ |\phinf(f)|^2 \le V_i |\phinf(0)|^2, 
& & \K \in \llbracket 1, l \rrbracket;\\
&
& & \int_{-\al_\K c W}^{\al_\K c W} |\phif|^2 df \ge \eta_\K,  
& & \K \in \llbracket 1, l \rrbracket;\\
&
& & \T > 0.
\end{aligned} \label{eq:optd}
\end{equation}
In the following, we show that $\Tsb/c$ is a feasible solution of (P$1'_\B$) in the direct part, and any value greater than $\Tsb/c$ is not feasible in the converse part. In the following we abbreviate $\Tsb$ as $\Tabb$ for convenience.

\underline{\it 1) Direct part:} Since $\Tabb$ is a feasible solution of the optimization problem (P$1_\B$), we have
\begin{align}
&\forall f \ge \al_\K W, \ |\widehat{\phi}_{\Tabb,\B}(f)|^2 \le V_i |\widehat{\phi}_{\Tabb,\B}(0)|^2, &\ \K \in \llbracket 1, l \rrbracket; \label{eq:cond1} \\
&\int_{-\al_\K W}^{\al_\K W} |\widehat{\phi}_{\Tabb,\B}(f)|^2 df \ge \eta_\K, &\ \K \in \llbracket 1, l \rrbracket. \label{eq:cond2}
\end{align} 
To prove that $\Tabb/c$ is a feasible solution of the optimization problem (P$1'_\B$), we need to verify that
\begin{align}
&\forall f \ge \al_\K c W, \ \left|\widehat{\phi}_{\frac{1}{c}\Tabb, \B}(f)\right|^2 \le V_i \left|\widehat{\phi}_{\frac{1}{c}\Tabb, \B}(0)\right|^2, 
& \K \in \llbracket 1, l \rrbracket; \label{eq:cons1} \\
&\int_{-\al_\K c W}^{\al_\K c W} \left|\widehat{\phi}_{\frac{1}{c}\Tabb, \B}(f)\right|^2 df \ge \eta_\K,  
& \K \in \llbracket 1, l \rrbracket; \label{eq:cons2}
\end{align}   
By the definition of RRC pulses, we have $\phi_{\frac{1}{c}\Tabb, \B}(t) = \sqrt{c} \cdot \phi_{\Tabb, \B}(c t)$, hence in the frequency domain, by the scaling property of the Fourier Transform, we obtain 
\begin{align}
\widehat{\phi}_{\frac{1}{c}\Tabb, \B}(f) = \frac{1}{\sqrt{c}} \cdot \widehat{\phi}_{\Tabb, \B}\left(\frac{f}{c}\right). \label{eq:relation}
\end{align}
The first constraint~\eqref{eq:cons1} is satisfied since for every $\K \in \llbracket 1,l \rrbracket$,
\begin{align}
\forall f \ge \al_\K c W, \ \left|\widehat{\phi}_{\frac{1}{c}\Tabb, \B}(f)\right|^2 = \left|\frac{1}{\sqrt{c}} \cdot \widehat{\phi}_{\Tabb, \B}\left(\frac{f}{c}\right) \right|^2\le \frac{1}{c} \cdot V_\K \cdot \left|\widehat{\phi}_{\Tabb,\B}(0)\right|^2  = V_\K \cdot \left|\widehat{\phi}_{\frac{1}{c}\Tabb,\B}(0)\right|^2. \label{eq:bing3}
\end{align}
The second constraint~\eqref{eq:cons2} is also satisfied since for every $\K \in \llbracket 1,l \rrbracket$, 
\begin{align}
\int_{-\al_\K c W}^{\al_\K c W} \left|\widehat{\phi}_{\frac{1}{c}\Tabb, \B}(f)\right|^2 df &= \frac{1}{c} \int_{-\al_\K c W}^{\al_\K c W} \left|\widehat{\phi}_{\Tabb, \B}\left(\frac{f}{c}\right)\right|^2 df \label{eq:bing4} \\
& \stackrel{\tau = f/c}{=} \frac{1}{c} \int_{-\al_\K W}^{\al_\K W} c \cdot \left|\widehat{\phi}_{\Tabb, \B}(\tau)\right|^2  d\tau \ge \eta_\K, \label{eq:bing6}
\end{align}
where inequality~\eqref{eq:bing6} follows from~\eqref{eq:cond2}. This completes the proof of the direct part.

\underline{\it 2) Converse part:} Suppose there exists a $\kappa > \frac{1}{c}\Tabb$ such that $\kappa$ is a feasible solution of the optimization problem (P$1'_\B$), then one can show that $c \kappa$ is also a feasible solution of the optimization problem (P$1_\B$). The idea of the proof is similar to the direct part, hence we omit the detail here. This fact contradicts our assumption since $c \kappa$ is greater than $\Tabb$.

Combining the direct part and the converse part together, we complete the proof of Lemma~\ref{lemma:obs}.

\section{} \label{appendix:berry}
In the following, we provide the detailed calculations for equations~\eqref{eq:jia1}-\eqref{eq:jia3} in Section~\ref{sec:covert}. By a Taylor series expansion, we have 
\begin{align*}
\log\left(\frac{1}{2}e^{\frac{-2\an z - \an^2}{N_w}} + \frac{1}{2}e^{\frac{2\an z - \an^2}{N_w}}\right) = \frac{1}{2}\left(\frac{2\an z}{N_w}\right)^2 - \frac{1}{12}\left(\frac{2\an z}{N_w}\right)^4 + \mathcal{O}(\an^6).
\end{align*}
Also, note that
\begin{align*}
\E_{\Q}\left(\log \frac{\Qt(Z)}{\Q(Z)} \right)
& = \int_{-\infty}^{\infty}Q_0(z) \log\left(\frac{1}{2}e^{\frac{-2\an z - \an^2}{N_w}} + \frac{1}{2}e^{\frac{2\an z - \an^2}{N_w}}\right) dz \\
&=-\frac{\an^2}{N_w}\int_{-\infty}^{\infty}Q_0(z) dz + \int_{-\infty}^{\infty}Q_0(z) \log\left(\frac{1}{2}e^{\frac{-2\an z}{N_w}} + \frac{1}{2}e^{\frac{2\an z}{N_w}}\right) dz \\
&= -\frac{\an^2}{N_w} + \int_{-\infty}^{\infty}Q_0(z)\left(\frac{1}{2}\left(\frac{2\an z}{N_w}\right)^2 - \frac{1}{12}\left(\frac{2\an z}{N_w}\right)^4 + \mathcal{O}(\an^6) \right)dz \\
& = -\frac{\an^2}{N_w} + \frac{\an^2}{N_w} -\frac{\an^4}{N_w^2} + \mathcal{O}(\an^6) = -\frac{\an^4}{N_w^2} + \mathcal{O}(\an^6), \\
\E_{\Q}\left(\log^2 \frac{\Qt(Z)}{\Q(Z)} \right) &= \int_{-\infty}^{\infty}Q_0(z) \log^2\left(\frac{1}{2}e^{\frac{-2\an z - \an^2}{N_w}} + \frac{1}{2}e^{\frac{2\an z - \an^2}{N_w}}\right) \\
&= \int_{-\infty}^{\infty}Q_0(z) \left(-\frac{\an^2}{N_w} + \log\left(\frac{1}{2}e^{\frac{-2\an z}{N_w}} + \frac{1}{2}e^{\frac{2\an z}{N_w}}\right)\right)^2 dz \\
&= \frac{\an^4}{N_w^2} - \frac{2\an^2}{N_w}\int_{-\infty}^{\infty}Q_0(z) \log\left(\frac{1}{2}e^{\frac{-2\an z}{N_w}} + \frac{1}{2}e^{\frac{2\an z}{N_w}}\right) dz \notag \\
&\qquad\qquad\qquad\qquad\qquad\qquad +\int_{-\infty}^{\infty}Q_0(z) \log^2\left(\frac{1}{2}e^{\frac{-2\an z}{N_w}} + \frac{1}{2}e^{\frac{2\an z}{N_w}}\right) dz \\
&= \frac{\an^4}{N_w^2} - \frac{2\an^2}{N_w}\left(\frac{\an^2}{N_w} -\frac{\an^4}{N_w^2} + \mathcal{O}(\an^6)\right) + \int_{-\infty}^{\infty}Q_0(z) \left( \frac{1}{4}\left(\frac{2\an z}{N_w}\right)^4 + \mathcal{O}(\an^6) \right) dz \\
&= \frac{2\an^4}{N_w^2} + \mathcal{O}(\an^6),\\
\text{Var}_{\Q}\left(\log \frac{\Qt(Z)}{\Q(Z)} \right) &= \E_{\Q}\left(\log^2 \frac{\Qt(Z)}{\Q(Z)} \right) - \left(\E_{\Q}\left(\log \frac{\Qt(Z)}{\Q(Z)} \right)\right)^2 = \frac{2\an^4}{N_w^2} + \mathcal{O}(\an^6), \\
\E_{\Q}\left(\log^3 \frac{\Qt(Z)}{\Q(Z)} \right) &= \int_{-\infty}^{\infty}Q_0(z) \log^3\left(\frac{1}{2}e^{\frac{-2\an z - \an^2}{N_w}} + \frac{1}{2}e^{\frac{2\an z - \an^2}{N_w}}\right) \\
&= \int_{-\infty}^{\infty}Q_0(z) \left(-\frac{\an^2}{N_w} + \log\left(\frac{1}{2}e^{\frac{-2\an z}{N_w}} + \frac{1}{2}e^{\frac{2\an z}{N_w}}\right)\right)^3 dz \\
&= \int_{-\infty}^{\infty}Q_0(z) \left(-\frac{\an^2}{N_w} + 
\frac{1}{2}\left(\frac{2\an z}{N_w}\right)^2 - \frac{1}{12}\left(\frac{2\an z}{N_w}\right)^4 + \mathcal{O}(\an^6)\right)^3 dz  = \mathcal{O}(\an^6).
\end{align*}
Similarly, the detailed calculations for equations~\eqref{eq:jia4}-\eqref{eq:jia6} are provided as follows:
\begin{align*}
&\E_{\Qt}\left(\log \frac{\Qt(Z_i)}{\Q(Z_i)} \right) \\
&= \frac{1}{2}\int_{-\infty}^{\infty}\Qa(z) \log\left(\frac{1}{2}e^{\frac{-2\an z - \an^2}{N_w}} + \frac{1}{2}e^{\frac{2\an z - \an^2}{N_w}}\right) dz + \frac{1}{2}\int_{-\infty}^{\infty}Q_{-a}(z) \log\left(\frac{1}{2}e^{\frac{-2\an z - \an^2}{N_w}} + \frac{1}{2}e^{\frac{2\an z - \an^2}{N_w}}\right) dz \\
&=\int_{-\infty}^{\infty}\Qa(z) \log\left(\frac{1}{2}e^{\frac{-2\an z - \an^2}{N_w}} + \frac{1}{2}e^{\frac{2\an z - \an^2}{N_w}}\right) dz \\
&=\left(-\frac{\an^2}{N_w}\right)\int_{-\infty}^{\infty}\Qa(z) dz + \int_{-\infty}^{\infty}\Qa(z) \log\left(\frac{1}{2}e^{\frac{-2\an z}{N_w}} + \frac{1}{2}e^{\frac{2\an z}{N_w}}\right) dz \\
&= -\frac{\an^2}{N_w} + \int_{-\infty}^{\infty}\Qa(z) \left(\frac{1}{2}\left(\frac{2\an z}{N_w}\right)^2 - \frac{1}{12}\left(\frac{2\an z}{N_w}\right)^4 + \mathcal{O}(\an^6) \right) dz = \frac{\an^4}{N_w^2} + \mathcal{O}(\an^6), \\
&\E_{\Qt}\left(\log^2 \frac{\Qt(Z_i)}{\Q(Z_i)} \right) \\
&= \int_{-\infty}^{\infty}\Qa(z) \log^2\left(\frac{1}{2}e^{\frac{-2\an z - \an^2}{N_w}} + \frac{1}{2}e^{\frac{2\an z - \an^2}{N_w}}\right) dz \\
&= \int_{-\infty}^{\infty}\Qa(z) \left(-\frac{\an^2}{N_w} + \log\left(\frac{1}{2}e^{\frac{-2\an z}{N_w}} + \frac{1}{2}e^{\frac{2\an z}{N_w}}\right)\right)^2 dz \\
&= \frac{\an^4}{N_w^2} - \frac{2\an^2}{N_w}\int_{-\infty}^{\infty}\Qa(z)\log\left(\frac{1}{2}e^{\frac{-2\an z}{N_w}} + \frac{1}{2}e^{\frac{2\an z}{N_w}}\right)dz + \int_{-\infty}^{\infty}\Qa(z)\log^2\left(\frac{1}{2}e^{\frac{-2\an z}{N_w}} + \frac{1}{2}e^{\frac{2\an z}{N_w}}\right)dz \\
&= \frac{\an^4}{N_w^2} - \frac{2\an^2}{N_w} \int_{-\infty}^{\infty}\Qa(z) \left(\frac{1}{2}\left(\frac{2\an z}{N_w}\right)^2 + \mathcal{O}(\an^4)\right)dz + \int_{-\infty}^{\infty}\Qa(z) \left(\frac{1}{4}\left(\frac{2\an z}{N_w}\right)^4 + \mathcal{O}(\an^6)\right)dz \\
& =  \frac{2\an^4}{N_w^2} + \mathcal{O}(\an^6), \\
&\text{Var}_{\Qt}\left(\log \frac{\Qt(Z)}{\Q(Z)} \right) = \E_{\Qt}\left(\log^2 \frac{\Qt(Z)}{\Q(Z)} \right) - \left(\E_{\Qt}\left(\log \frac{\Qt(Z)}{\Q(Z)} \right)\right)^2 = \frac{2\an^4}{N_w^2} + \mathcal{O}(\an^6), \\
&\E_{\Qt}\left(\log^3 \frac{\Qt(Z)}{\Q(Z)} \right) \\
&= \int_{-\infty}^{\infty}\Qa(z) \log^3\left(\frac{1}{2}e^{\frac{-2\an z - \an^2}{N_w}} + \frac{1}{2}e^{\frac{2\an z - \an^2}{N_w}}\right) \\
&= \int_{-\infty}^{\infty}\Qa(z) \left(-\frac{\an^2}{N_w} + \log\left(\frac{1}{2}e^{\frac{-2\an z}{N_w}} + \frac{1}{2}e^{\frac{2\an z}{N_w}}\right)\right)^3 dz \\
&= \int_{-\infty}^{\infty}\Qa(z) \left(-\frac{\an^2}{N_w} + 
\frac{1}{2}\left(\frac{2\an z}{N_w}\right)^2 - \frac{1}{12}\left(\frac{2\an z}{N_w}\right)^4 + \mathcal{O}(\an^6)\right)^3 dz = \mathcal{O}(\an^6).
\end{align*}

\section{Proof of Lemma~\ref{lemma:reliability}} \label{appendix:reliability}
Suppose a key $s$ is shared between Alice and Bob. According to the decoding rule, three types of error may occur:
\begin{enumerate}
	\item $\x_{ms}$ is transmitted but $(\x_{ms},\y) \notin \Aset$,
	\item $\x_{ms}$ is transmitted but there exists a $m' \ne m$ such that $(\x_{m's},\y) \in \Aset$,
	\item no transmission occurs but there exists a $m$ such that $(\x_{ms},\y) \in \Aset$.
\end{enumerate}
We bound the probability of error $P_{\text{err}}^{\text{(avg)}}$ averaged over the random code ensemble as follows:
\begin{align}
&\E(P_{\text{err}}^{\text{(avg)}}) \\
&= \E\left(\sum_{m=1}^M \sum_{s=1}^K \frac{1}{MK} \int_{\y} \wyxn(\y|\X_{ms}) \mathbbm{1}\left\{(\X_{ms},\y) \notin \Aset \text{ or } \exists m' \ne m \text{ s.t. } (\X_{m's},\y) \in \Aset  \right\} d\y \right)  \notag \\
&+ \E\left(\int_{\y} \Pn(\y)\mathbbm{1}\left\{\exists m \text{ s.t. } (\X_{ms},\y) \in \Aset \right\} d\y \right) \\
&= \E\left(\int_{\y} \wyxn(\y|\X_{11}) \mathbbm{1}\left\{(\X_{11},\y) \notin \Aset \text{ or } \exists m' \ne 1 \text{ s.t. } (\X_{m'1},\y) \in \Aset  \right\} d\y \right)  \notag \\
&+ \E\left(\int_{\y} \Pn(\y)\mathbbm{1}\left\{\exists m \text{ s.t. } (\X_{ms},\y) \in \Aset \right\} d\y \right) \\
&\le \E\left(\int_{\y} \wyxn(\y|\X_{11}) \mathbbm{1}\left\{(\X_{11},\y) \notin \Aset \right\} d\y\right) + \sum_{m' \ne 1}\E\left( \int_{\y} \wyxn(\y|\X_{11}) \mathbbm{1}\left\{(\X_{m'1},\y) \in \Aset \right\} d\y\right) \notag \\
&+ \sum_{m=1}^M \E\left(\int_{\y} \Pn(\y)\mathbbm{1}\left\{(\X_{ms},\y) \in \Aset \right\} d\y \right). \label{eq:error2}
\end{align}
The first term of~\eqref{eq:error2} is given by 
\begin{align}
\E\left(\int_{\y} \wyxn(\y|\X_{11}) \mathbbm{1}\left\{(\X_{11},\y) \notin \Aset \right\} d\y\right)
&= \PP_{\Pxn\wyxn}\left(\log \frac{\wyxn(\Y|\X)}{\Pn(\Y)}  \le \gamma \right).
\end{align}
Now we consider the second term of~\eqref{eq:error2}. For $m' \ne 1$, we have 
\begin{align}
&\E\left( \int_{\y} \wyxn(\y|\X_{11}) \mathbbm{1}\left\{(\X_{m'1},\y) \in \Aset \right\} d\y\right) \\
&= \int_{\y} \sum_{\x_{m'1}}\Pxn(\x_{m'1}) \mathbbm{1}\left\{(\x_{m'1},\y) \in \Aset \right\} \sum_{\x_{11}}\Pxn(\x_{11})\wyxn(\y|\x_{11}) d\y \\
&= \int_{\y} \sum_{\x_{m'1}}\Pxn(\x_{m'1}) \Ptn(\y) \mathbbm{1}\left\{(\x_{m'1},\y) \in \Aset \right\} d\y \\
&\le e^{-\gamma} \int_{\y} \sum_{\x_{m'1}}\Pxn(\x_{m'1}) \wyxn(\y|\x) \frac{\Ptn(\y)}{\Pn(\y)} \mathbbm{1}\left\{(\x_{m'1},\y) \in \Aset \right\} d\y \label{eq:gamma} \\
&\le e^{-\gamma} \cdot \E_{\Ptn} \left( \frac{\Ptn(\y)}{\Pn(\y)}\right), \label{eq:house}
\end{align}
where~\eqref{eq:gamma} holds since $\Pn(\y) \le e^{-\gamma}\wyxn(\y|\x)$ for $(\x,\y) \in \Aset$.
Moreover, we have 
\begin{align}
&\E_{\Ptn} \left( \frac{\Ptn(\y)}{\Pn(\y)}\right) = \left(\E_{\Pt} \left( \frac{\Pt(Y)}{P_0(Y)}\right) \right)^n, \label{eq:house2}\\
&\E_{\Pt} \left( \frac{\Pt(Y)}{P_0(Y)}\right) \\
&= \int_{-\infty}^{\infty} \frac{1}{\sqrt{\pi N_b}}\left[\frac{1}{2}\exp\left(-\frac{(y-\an)^2}{N_b}\right) + \frac{1}{2}\exp\left(-\frac{(y+\an)^2}{N_b}\right) \right]\frac{\frac{1}{2}\exp\left(-\frac{(y-\an)^2}{N_b}\right) + \frac{1}{2}\exp\left(-\frac{(y+\an)^2}{N_b}\right)}{\exp\left(\-\frac{y^2}{N_b}\right)} dy \notag \\
&= \frac{1}{4\sqrt{\pi N_b}} \int_{-\infty}^{\infty} \exp\left(\frac{-y^2-2\an^2+4\an y}{N_b}\right) + 2\exp\left(\frac{-y^2-2\an^2}{N_b}\right) + \exp\left(\frac{-y^2-2\an^2-4\an y}{N_b}\right) dy \\
&= \frac{1}{2}\left[\exp\left(\frac{2\an^2}{N_b}\right) + \exp\left(-\frac{2\an^2}{N_b}\right) \right] = 1 + \frac{1}{2}\left(\frac{2\an^2}{N_b}\right)^2 + \mathcal{O}\left(n^{-2}\right) \le 1 + \frac{4\an^4}{N_b^2}, \label{eq:taylor2}
\end{align}
where~\eqref{eq:taylor2} follows from Taylor series expansion, and holds for sufficiently large $n$. Combining~\eqref{eq:house},~\eqref{eq:house2}, and~\eqref{eq:taylor2}, we can bound the second term of~\eqref{eq:error2} from above as 
\begin{align}
\sum_{m' \ne 1}\E\left( \int_{\y} \wyxn(\y|\X_{11}) \mathbbm{1}\left\{(\X_{m'1},\y) \in \Aset \right\} d\y\right)
& \le M e^{-\gamma} \left(1 + \frac{4\an^4}{N_b^2}\right)^n \le  M e^{-\gamma}\left(\frac{4\an^4 n}{N_b^2}\right). 
\end{align} 
Finally, the third term of~\eqref{eq:error2} can be bounded from above as 
\begin{align}
\sum_{m=1}^M \E\left(\int_{\y} \Pn(\y)\mathbbm{1}\left\{(\X_{ms},\y) \in \Aset \right\} d\y \right) &\le M \cdot \int_{\y} \Pn(\y) \sum_{\x_{ms}}\Pxn(\x_{ms})\mathbbm{1}\left\{(\x_{ms},\y) \in \Aset \right\} d\y \\
&\le M e^{-\gamma} \cdot \int_{\y} \sum_{\x_{ms}}\Pxn(\x_{ms})\wyxn(\y|\x_{ms}) d\y \\
&= M e^{-\gamma}.
\end{align}
Therefore, we conclude that 
\begin{align}
\E(P_{\text{err}}^{\text{(avg)}}) &\le \PP_{\Pxn\wyxn}\left(\log \frac{\wyxn(\Y|\X)}{\Pn(\Y)}  \le \gamma \right) + \left(1+\frac{4\an^4 n}{N_b^2} \right) M e^{-\gamma} \\
&\le \PP_{\Pxn\wyxn}\left(\log \frac{\wyxn(\Y|\X)}{\Pn(\Y)} \le \gamma \right) + C_1 M e^{-\gamma},
\end{align}
for some constant $C_1 > 0$, since $4\an^4 n/N_b^2$ scales as $\mathcal{O}(1)$.

%%%%%%%%%%%%%%%%%%%%%%%%%%%%%%%%%%%%%%%%%%%%%%%%%%%%%%%%%%%%%%%%%%%%%%%%
%%%%%% Detailed proof of converse for continuous-time channels %%%%%%%%%
%%%%%%%%%%%%%%%%%%%%%%%%%%%%%%%%%%%%%%%%%%%%%%%%%%%%%%%%%%%%%%%%%%%%%%%%

\section{Converse for continuous-time channels}

To derive a converse for continuous-time channels indicating that one can transmit at most $\mathcal{O}(\sqrt{WT})$ bits of information reliably and covertly over time $T$, we require that the energy of each codeword in the code does not differ too much. Hence, we add one more constraint in the definition of spectral mask as follows.
\begin{definition}[Spectral mask] \label{def:mask2} Let $l \in \mathbb{N}^{\ast}$ be the number of constraints, $W > 0$ be the bandwidth of interest, $\{\U_\K\}_{\K=1}^l$, $\{\al_\K\}_{\K=1}^l$, and $\{\eta_\K \}_{\K=1}^l$ be non-decreasing real-valued sequences, and $V_\K = 10^{-\frac{U_\K}{10}}$ for each $\K$. A code $\C$ with ESD $\widehat{E}(f)$ is said to fit into the spectral mask $\mathcal{S}(W, \{\U_\K\}_{\K=1}^l, \{\al_\K\}_{\K=1}^l, \{\eta_\K\}_{\K=1}^l)$ if for every $\K \in \llbracket 1,l \rrbracket$:
	\begin{enumerate}
		\item its $U_\K$-\emph{dB} bandwidth is at most $\al_\K W$, i.e., $\forall f \ge \al_\K W, \ \ \widehat{E}(f) < V_{\K} \big[\widehat{E}(f)\big]_{\max};$
		\item the energy allocated in $f \in [-\al_\K W, \al_\K W]$ satisfies $\int_{-\al_\K W}^{\al_\K W} \widehat{E}(f) df \ge \eta_\K \int_{-\infty}^{+\infty} \widehat{E}(f) df$.
		\item $E_{\max}/E_{\min} < \frac{\eta_l}{1-\eta_l}\cdot \frac{(1-\delta)^2}{2(1+\delta)-(1-\delta^2)}$, where $E_{\max}$ and $E_{\min}$ are respectively the maximum and minimum energy of codewords in the code.
	\end{enumerate}
\end{definition}

Note that the achievability scheme in Section~\ref{sec:achievability} is still valid since every codeword in our code construction has the same energy.  

For specific choices of bandwidth $\al_l W$ and time $T$, we denote the corresponding set of prolate spheroidal wave functions (PSWFs) by $\{\psi_i(t) \}_{i=1}^{\infty}$, and the truncated PSWFs (time-limited to $[0,T]$) by 
\begin{align}
\dtsi = \begin{cases} \psi_i(t), \ 0 \le t \le T, \\
0, \ \ \ \ \ \ \text{otherwise}.
\end{cases}
\end{align}  
The set of PSWFs has the properties that (i) $\{\dtsi \}_{i=1}^{\infty}$ are complete over $\mathcal{L}_2$ functions that are time-limited to $[0,T]$, and (ii) there exist an $\epsilon_1 > 0$ and a decreasing real-valued sequence $\{\lambda_i\}_{i=1}^{\infty}$ ($\lambda_i \in (0,1)$ for every $i$) such that 
\begin{align}
&\int_{-\infty}^\infty \dtsi^2 dt = \lambda_i, \ \text{ where } \lim_{\al_l WT \to \infty} \lambda_i = \begin{cases}
1, \ \ 1 \le i \le 2\al_l WT(1-\epsilon_1), \\
0, \ \ i \ge 2\al_l WT(1+\epsilon_1). 
\end{cases} \label{eq:ma1} \\
&\int_{-\infty}^\infty \dtsi D_T[\psi_j(t)] dt = 0, \ \forall i \ne j.\label{eq:ma2}
\end{align} 
Hence, every CT-codeword $\xt$, which is time-limited to $[0,T]$, can be expressed as a linear combination of the orthonormal basis $\{\frac{1}{\sqrt{\lambda_i}}\dtsi \}_{i=1}^{\infty}$, i.e.,
\begin{align}
\xt = \sum_{i=1}^\infty x_i \frac{1}{\sqrt{\lambda_i}} \dtsi, \ \text{ where } x_i = \frac{1}{\sqrt{\lambda_i}} \int_0^T \xt \dtsi dt.   
\end{align}
For every $\xt$, we define the bandlimited version of $\xt$ as
\begin{align}
 \bwxt \triangleq \int_{-\al W}^{\al W} \hat{x}(f) e^{\jj2\pi ft} df, \ \ t \in \mathbb{R}.
\end{align}

\begin{claim}
    For every $\xt$ that is time-limited to $[0,T]$, its energy $E(\xt) = \sum_{i=1}^\infty x_i^2$, and the energy of $\bwxt$ (i.e., the energy of $\xt$ allocated in $[-\al_l W,\al_l W]$ in the frequency domain) is $E_{\al_l W}(\xt) = \sum_{i=1}^\infty x_i^2 \lambda_i$. 
\end{claim}
\noindent{\it Proof:} By recalling the properties of PSWFs in~\eqref{eq:ma1} and~\eqref{eq:ma2}, we have
\begin{align}
E(\xt) = \int_{-\infty}^{\infty}x^2(t) dt = \int_{-\infty}^{\infty}\left(\sum_{i=1}^\infty \frac{x_i}{\sqrt{\lambda_i}}\dtsi\right)^2 dt = \sum_{i=1}^\infty x_i^2,
\end{align}
and the energy of $\xt$ allocated in $[-\al_l W, \al_l W]$ in the frequency domain is 
\begin{align}
E_{\al_l W}(\xt) &= \int_{-\infty}^{\infty}\bwxt^2 dt \\
&= \int_{-\infty}^{\infty}\left(\sum_{i=1}^\infty \frac{x_i}{\sqrt{\lambda_i}}B_{\al_l W}[\dtsi]\right)^2 dt \\
&= \sum_{i=1}^\infty \frac{x_i^2}{\lambda_i} \int_{-\infty}^{\infty}B_{\al_l W}[\dtsi]^2 dt + \sum_{i \ne j}\frac{x_ix_j}{\sqrt{\lambda_i \lambda_j}}\int_{-\infty}^{\infty}B_{\al_l W}[\dtsi] B_{\al_l W}[D_T[\psi_j(t)]] dt \\
&= \sum_{i=1}^\infty x_i^2 \lambda_i.
\end{align}
\qed 

\noindent{\it \underline{Willie's estimator:}} Willie maps the received signal $Z(t)$ to the PSWFs $\psi_1(t), \ldots, \psi_\NN(t)$, where $\NN$ is chosen to be $2\al_l WT(1+\epsilon_1)$, and his estimator takes the form $\Phi(Z(t)) \triangleq \sum_{i=1}^\NN [\int_{-\infty}^{\infty}Z(t)\psi_i(t) dt ]^2$. The threshold $\tau$ is set to be $\NN(N_w/2+d/\sqrt{\NN})$ with $d > \frac{N_w}{\sqrt{1-\delta}}$. His estimator outputs $\widehat{\Lambda} = 0$ if $\Phi(Z(t)) < \tau$, and $\widehat{\Lambda} = 1$ if $\Phi(Z(t)) \ge \tau$. 

When Alice is silent, the received signal $Z(t) = W(t)$, and 
\begin{align}
\sum_{i=1}^\NN \left[\int_{-\infty}^{\infty}Z(t)\psi_i(t) dt \right]^2 = \sum_{i=1}^\NN \left[\int_{-\infty}^{\infty}W(t)\psi_i(t) dt \right]^2 = \sum_{i=1}^\NN W_i^2, \ \text{ where } W_i \sim \mathcal{N}\left(0,\frac{N_w}{2}\right).
\end{align}
Note that $\{W_i\}_{i=1}^{\infty}$ are i.i.d. random variables with $\E(W_i^2) = N_w/2$ and $\text{Var}(W_i^2) = N_w^2/2$. By using the Chebyshev's inequality, we have 
\begin{align}
P_{\text{FA}} = \PP_{H_0}\left(\sum_{i=1}^\NN W_i^2 \ge \tau\right) \le \frac{N_w^2}{2d^2}.
\end{align}
When Alice transmits a CT-codeword $\xt$, we have  
\begin{align}
\sum_{i=1}^\NN \left[\int_{-\infty}^{\infty}Z(t)\psi_i(t) dt \right]^2 = \sum_{i=1}^\NN \left[\int_{-\infty}^{\infty}(\xt + W(t))\psi_i(t) dt \right]^2 = \sum_{i=1}^\NN W_i^2 + 2x_i\sqrt{\lambda_i}W_i + x_i^2\lambda_i,
\end{align}
since $\int_{-\infty}^{\infty} x(t)\psi_i(t) dt = x_i\sqrt{\lambda_i}$. One can show that $\{W_i^2 + 2x_i\sqrt{\lambda_i}W_i\}_{i=1}^\infty$ are i.i.d. random variables with 
\begin{align}
\E\left(W_i^2 + 2x_i\sqrt{\lambda_i}W_i\right) = \frac{N_w}{2}, \ \text{Var}\left(W_i^2 + 2x_i\sqrt{\lambda_i}W_i \right) = \frac{N_w^2}{2}+2x_i^2\lambda_i N_w.
\end{align}
When $\xt$ is transmitted, we use the Chebyshev's inequality to bound the probability of missed detection from above as
\begin{align}
P_{\text{MD}}(\xt) = \PP\left(\sum_{i=1}^\NN W_i^2 + 2x_i\sqrt{\lambda_i}W_i + x_i^2\lambda_i < \tau\right) \le \frac{\NN N_w^2 + 4N_w \sum_{i=1}^{\NN} x_i^2\lambda_i}{2\left(d\sqrt{\NN} - \sum_{i=1}^\NN x_i^2\lambda_i\right)^2}.
\end{align}
Let $\Dd \triangleq \{(m,s) \in \llbracket 1,M\rrbracket \times \llbracket 1,K\rrbracket: \sum_{i=1}^\NN x_{ms,i}^2\lambda_i \le 2d\sqrt{\NN} \}$ be a subset of codewords such that the energy allocated in  $f\in [-\al_l W, \al_l W]$ is small, and let $\gamma \triangleq (1-\delta-\frac{N_w^2}{d^2})/(1-\frac{N_w^2}{2d^2})$ be a constant depending on $N_w, \delta$ and $d$. Note that $\gamma > 0$ since we set $d > \frac{N_w}{\sqrt{1-\delta}}$.
\begin{claim} \label{claim:dd}
	For any code $\C$ that ensures $(1-\delta)$-covertness, we have 
	\begin{align}
	\frac{|\Dd|}{MK} \ge \gamma + \mathcal{O}\left(\NN^{-1/2}\right). \label{eq:soc}
	\end{align}
\end{claim}
\noindent{\it Proof:} By using the estimator $\Phi(\cdot)$ described above, Willie can bound the sum of $P_{\text{FA}}$ and $P_{\text{MD}}$ as
\begin{align} 
P_{\text{FA}} + P_{\text{MD}} &= P_{\text{FA}} + \sum_{m=1}^{\N}\sum_{s=1}^K \frac{1}{MK} P_{\text{MD}}(x_{ms}(t)) \\
&= P_{\text{FA}}+\frac{1}{MK}\sum_{(m,s)\in \Dd}P_{\text{MD}}(x_{ms}(t)) + \frac{1}{MK} \sum_{(m,s)\notin \Dd}P_{\text{MD}}(x_{ms}(t)) \\
&\le \frac{N_w^2}{2d^2} + \frac{|\Dd|}{MK} + \frac{1}{MK}\sum_{(m,s)\notin \Dd}\frac{\NN N_w^2 + 4N_w \sum_{i=1}^{\NN} x_{ms,i}^2\lambda_i}{2\left(d\sqrt{\NN} - \sum_{i=1}^\NN x_{ms,i}^2\lambda_i\right)^2} \\
&\le \frac{N_w^2}{2d^2} + \frac{|\Dd|}{MK} + \frac{MK-|\Dd|}{MK} \frac{\NN N_w^2 + 8N_w d\sqrt{\NN}}{2d^2\NN}.
\end{align} 
Suppose inequality~\eqref{eq:soc} does not hold, then we have
\begin{align}
P_{\text{FA}}+P_{\text{MD}} &\le \frac{N_w^2}{2d^2} + \frac{\NN N_w^2 + 8N_w d\sqrt{\NN}}{2d^2\NN} + \frac{|\Dd|}{MK}\left(1 - \frac{\NN N_w^2 + 8N_w d\sqrt{\NN}}{2d^2\NN}\right) \\
& \le \frac{N_w^2}{2d^2} + \frac{\NN N_w^2 + 8N_w d\sqrt{\NN}}{2d^2\NN} + 1 - \delta - \frac{N_w^2}{2d^2} - \frac{\NN N_w^2 + 8N_w d\sqrt{\NN}}{2d^2\NN} \\
&\le 1 - \delta,
\end{align}
which contradicts the assumption that the code ensures $(1-\delta)$-covertness. This completes the proof of  Claim~\ref{claim:dd}. \qed 

Furthermore, we define $\Dd^s \triangleq \Dd \cap \C_s$ as the collection of CT-codewords lying in the intersection between $\Dd$ and the sub-code $\C_s$. For any $s \in \llbracket 1, K\rrbracket$, we say $\Dd^s$ is a {\it small sub-code} if $|\Dd^s| \le \gamma M/2$, and a {\it big sub-code} otherwise. Let $K_s$ and $K_b$ be the numbers of small and big sub-codes, respectively. As $T$ tends to infinity, we have 
\begin{align}
    \gamma MK \stackrel{T \to \infty}{\le} |\Dd| \le K_s \frac{\gamma M}{2} + K_b M,
\end{align}
which yields that $K_b \ge \frac{\gamma}{2 - \gamma} K$. We then focus on these big sub-codes. Let $\widetilde{\mathcal{D}}_d$ be the collection of CT-codewords in the big sub-codes, i.e., $$\widetilde{\mathcal{D}}_d \triangleq \left\{(m,s) \in \Dd: |\Dd^s| > \gamma M /2  \right\}.$$

Let $\phi \triangleq \eta_l - (1-\eta_l)\frac{(1+4\delta - \delta^2)E_{\max}}{(1-\delta)E_{\min}}$, which is positive due to the assumption $\frac{E_{\max}}{E_{\min}} < \frac{\eta_l}{1-\eta_l}\cdot \frac{(1-\delta)^2}{2(1+\delta)-(1-\delta^2)}$. Suppose every big sub-code $\Dd^s$ satisfies 
\begin{align}
    \frac{\sum_{(m,s) \in \Dd^s}E_{\al_l W}(\xmt)}{\sum_{(m,s) \in \Dd^s}E(\xmt)} < \phi,
\end{align}
then the set $\widetilde{\mathcal{D}}_d$ satisfies
\begin{align}
    \frac{\sum_{(m,s) \in \widetilde{\mathcal{D}}_d}E_{\al_l W}(\xmt)}{\sum_{(m,s) \in \widetilde{\mathcal{D}}_d}E(\xmt)} < \phi. \label{eq:dd}
\end{align}
Hence, the fractional energy allocated in $f \in [-\al_l W, \al_l W]$ of code $\C$ also satisfies  
\begin{align}
    \frac{\int_{-\al_l W}^{\al_l W} \widehat{E}(f) df}{\int_{-\infty}^{+\infty} \widehat{E}(f) df} &= \frac{\frac{1}{MK}\sum_{(m,s)}E_{\al_l W}(\xmt)}{\frac{1}{MK}\sum_{(m,s)}E(\xmt)} \\
    &= \frac{\sum_{(m,s) \in \widetilde{\mathcal{D}}_d}E_{\al_l W}(\xmt) + \sum_{(m,s) \notin \widetilde{\mathcal{D}}_d}E_{\al_l W}(\xmt)}{\sum_{(m,s) \in \widetilde{\mathcal{D}}_d}E(\xmt) + \sum_{(m,s) \notin \widetilde{\mathcal{D}}_d}E(\xmt)} \\
    &< \frac{ \phi \sum_{(m,s) \in \widetilde{\mathcal{D}}_d}E(\xmt) + \sum_{(m,s) \notin \widetilde{\mathcal{D}}_d}E(\xmt)}{\sum_{(m,s) \in \widetilde{\mathcal{D}}_d}E(\xmt) + \sum_{(m,s) \notin \widetilde{\mathcal{D}}_d}E(\xmt)} \label{eq:cc}\\
    &\le \frac{ \phi \frac{\gamma^2}{2(2-\gamma)}MKE_{\min} + \left(1-\frac{\gamma^2}{2(2-\gamma)}\right)MK E_{\max}}{\frac{\gamma^2}{2(2-\gamma)}MKE_{\min} + \left(1-\frac{\gamma^2}{2(2-\gamma)}\right)MK E_{\max}} \label{eq:bb}\\
    &= \frac{ \phi \frac{(1-\delta)^2}{2(1+\delta)}MKE_{\min} + \left(1-\frac{(1-\delta)^2}{2(1+\delta)}\right)MK E_{\max}}{\frac{(1-\delta)^2}{2(1+\delta)}MKE_{\min} + \left(1-\frac{(1-\delta)^2}{2(1+\delta)}\right)MK E_{\max}} \label{eq:aa} \\
    &= \eta_l. \label{eq:00}
\end{align}
where inequality~\eqref{eq:cc} follows from~\eqref{eq:dd}, inequality~\eqref{eq:bb} holds since $\sum_{(m,s) \in \widetilde{\mathcal{D}}_d}E(\xmt) \ge \frac{\gamma^2}{2(2-\gamma)}MK E_{\min}$ and $\sum_{(m,s) \notin \widetilde{\mathcal{D}}_d}E(\xmt) \le \left(1-\frac{\gamma^2}{2(2-\gamma)}\right)MK E_{\max}$, and~\eqref{eq:aa} is obtained since $\gamma \to (1-\delta)$ if we set $d$ to be large enough. Equation~\eqref{eq:00} is obtained by substituting the expression of $\phi$ into~\eqref{eq:aa}. Hence,~\eqref{eq:00} contradicts the requirement that $\int_{-\al_l W}^{\al_l W} \widehat{E}(f) df \ge \eta_l \int_{-\infty}^{+\infty} \widehat{E}(f) df$.
Therefore, there must exist a big sub-code $\Dd^{s_0}$ (satisfying $|\Dd^{s_0}| > \gamma M /2$) such that 
\begin{align}
\frac{\sum_{\xt \in \Dd^{s_0}}E_{\al_l W}(\xt)}{\sum_{\xt \in \Dd^{s_0}}E(\xt)} \ge \phi. \label{eq:nuss}
\end{align}

Recall that every $\xt \in \Dd^{s_0}$ satisfies $\sum_{i=1}^\NN x_i^2 \lambda_i \le 2d\sqrt{\NN}$. Hence, the average energy of codewords in $\Dd^{s_0}$ can be upper bounded as
\begin{align}
 \frac{\sum_{(m,s) \in \Dd^{s_0}}E(\xmt)}{|\Dd^{s_0}|} &= \frac{\sum_{(m,s) \in \Dd^{s_0}}E_{\al_l W}(\xmt) }{\phi \cdot |\Dd^{s_0}|}  \\
 &= \frac{\sum_{(m,s) \in \Dd^{s_0}} \sum_{i=1}^{\infty} x_{ms,i}^2 \lambda_i }{\phi \cdot |\Dd^{s_0}|} \\
 &= \frac{\sum_{i=1}^{N}\lambda_i \sum_{(m,s) \in \Dd^{s_0}} x_{ms,i}^2 }{\phi \cdot |\Dd^{s_0}|} + \frac{\sum_{i=N+1}^{\infty}\lambda_i \sum_{(m,s) \in \Dd^{s_0}} x_{ms,i}^2 }{\phi \cdot |\Dd^{s_0}|} \\
 & \stackrel{T \to \infty}{=} \frac{\sum_{i=1}^{N}\lambda_i \sum_{(m,s) \in \Dd^{s_0}} x_{ms,i}^2 }{\phi \cdot |\Dd^{s_0}|} \\
 &\le \frac{2d\sqrt{N}}{\phi}.
\end{align}
For notational convenience we define the average energy of codewords in $\Dd^{s_0}$ as
\begin{align}
    \sum_{i=1}^{\infty} \overline{x_i^2} \triangleq \sum_{i=1}^{\infty} \frac{ \sum_{(m,s) \in \Dd^{s_0}} x_{ms,i}^2}{|\Dd^{s_0}|} =  \frac{\sum_{(m,s) \in \Dd^{s_0}}\sum_{i=1}^{\infty} x_{ms,i}^2}{|\Dd^{s_0}|} = \frac{\sum_{(m,s) \in \Dd^{s_0}}E(\xt)}{|\Dd^{s_0}|} \le  \frac{2d\sqrt{N}}{\phi}.
\end{align}
Finally, by applying information inequalities~\cite[Chapter 8.2]{gallager1968information} on $\Dd^{s_0}$, we have 
\begin{align}
\log(\gamma/2) + \log\N = \log(\gamma \N/2) &\le \I(X(t);Y(t)) + 2\epsilon_n/\gamma \\
&= \lim_{n \to \infty} \I(X^n;Y^n) + 2\epsilon_n/\gamma \\
&= \lim_{n \to \infty} \sum_{i=1}^n \frac{1}{2}\log\left(1+ \frac{2\overline{x_i^2} }{N_b}\right) + 2\epsilon_n/\gamma \\
&= \lim_{n \to \infty} \frac{n}{2} \log\left(1+ \frac{\frac{1}{n}\sum_{i=1}^n 2 \overline{x_i^2} }{N_b}\right) + 2\epsilon_n/\gamma \\
&= \lim_{n \to \infty} \frac{\sum_{i=1}^n \overline{x_i^2} }{N_b} + 2\epsilon_n/\gamma\\
& < \frac{2d\sqrt{\NN}}{\phi N_b} + 2\epsilon_n/\gamma \\
& = \frac{2d\sqrt{2\al_l WT(1+\epsilon_1)}}{\phi N_b} + 2\epsilon_n/\gamma,
\end{align}
and 
\begin{align}
\lim_{T \to \infty} \frac{\log \N}{\sqrt{T}} \le \frac{2d\sqrt{2\al_l W(1+\epsilon_1)}}{\phi N_b}.
\end{align}

%%%%%%%%%%%%%%%%%%%%%%%%%%%%%%%%%%%%%%%%%%%%%%%%%%%%%%%%%%%%%%
%%%%%%%%%%% Appendix for KL-divergence %%%%%%%%%%%%%%%%%%%%%%%
%%%%%%%%%%%%%%%%%%%%%%%%%%%%%%%%%%%%%%%%%%%%%%%%%%%%%%%%%%%%%%
%\begin{comment}

\vspace{15pt}

\section{} \label{sec:calculation}
We first show that 
\begin{align}
\I(X;Z) &= \sum_{x \in \{-\an, \an \}}\Px(x) \int_{-\infty}^{\infty}\wzx(z|x)\log\frac{\wzx(z|x)}{\Qt(z)} dz \\
& = \frac{1}{2} \int_{-\infty}^{\infty}\Qa(z)\log\frac{\Qa(z)}{\Qt(z)} dz + \frac{1}{2} \int_{-\infty}^{\infty}Q_{-a}(z)\log\frac{Q_{-a}(z)}{\Qt(z)} dz \\
&= \int_{-\infty}^{\infty}\Qa(z)\log\frac{\Qa(z)}{\Qt(z)} dz \label{eq:sym} \\
&= \D(\Qa || \Qt),
\end{align}
where~\eqref{eq:sym} follows from the symmetry of Gaussian random variables. Note that
\begin{align}
\D(\Qa || \Qt) &= \int_{-\infty}^{\infty}\Qa(z)\log\left( \frac{\exp\left(\frac{-\an^2+2\an z}{N_w}\right)}{\frac{1}{2}\exp\left(\frac{-\an^2+2\an z}{N_w}\right) + \frac{1}{2}\exp\left(\frac{-\an^2-2\an z}{N_w}\right)}\right) dz \\
&= \int_{-\infty}^{\infty}\Qa(z)\left(\frac{-\an^2+2\an z}{N_w}\right)dz - \int_{-\infty}^{\infty} \Qa(z)\log\left[\frac{1}{2}e^{\frac{-\an^2+2\an z}{N_w}} + \frac{1}{2}e^{\frac{-\an^2-2\an z}{N_w}}\right]dz \\
&= \int_{-\infty}^{\infty}\Qa(z)\left(\frac{-\an^2+2\an z}{N_w}\right)dz - \int_{-\infty}^{\infty}\Qa(z)\left(-\frac{\an^2}{N_w}\right)dz \notag \\
&\qquad\qquad\qquad\qquad\qquad\qquad\qquad- \int_{-\infty}^{\infty} \Qa(z)\log\left[\frac{1}{2}e^{\frac{2\an z}{N_w}}+ \frac{1}{2}e^{\frac{-2\an z}{N_w}}\right]dz  \\
&\le \frac{\an^2}{N_w} + \frac{\an^2}{N_w} + \int_{-\infty}^{\infty} \Qa(z)\left(\frac{2\an^2 z^2}{N_w} - \frac{4\an^4z^4}{3N_w^2}\right) dz \label{eq:tt} \\
&= \frac{2\an^2}{N_w} - \left(\frac{\an^2}{N_w} + \frac{\an^4}{N_w^2} - \frac{4\an^6}{N_w^3} - \frac{4\an^8}{3N_w^4}\right) \\
&\le \frac{\an^2}{N_w},
\end{align}
where inequality~\eqref{eq:tt} follows from Taylor series expansion. 
Therefore, we have 
\begin{align}
\I(X;Z) = \D(\Qa || \Qt) \le \frac{\an^2}{N_w}.
\end{align}
Furthermore, we show that
\begin{align}
&\E_{\Px\wzx}\left(\log^2\left(\frac{\wzx(Z|X)}{\Qt(Z)}\right)\right) \\
&=  \sum_{x \in \{-\an,\an\} }\Px(x) \int_{-\infty}^{\infty}\wzx(z|x)\log^2\left(\frac{\wzx(z|x)}{\Qt(z)}\right) dz \\
&= \frac{1}{2}\int_{-\infty}^{\infty} \Qa(z)\log^2\left(\frac{\Qa(z)}{\Qt(z)}\right)dz + \frac{1}{2}\int_{-\infty}^{\infty} Q_{-a}(z)\log^2\left(\frac{Q_{-a}(z)}{\Qt(z)}\right)dz \\
& = \int_{-\infty}^{\infty} \Qa(z)\log^2\left(\frac{\Qa(z)}{\Qt(z)}\right)dz \\
&= \int_{-\infty}^{\infty} \Qa(z) \frac{4\an^2z^2}{N_w^2}dz - \int_{-\infty}^{\infty} \Qa(z) \frac{4\an z}{N_w}\log\left(\frac{1}{2}\exp\left(\frac{2\an z}{N_w}\right) + \frac{1}{2}\exp\left(-\frac{2\an z}{N_w}\right) \right) dz \notag \\
& \qquad\qquad\qquad\qquad\qquad\qquad +\int_{-\infty}^{\infty} \Qa(z)\log^2\left(\frac{1}{2}\exp\left(\frac{2\an z}{N_w}\right) + \frac{1}{2}\exp\left(-\frac{2\an z}{N_w}\right) \right) dz \\
& = \frac{2\an^2}{N_w} + \mathcal{O}(\an^4).
\end{align}

\section{Proof of Lemma~\ref{lemma:kl3}} \label{appendix:kl3}
Note that
\newcommand{\zi}{\mathbf{z}_{\sim i}}
\begin{align}
\int_{\z}\left(\Qh(\z)-\Qtn(\z)\right)\log \frac{\Qtn(\z)}{\Qn(\z)} d\z &= \int_{\z}\left(\Qh(\z)-\Qtn(\z)\right)\left(\sum_{i=1}^n \log \frac{\Qt(z_i)}{\Q(z_i)}\right) d\z \\
&=\sum_{i=1}^n \int_{z_i} \left(\log \frac{\Qt(z_i)}{\Q(z_i)}\right) \int_{\zi} \left(\Qh(\z)-\Qtn(\z)\right) d\zi dz_i,
\end{align}
where $\zi = [z_1,\ldots,z_{i-1},z_{i+1},\ldots,z_n]$. Let $\pm \nun$ be the roots of the function  
\begin{align}
f_n(x) = \frac{1}{2}\exp\left(-\frac{(x-\an)^2}{N_w}\right) + \frac{1}{2}\exp\left(-\frac{(x+\an)^2}{N_w}\right) - \exp\left(-\frac{x^2}{N_w}\right).
\end{align}
One may check that $f_n(x)$ is odd and only has two roots (by checking the derivative of $f_n(x)$). Hence, 
\begin{itemize}
	\item When $z_i \in (-\nun,\nun)$, $\Qt(z_i) < \Q(z_i)$ and $\log(\Qt(z_i)/\Q(z_i)) < 0$;
	\item When $z_i \in (-\infty,-\nun] \cup [\nun,\infty)$, $\Qt(z_i) \ge \Q(z_i)$ and $\log(\Qt(z_i)/\Q(z_i)) \ge 0$.
\end{itemize}
Note that 
\begin{align}
&\int_{z_i} \left(\log \frac{\Qt(z_i)}{\Q(z_i)}\right) \int_{\zi} \left(\Qh(\z)-\Qtn(\z)\right) d\zi dz_i \\
&=\int_{\nun}^{\infty} \left(\log \frac{\Qt(z_i)}{\Q(z_i)}\right) \int_{\zi} \left(\Qh(\z)-\Qtn(\z)\right) d\zi dz_i + \int_{-\infty}^{-\nun} \left(\log \frac{\Qt(z_i)}{\Q(z_i)}\right) \int_{\zi} \left(\Qh(\z)-\Qtn(\z)\right) d\zi dz_i \notag\\
&\qquad\qquad\qquad\qquad\qquad\qquad\qquad\qquad\qquad\qquad+ \int_{-\nun}^{\nun} \left(\log \frac{\Qt(z_i)}{\Q(z_i)}\right) \int_{\zi} \left(\Qh(\z)-\Qtn(\z)\right) d\zi dz_i. \label{eq:moo2}
\end{align}

\begin{claim} \label{claim:1}
	With probability at least $1 - 2\exp\left(-\frac{\varepsilon^2 \N K}{6}\right)$ over the code design, a randomly chosen code $\C$ satisfies
	\begin{align}
	\forall z_i \in \mathbb{R}, \ \left| \int_{\zi} \left(\Qh(\z)-\Qtn(\z)\right) d\zi \right| \le \varepsilon\Qt(z_i) \le \varepsilon \Qt(\an).
	\end{align}
\end{claim}
\noindent{\it Proof:} First note that for any $z_i \in \mathbb{R}$,
\begin{align}
\int_{\zi}\Qh(\z)d\zi &= \int_{\zi}\frac{1}{\N K}\sum_{m=1}^\N \sum_{s=1}^K \wzxn(\z|\x_{ms}) d\zi \\
&= \frac{1}{\N K}\sum_{m=1}^\N \sum_{s=1}^K \int_{\zi} \wzxn(\z|\x_{ms}) d\zi \\
&= \frac{1}{\N K}\sum_{m=1}^\N \sum_{s=1}^K \wzx(z_i|x_{ms,i}) \\
&= \frac{\wzx(z_i|\an)\left(\sum_{m=1}^\N \sum_{s=1}^K \mathbbm{1}\left\{x_{ms,i}=\an \right\} \right)}{\N K} + \frac{\wzx(z_i|-\an)\left(\sum_{m=1}^\N \sum_{s=1}^K \mathbbm{1}\left\{x_{ms,i}=-\an\right\} \right)}{\N K}.
\end{align}
Since
\begin{align}
\E\left(\sum_{m=1}^\N \sum_{s=1}^K \mathbbm{1}\left\{x_{ms,i}=\an\right\}\right) = \E\left(\sum_{m=1}^\N \sum_{s=1}^K \mathbbm{1}\left\{x_{ms,i}=-\an\right\}\right) = \frac{\N K}{2},
\end{align}
by the Chernoff bound, we have
\begin{align}
\PP\left(\frac{\N K(1-\varepsilon)}{2} \le \sum_{m=1}^\N \sum_{s=1}^K \mathbbm{1}\left\{x_{ms,i}=\an \right\} \le \frac{\N K(1+\varepsilon)}{2} \right) \ge 1 - 2\exp\left(-\frac{\varepsilon^2 \N K}{6}\right). \label{eq:moo3}
\end{align} 
Equation~\eqref{eq:moo3} further implies that with probability at least $1 - 2\exp\left(-\frac{\varepsilon^2 \N K}{6}\right)$ over the code design, a randomly chosen code $\C$ satisfies
\begin{align}
\int_{\zi}\Qh(\z)d\zi &= \frac{\wzx(z_i|\an)\left(\sum_{m=1}^\N \sum_{s=1}^K \mathbbm{1}\left\{x_{ms,i}=\an\right\} \right)}{\N K} + \frac{\wzx(z_i|-\an)\left(\sum_{m=1}^\N \sum_{s=1}^K \mathbbm{1}\left\{x_{m,i}=-\an\right\} \right)}{\N K} \\
&\le \frac{(1+\varepsilon)}{2} \wzx(z_i|\an) + \frac{(1+\varepsilon)}{2} \wzx(z_i|-\an) \\
& \le (1+\varepsilon)\left(\frac{1}{2} \Qa(z_i) + \frac{1}{2} Q_{-a}(z_i) \right) \\
& = (1+\varepsilon) \Qt(z_i), \\
\int_{\zi}\Qh(\z)d\zi &\ge (1-\varepsilon) \Qt(z_i).
\end{align}
By noting that $\int_{\zi}\Qtn(\z)d\zi = \Qt(z_i)$ for any $z_i \in \mathbb{R}$, we have
\begin{align}
\left| \int_{\zi} \left(\Qh(\z)-\Qtn(\z)\right) d\zi \right| = \left| \int_{\zi}\Qh(\z) d\zi - \int_{\zi}\Qtn(\z)d\zi \right| \stackrel{\text{w.h.p.}}{\le} \varepsilon \Qt(z_i) \le \varepsilon \Qt(\an),
\end{align}
where the last inequality holds since the maximum value of $\Qt(z_i)$ is achieved when $z_i = \pm \an$.
\qed

With the help of Claim~\ref{claim:1}, we are now able to bound~\eqref{eq:moo2}. Because of the symmetry property, the techniques used to bound the first and the second terms in~\eqref{eq:moo2} are the same, hence in the following we only focus on the first and the third terms.     

\noindent{1) \it Bounding the first term of~\eqref{eq:moo2}:} 
\begin{align}
&\int_{\nun}^{\infty} \left(\log \frac{\Qt(z_i)}{\Q(z_i)}\right) \int_{\zi} \left(\Qh(\z)-\Qtn(\z)\right) d\zi dz_i \\
&\le \int_{\nun}^{\infty} \left(\log \frac{\Qt(z_i)}{\Q(z_i)}\right) \left|\int_{\zi} \left(\Qh(\z)-\Qtn(\z)\right) d\zi \right| dz_i \\
& \le \int_{\nun}^{\infty} \frac{\Qt(z_i)}{\Q(z_i)} \left|\int_{\zi} \left(\Qh(\z)-\Qtn(\z)\right) d\zi \right| dz_i \\
& = \int_{\nun}^{\fn} \frac{\Qt(z_i)}{\Q(z_i)} \left|\int_{\zi} \left(\Qh(\z)-\Qtn(\z)\right) d\zi \right| dz_i + \int_{\fn}^{\infty} \frac{\Qt(z_i)}{\Q(z_i)} \left|\int_{\zi} \left(\Qh(\z)-\Qtn(\z)\right) d\zi \right| dz_i \\
&\stackrel{\text{w.h.p.}}{\le} \varepsilon \Qt(\an) \int_{\nun}^{\fn} \frac{\Qt(z_i)}{\Q(z_i)} dz_i + \int_{\fn}^{\infty} \frac{\Qt(z_i)}{\Q(z_i)}\left(\int_{\zi} \Qh(\z) d\zi + \int_{\zi}\Qtn(\z) d\zi \right) dz_i .
\end{align}
By choosing $\varepsilon = \exp\left(-\mathcal{O}(\sqrt{n})\right)$, we have
\begin{align}
&\varepsilon \Qt(\an) \int_{\nun}^{\fn} \frac{\Qt(z_i)}{\Q(z_i)} dz_i \\
&\le \varepsilon \Qt(\an) \int_{0}^{\fn} \frac{\Qt(z_i)}{\Q(z_i)} dz_i \\
&= \varepsilon \Qt(\an) \int_{0}^{\fn}\frac{1}{2}\exp\left(\frac{-\an^2+2\an z_i}{N_w}\right) + \frac{1}{2}\exp\left(\frac{-\an^2-2\an z_i}{N_w}\right) dz_i \\
&=  \frac{\varepsilon \Qt(\an)N_w}{4\an}\left\{\left[\exp\left(\frac{2\an \fn-\an^2}{N_w}\right)-\exp\left(\frac{-\an^2}{N_w}\right)\right] - \left[\exp\left(\frac{-2\an \fn-\an^2}{N_w}\right) - \exp\left(\frac{-\an^2}{N_w}\right)\right]\right\} \\
&= \exp\left(-\mathcal{O}(\sqrt{n})\right).
\end{align}
Further, for any $z_i > \fn$, 
\begin{align}
\int_{\zi} \Qh(\z) d\zi + \int_{\zi}\Qtn(\z) d\zi = \frac{1}{\N K}\sum_{m=1}^\N \sum_{s=1}^K\wzx(z_i|x_{ms,i}) + \Qt(z_i) \le 2\Qa(z_i),
\end{align}
hence, we have
\begin{align}
&\int_{\fn}^{\infty} \frac{\Qt(z_i)}{\Q(z_i)}\left(\int_{\zi} \Qh(\z) d\zi + \int_{\zi}\Qtn(\z) d\zi \right) dz_i\\
& \le 2\int_{\fn}^{\infty} \frac{\Qt(z_i)\Qa(z_i)}{\Q(z_i)} dz_i \\
&\le \frac{1}{\sqrt{\pi N_w}} \int_{\fn}^{\infty} \exp\left(\frac{-z_i^2 - 2\an^2 + 4\an z_i}{N_w}\right) dz_i + \frac{1}{\sqrt{\pi N_w}} \int_{\fn}^{\infty} \exp\left(\frac{-z_i^2 - 2\an^2}{N_w}\right) dz_i \\
&\le \frac{1}{2}\exp\left(\frac{2\an^2}{N_w}\right)\exp\left(-\frac{\fn^2}{N_w} + \frac{4\an \fn}{N_w} - \frac{4\an}{N_w}\right) + \frac{1}{2}\exp\left(\frac{2\an^2}{N_w}\right)\exp\left(-\frac{\fn^2}{N_w}\right) \\
& = \exp\left(-\mathcal{O}(n)\right).
\end{align}
Therefore, with high probability the first term of~\eqref{eq:moo2} is bounded from above as 
\begin{align}
\int_{\nun}^{\infty} \left(\log \frac{\Qt(z_i)}{\Q(z_i)}\right) \int_{\zi} \left(\Qh(\z)-\Qtn(\z)\right) d\zi dz_i \le \exp\left(-\mathcal{O}(\sqrt{n})\right).
\end{align}
By symmetry, with high probability the second term of~\eqref{eq:moo2} is also bounded from above as
\begin{align}
\int_{-\infty}^{-\nun} \left(\log \frac{\Qt(z_i)}{\Q(z_i)}\right) \int_{\zi} \left(\Qh(\z)-\Qtn(\z)\right) d\zi dz_i \le \exp\left(-\mathcal{O}(\sqrt{n})\right).
\end{align}

\noindent{2) \it Bounding the third term of~\eqref{eq:moo2}:} Since $\Qt(z_i) < \Q(z_i)$ when $z_i \in (-\nun,\nun)$, we have
\begin{align}
&\int_{-\nun}^{\nun} \left(\log \frac{\Qt(z_i)}{\Q(z_i)}\right) \int_{\zi} \left(\Qh(\z)-\Qtn(\z)\right) d\zi dz_i \\
&\le \int_{-\nun}^{\nun} -\left(\log \frac{\Qt(z_i)}{\Q(z_i)}\right)\left| \int_{\zi} \left(\Qh(\z)-\Qtn(\z)\right) d\zi \right| dz_i \\
& = \int_{-\nun}^{\nun} \left(\log \frac{\Q(z_i)}{\Qt(z_i)}\right)\left| \int_{\zi} \left(\Qh(\z)-\Qtn(\z)\right) d\zi \right| dz_i \\
& \le \int_{-\nun}^{\nun} \frac{\Q(z_i)}{\Qt(z_i)}\left| \int_{\zi} \left(\Qh(\z)-\Qtn(\z)\right) d\zi \right| dz_i \\
& \stackrel{\text{w.h.p.}}{\le} \varepsilon \Qt(z_i) \int_{-\nun}^{\nun} \frac{\Q(z_i)}{\Qt(z_i)} dz_i \\
&\le \varepsilon \Qt(z_i) \frac{2\nun \Q(0)}{\Qt(0)}. \label{eq:mei1}
\end{align}
Though we do not obtain the value of $\nun$ analytically, we are still able to bound $\nun$ from above as 
\begin{align}
\nun \le \frac{\an}{2} + \frac{N_w}{2\an}. \label{eq:mei2}
\end{align}
Combining~\eqref{eq:mei1} and~\eqref{eq:mei2}, we have 
\begin{align}
\int_{-\nun}^{\nun} \left(\log \frac{\Qt(z_i)}{\Q(z_i)}\right) \int_{\zi} \left(\Qh(\z)-\Qtn(\z)\right) d\zi dz_i = \exp\left(-\mathcal{O}(\sqrt{n})\right).
\end{align}

\bibliographystyle{IEEEtran}
\bibliography{mainfunction,steg_mayank}

\end{document}

%% file: system-model.tex
{\pgfkeys{/pgf/fpu/.try=false}%
\ifx\XFigwidth\undefined\dimen1=0pt\else\dimen1\XFigwidth\fi
\divide\dimen1 by 6602
\ifx\XFigheight\undefined\dimen3=0pt\else\dimen3\XFigheight\fi
\divide\dimen3 by 2704
\ifdim\dimen1=0pt\ifdim\dimen3=0pt\dimen1=3946sp\dimen3\dimen1
  \else\dimen1\dimen3\fi\else\ifdim\dimen3=0pt\dimen3\dimen1\fi\fi
\tikzpicture[x=+\dimen1, y=+\dimen3]
{\ifx\XFigu\undefined\catcode`\@11
\def\temp{\alloc@1\dimen\dimendef\insc@unt}\temp\XFigu\catcode`\@12\fi}
\XFigu3946sp
% Uncomment to scale line thicknesses with the same
% factor as width of the drawing.
%\pgfextractx\XFigu{\pgfqpointxy{1}{1}}
\ifdim\XFigu<0pt\XFigu-\XFigu\fi
\pgfdeclarearrow{
  name = xfiga1,
  parameters = {
    \the\pgfarrowlinewidth \the\pgfarrowlength \the\pgfarrowwidth\ifpgfarrowopen o\fi},
  defaults = {
	  line width=+7.5\XFigu, length=+120\XFigu, width=+60\XFigu},
  setup code = {
    % miter protrusion = thk * sqrt(wd^2 + (tipmv*len)^2) / (2 * wd)
    \dimen7 2.1\pgfarrowlength\pgfmathveclen{\the\dimen7}{\the\pgfarrowwidth}
    \dimen7 2\pgfarrowwidth\pgfmathdivide{\pgfmathresult}{\the\dimen7}
    \dimen7 \pgfmathresult\pgfarrowlinewidth
    \pgfarrowssettipend{+\dimen7}
    \pgfarrowssetbackend{+-\pgfarrowlength}
    \dimen9 -\pgfarrowlength\advance\dimen9 by-0.45\pgfarrowlinewidth
    \pgfarrowssetlineend{+\dimen9}
    \dimen9 -\pgfarrowlength\advance\dimen9 by-0.5\pgfarrowlinewidth
    \pgfarrowssetvisualbackend{+\dimen9}
    \pgfarrowshullpoint{+\dimen7}{+0pt}
    \pgfarrowsupperhullpoint{+-\pgfarrowlength}{+0.5\pgfarrowwidth}
    \pgfarrowssavethe\pgfarrowlinewidth
    \pgfarrowssavethe\pgfarrowlength
    \pgfarrowssavethe\pgfarrowwidth
  },
  drawing code = {\pgfsetdash{}{+0pt}
    \ifdim\pgfarrowlinewidth=\pgflinewidth\else\pgfsetlinewidth{+\pgfarrowlinewidth}\fi
    \pgfpathmoveto{\pgfqpoint{-\pgfarrowlength}{-0.5\pgfarrowwidth}}
    \pgfpathlineto{\pgfqpoint{0pt}{0pt}}
    \pgfpathlineto{\pgfqpoint{-\pgfarrowlength}{0.5\pgfarrowwidth}}
    \pgfpathclose
    \ifpgfarrowopen\pgfusepathqstroke\else\pgfsetfillcolor{.}
	\ifdim\pgfarrowlinewidth>0pt\pgfusepathqfillstroke\else\pgfusepathqfill\fi\fi
  }
}
\clip(2849,-5103) rectangle (9451,-2399);
\tikzset{inner sep=+0pt, outer sep=+0pt}
\pgfsetlinewidth{+7.5\XFigu}
\pgfsetdash{{+60\XFigu}{+60\XFigu}}{++0pt}
\pgfsetstrokecolor{black}
\pgfsetfillcolor{black!5}
\filldraw (5250,-3975) rectangle (7200,-4800);
\filldraw (5250,-2775) rectangle (7200,-3600);
\pgfsetlinewidth{+15\XFigu}
\pgfsetdash{}{+0pt}
\pgfsetfillcolor{white}
\filldraw  (6148,-4649) circle [radius=+100];
\filldraw  (6148,-3449) circle [radius=+100];
\pgfsetarrows{[line width=7.5\XFigu]}
\pgfsetarrowsstart{xfiga1}
\draw (7800,-4650)--(6300,-4650);
\draw (9225,-4650)--(8850,-4650);
\draw (8850,-4500) rectangle (7800,-4800);
\draw (6150,-4575)--(6150,-4275);
\pgfsetfillcolor{black}
\pgftext[base,at=\pgfqpointxy{6150}{-5025}] {\fontsize{12}{14.4}\normalfont AWGN($N_w/2$)}
\pgftext[base,at=\pgfqpointxy{7500}{-4575}] {\fontsize{12}{14.4}\normalfont $Z(t)$}
\pgftext[base,at=\pgfqpointxy{8325}{-4725}] {\fontsize{12}{14.4}\normalfont $\Phi$}
\pgftext[base,left,at=\pgfqpointxy{9225}{-4725}] {\fontsize{12}{14.4}\normalfont $\widehat{\Lambda}$}
\pgftext[base,at=\pgfqpointxy{6150}{-4200}] {\fontsize{12}{14.4}\normalfont $W(t)$}
\draw (6000,-4650)--(4950,-4650);
\draw (6000,-3450)--(4500,-3450);
\pgfsetarrowsstart{}
\draw (4950,-4650)--(4950,-3450);
\pgfsetarrowsstart{xfiga1}
\draw (7800,-3450)--(6300,-3450);
\pgfsetarrowsend{xfiga1}
\draw (4050,-3225)--(4050,-2700)--(8100,-2700)--(8100,-3300);
\pgfsetarrowsend{}
\draw (3525,-3600)--(3075,-3600);
\draw (3525,-3300)--(3075,-3300);
\draw (9225,-3450)--(8850,-3450);
\draw (8850,-3300) rectangle (7800,-3600);
\draw (4500,-3225) rectangle (3525,-3675);
\draw (6150,-3375)--(6150,-3075);
\pgftext[base,at=\pgfqpointxy{6150}{-3825}] {\fontsize{12}{14.4}\normalfont AWGN($N_b/2$)}
\pgftext[base,at=\pgfqpointxy{4875}{-3375}] {\fontsize{12}{14.4}\normalfont $X(t)$}
\pgftext[base,at=\pgfqpointxy{7500}{-3375}] {\fontsize{12}{14.4}\normalfont $Y(t)$}
\pgftext[base,at=\pgfqpointxy{3975}{-3525}] {\fontsize{12}{14.4}\normalfont $\Psi$}
\pgftext[base,at=\pgfqpointxy{8325}{-3525}] {\fontsize{12}{14.4}\normalfont $\Gamma$}
\pgftext[base,left,at=\pgfqpointxy{9225}{-3525}] {\fontsize{12}{14.4}\normalfont $\widehat{M}$}
\pgftext[base,right,at=\pgfqpointxy{3075}{-3375}] {\fontsize{12}{14.4}\normalfont $M$}
\pgftext[base,right,at=\pgfqpointxy{3075}{-3675}] {\fontsize{12}{14.4}\normalfont $\Lambda$}
\pgftext[base,at=\pgfqpointxy{6150}{-3000}] {\fontsize{12}{14.4}\normalfont $B(t)$}
\pgftext[base,at=\pgfqpointxy{6150}{-2625}] {\fontsize{12}{14.4}\normalfont $S$}
\pgfsetarrowsstart{}
\draw (6150,-3400)--(6150,-3500);
\draw (6100,-3450)--(6200,-3450);
\draw (6150,-4600)--(6150,-4700);
\draw (6100,-4650)--(6200,-4650);
\endtikzpicture}%

%% file: main.bbl
\begin{thebibliography}{10}
\providecommand{\url}[1]{#1}
\csname url@samestyle\endcsname
\providecommand{\newblock}{\relax}
\providecommand{\bibinfo}[2]{#2}
\providecommand{\BIBentrySTDinterwordspacing}{\spaceskip=0pt\relax}
\providecommand{\BIBentryALTinterwordstretchfactor}{4}
\providecommand{\BIBentryALTinterwordspacing}{\spaceskip=\fontdimen2\font plus
\BIBentryALTinterwordstretchfactor\fontdimen3\font minus
  \fontdimen4\font\relax}
\providecommand{\BIBforeignlanguage}[2]{{%
\expandafter\ifx\csname l@#1\endcsname\relax
\typeout{** WARNING: IEEEtran.bst: No hyphenation pattern has been}%
\typeout{** loaded for the language `#1'. Using the pattern for}%
\typeout{** the default language instead.}%
\else
\language=\csname l@#1\endcsname
\fi
#2}}
\providecommand{\BIBdecl}{\relax}
\BIBdecl

\bibitem{bash2013limits}
B.~A. Bash, D.~Goeckel, and D.~Towsley, ``Limits of reliable communication with
  low probability of detection on {AWGN} channels,'' \emph{IEEE J. Sel. Areas
  Commun.}, vol.~31, no.~9, pp. 1921--1930, 2013.

\bibitem{CheBJ:13}
P.~H. Che, M.~Bakshi, and S.~Jaggi, ``Reliable deniable communication: Hiding
  messages in noise,'' in \emph{Proc. IEEE Int. Symp. Inf. Theory}, 2013, pp.
  2945--2949.

\bibitem{7407378}
M.~R. Bloch, ``Covert communication over noisy channels: A resolvability
  perspective,'' \emph{IEEE Trans. Inf. Theory}, vol.~62, no.~5, pp.
  2334--2354, May 2016.

\bibitem{wang2016fundamental}
L.~Wang, G.~W. Wornell, and L.~Zheng, ``Fundamental limits of communication
  with low probability of detection,'' \emph{IEEE Trans. Inf. Theory}, vol.~62,
  no.~6, pp. 3493--3503, 2016.

\bibitem{tahmasbi2018first}
M.~Tahmasbi and M.~R. Bloch, ``First-and second-order asymptotics in covert
  communication,'' \emph{IEEE Trans. Inf. Theory}, vol.~65, no.~4, pp.
  2190--2212, 2018.

\bibitem{arumugam2018covert}
K.~S.~K. Arumugam and M.~R. Bloch, ``Covert communication over a {K}-user
  multiple-access channel,'' \emph{IEEE Trans. Inf. Theory}, vol.~65, no.~11,
  pp. 7020--7044, 2019.

\bibitem{arumugam2019embedding}
------, ``Embedding covert information in broadcast communications,''
  \emph{IEEE Trans. Inf. Forensics Secur.}, vol.~14, no.~10, pp. 2787--2801,
  2019.

\bibitem{tan2018time}
V.~Y. Tan and S.-H. Lee, ``Time-division is optimal for covert communication
  over some broadcast channels,'' \emph{IEEE Trans. Inf. Forensic Secur.},
  2018.

\bibitem{kibloff2019embedding}
D.~Kibloff, S.~Perlaza, and L.~Wang, ``Embedding covert information on a given
  broadcast code,'' in \emph{Proc. IEEE Int. Symp. Inf. Theory}, 2019.

\bibitem{lee2018covert}
S.-H. Lee, L.~Wang, A.~Khisti, and G.~W. Wornell, ``Covert communication with
  channel-state information at the transmitter,'' \emph{IEEE Trans. Inf.
  Forensics Secur.}, vol.~13, no.~9, pp. 2310--2319, 2018.

\bibitem{zivari2019keyless}
H.~Zivari-Fard, M.~Bloch, and A.~Nosratinia, ``Keyless covert communication in
  the presence of non-causal channel state information,'' in \emph{Proc. IEEE
  Inf. Theory Workshop}, 2019.

\bibitem{compound}
S.~Salehkalaibar, Y.~M., and V.~Tan, ``Covert communication over a compound
  channel,'' \emph{ArXiv preprint arXiv:1906.06675}, 2018.

\bibitem{zhang2018covert}
Q.~Zhang, M.~Bakshi, and S.~Jaggi, ``Covert communication over adversarially
  jammed channels,'' in \emph{Proc. IEEE Inf. Theory Workshop (ITW)}, 2018, pp.
  1--5.

\bibitem{sobers2017covert}
T.~V. Sobers, B.~A. Bash, S.~Guha, D.~Towsley, and D.~Goeckel, ``Covert
  communications on continuous-time channels in the presence of jamming,'' in
  \emph{Proc. 51st Asilomar Conf. Signals, Systems and Computers}, 2017, pp.
  1697--1701.

\bibitem{wang2018covert}
L.~Wang, ``On covert communication over infinite-bandwidth gaussian channels,''
  in \emph{IEEE 19th International Workshop on Signal Processing Advances in
  Wireless Communications (SPAWC)}, 2018, pp. 1--5.

\bibitem{wang2019gaussian}
------, ``On gaussian covert communication in continuous time,'' \emph{EURASIP
  Journal on Wireless Communications and Networking}, vol. 2019, no.~1, pp.
  1--10, 2019.

\bibitem{wang2018continuous}
------, ``The continuous-time poisson channel has infinite covert communication
  capacity,'' in \emph{Proc. IEEE Int. Symp. Inf. Theory}, 2018.

\bibitem{gallager2008principles}
R.~G. Gallager, \emph{Principles of digital communication}.\hskip 1em plus
  0.5em minus 0.4em\relax Cambridge University Press, 2008, vol.~1.

\bibitem{Zhang_supp}
\BIBentryALTinterwordspacing
Q.~Zhang, M.~R. Bloch, M.~Bakshi, and S.~Jaggi, ``A supplemental document for
  covert communication over continuous-time {AWGN} channels,'' 2019. [Online].
  Available: \url{http://personal.ie.cuhk.edu.hk/~zq015/supplement.pdf}
\BIBentrySTDinterwordspacing

\bibitem{lapidoth2009foundation}
A.~Lapidoth, \emph{A foundation in digital communication}.\hskip 1em plus 0.5em
  minus 0.4em\relax Cambridge University Press, 2009.

\bibitem{lehmann2006testing}
E.~L. Lehmann and J.~P. Romano, \emph{Testing statistical hypotheses}.\hskip
  1em plus 0.5em minus 0.4em\relax New York: Springer, 2005.

\bibitem{cuff2013distributed}
P.~Cuff, ``Distributed channel synthesis,'' \emph{IEEE Trans. Inf. Theory},
  vol.~59, no.~11, pp. 7071--7096, 2013.

\bibitem{slepian1961prolate}
D.~Slepian and H.~O. Pollak, ``Prolate spheroidal wave functions, {F}ourier
  analysis and uncertainty—{I},'' \emph{Bell System Technical Journal},
  vol.~40, no.~1, pp. 43--63, 1961.

\bibitem{hayashi2006general}
M.~Hayashi, ``General nonasymptotic and asymptotic formulas in channel
  resolvability and identification capacity and their application to the
  wiretap channel,'' \emph{IEEE Transactions on Information Theory}, vol.~52,
  no.~4, pp. 1562--1575, 2006.

\bibitem{hou_thesis}
H.~Jie, ``Coding for relay networks and effective secrecy for wire-tap
  channels,'' \emph{\it Ph.D. dissertation, Technischen Universit\"at
  M\"unchen}, 2014.

\bibitem{gallager1968information}
R.~G. Gallager, \emph{Information theory and reliable communication}.\hskip 1em
  plus 0.5em minus 0.4em\relax Springer, 1968, vol.~2.

\end{thebibliography}
